%% file: ms.tex
\documentclass{aastex62}

\input{macro}
\usepackage[english]{babel}
\usepackage[utf8]{inputenc}
\usepackage{amsmath}
\usepackage{graphicx}
\usepackage[colorinlistoftodos,backgroundcolor=orange!20,textsize=small]{todonotes}

\begin{document}

\title{Candidate List of Edge-on Galaxies with Substantial Extraplanar Dust}

\author{Jong-Ho Shinn}
\affiliation{Korea Astronomy and Space Science Institute, 776 Daeduk-daero, Yuseong-gu, Daejeon, 305-348, the Republic of Korea}

\email{jhshinn@kasi.re.kr}

\begin{abstract}
We present a list of edge-on galaxies that might have substantial extraplanar dust.
Twenty-three edge-on galaxies were selected as target galaxies from an edge-on galaxy catalog, and their \galexfull{} far-ultraviolet images were fitted with three dimensional radiative transfer galaxy model.
The galaxy model is described by two disks: one for the light source and the other for the dust.
The best-fit parameters were found by employing a global optimization method, called differential evolution.
To find the galaxies with substantial extraplanar dust using the best-fit parameters, we plotted the ratio of scale-height to galactic diameter: $z_s/D_{25,ph}$ (light source) vs $z_d/D_{25,ph}$ (dust).
We found that 17 and 6 galaxies fall on the region of $(z_s/D_{25,ph}\times100)>0.2$ and $(z_s/D_{25,ph}\times100)<0.2$, respectively.
The former is named as ``high-group'' and the latter is named as ``low-group.''
We conclude that ``high-group'' is likely to be the galaxies with substantial extraplanar dust, while ``low-group'' is likely to be the ones with little extraplanar dust, i.e.~typical galactic thin disk, based on the following points: (1) the relative positions of ``high-group'' and ``low-group'' on the plot $z_s/D_{25,ph}$ vs $z_d/D_{25,ph}$ with respect to the reference values from optical radiative transfer studies; (2) the lower scale-height of the young stellar population than the old stellar population; and (3) a test result that shows the existence of extraplanar dust makes $z_s$ and $z_d$ overestimated in the fitting results.
We also examined the dependence of the group separation on the surface density of far-ultraviolet luminosity ($L_{FUV}/D^2_{25,ph}$), but found no strong dependence.
\end{abstract}

\keywords{dust, extinction --- galaxies: halos --- galaxies: spiral --- galaxies: structure --- radiative transfer}


\section{Introduction \label{intro}}
Galaxies evolve ever since their formation.
Stars are forming, aging, and dying, while mediums are being expelled, accreted, and circulated.
These mediums hold lots of information.
They reveal the energy flow (e.g.~galactic winds, \citealt{Veilleux_2005_ARA&A_43_769}), and transform the spectral shape of intrinsic stellar lights (see \citealt{Hollenbach_1999_RvMP_71_173} and  \citealt{Conroy_2013_ARA&A_51_393}).
Also, they themselves are the raw materials for the future star formation.
All these facts have close relations with the galactic energy distribution and hence the galaxy evolution.
Dust is one constituent of the mediums, and it plays important roles in many ways despite its mere $\sim$0.1\% occupation of the baryonic mass (\citealt{Blanton_2009_ARA&A_47_159}).

The extraplanar dust, located above the galactic plane, is of importance in diverse aspects as noted in \cite{Shinn_2015_ApJ_815_133}. 
It manifests that some mechanism makes the dust to be located away from the galactic plane defying the gravity.
It might also hold some hints on the connection between star-forming and star-bursting galaxies (see \citealt{Kennicutt_2012_ARA&A_50_531}) and between interstellar and intergalactic mediums (see \citealt{Meiksin_2009_RvMP_81_1405,Putman_2012_ARA&A_50_491}), since dust can move between the galactic plane and halo through the accretion and ejection processes.
Additionally, the extraplanar dust is worthy of meticulous study in the sense that the dust-scattered \Ha{} could substantially contaminate the extraplanar \Ha{} features \citep{Seon_2012_ApJ_758_109} which is  attributed to the diffuse ionized gas (see \citealt{Haffner_2009_RvMP_81_969}).
Many studies on the extraplanar dust, therefore, had been carried out at diverse wavelengths \citep[e.g.][]{Howk_1997_AJ_114_2463,Howk_1999_AJ_117_2077,Howk_1999_Ap&SS_269_293,Thompson_2004_AJ_128_662,Popescu_2004_A&A_414_45,Irwin_2006_A&A_445_123,Irwin_2007_A&A_474_461,Seon_2014_ApJ_785_L18,Hodges-Kluck_2014_ApJ_789_131,Shinn_2015_ApJ_815_133,Hodges-Kluck_2016_ApJ_833_58}.

To better understand how the extraplanar dust works in the context of galaxy evolution, three-dimensional physical modeling of the galaxy is useful as much as phenomenological analyses of observable quantities, given that one can extract more fundamental galactic parameters such as scale-height.
\cite{Seon_2014_ApJ_785_L18} had first attempted this approach focusing on the extraplanar dust, and successfully modeled the dust-scattered ultraviolet halo of an edge-on galaxy, NGC 891.
They used a galaxy model including a geometrically-thick dust disk in addition to the thin disk, and obtained several galactic parameters such as star formation rate, dust scale-height, face-on dust optical depth, light source scale-height, etc.
In order to determine whether there is any relation between the properties of extraplanar dust and host galaxy, 
\cite{Shinn_2015_ApJ_815_133} extended the approach of \cite{Seon_2014_ApJ_785_L18} to six highly-inclined galaxies that were reported to have ultraviolet halos \citep{Hodges-Kluck_2014_ApJ_789_131}.
However, \cite{Shinn_2015_ApJ_815_133} found that the additional geometrically-thick dust disk might not be required for three out of those six galaxies to reproduce the observed ultraviolet halos.

This finding suggests that the visual identification of galactic ultraviolet halos, without the aid of three-dimensional physical modeling, is less reliable in judging the existence of substantial extraplanar dust.
Obviously, a reliable target list is crucial for the investigation of the extraplanar dust.
Here we present a list of edge-on galaxies that seem to have substantial extraplanar dust, as a preparation for the future study on their extraplanar dust.
The list was produced from the edge-on disk galaxy catalog of \cite{Bizyaev_2014_ApJ_787_24} by performing model fittings to ultraviolet images from \galexfull{} (\galex).
We found that 17 out of 23 target galaxies seem to have substantial extraplanar dust.

\section{Data and Target Selection \label{data-tg}} 
\galex{} ultraviolet images were used for our study on the galactic extraplanar dust as in \cite{Shinn_2015_ApJ_815_133}.
Ultraviolet wavelength has an advantage in exposing the extraplanar dust through scattering processes, since the dominant ultraviolet light sources---massive stars---reside in the galactic plane with a \emph{smaller} scale-height than the old stellar population \citep{Bahcall_1980_ApJS_44_73,Wainscoat_1992_ApJS_83_111,Martig_2014_MNRAS_442_2474}.
Indeed, several studies using ultraviolet data have claimed the detection of dust-scattered ultraviolet halos around edge-on galaxies \citep{Seon_2014_ApJ_785_L18,Hodges-Kluck_2014_ApJ_789_131,Shinn_2015_ApJ_815_133,Hodges-Kluck_2016_ApJ_833_58}.
Besides, one of the studies provided the galactic parameters that excellently reproduce a panchromatic galactic spectral energy distribution \citep{Baes_2016_A&A_587_A86}.

\galex{} provides wide-field ($\sim1.25\degr$) images at two ultraviolet bands (FUV: 1344 -- 1786 \AA{}; NUV: 1771 -- 2831 \AA) with imaging resolutions of $\sim4''-5''$ \citep{Morrissey_2007_ApJS_173_682}.
It covered almost all sky at both bands except the Galactic plane \citep{Bianchi_2014_AdSpR_53_900}, hence GALEX data are well-suited for statistical studies of a large sample.
We used the archival data of GR6/7 (\url{http://galex.stsci.edu/GR6/}) and no additional data reduction procedure was applied except cropping and masking.
The standard pipeline procedures are described in \cite{Morrissey_2007_ApJS_173_682}.
Only the FUV images were analyzed for several reasons as mentioned in \cite{Shinn_2015_ApJ_815_133}: narrower bandwidth, smoother albedo variation over the bandwidth, and fewer foreground point sources.

We chose our target galaxies from the edge-on galaxy catalog of \cite{Bizyaev_2014_ApJ_787_24}.
The catalog is based on the optical images from the Seventh Data Release (DR7; \citealt{Abazajian_2009_ApJS_182_543}) of \sdssfull{} (\sdss; \citealt{York_2000_AJ_120_1579}), and the genuine edge-on galaxies were selected through both automatic and visual inspections.
The catalog lists 5794 edge-on galaxies as of November 2014, along with their structural parameters for stellar disks derived from 1D (ignoring dust extinction) and 3D (including dust absorption only) analyses, respectively.

We first narrowed down the catalog to 119 target galaxies by applying the following constraints one after another.
To select galaxies with large enough angular sizes, the scale-length from the 1D analysis is limited to be $\geq 10''$ in any of three bands ($g$, $r$, $i$). 
We then secured high signal-to-noise ratios (S/Ns) by limiting the \galex{} FUV exposure time to be $\geq 1000$ s.
The distance to the galaxies are essential for our analysis, therefore we excluded the galaxies if their distance information is not available from the NASA/IPAC Extragalactic Database (\url{https://ned.ipac.caltech.edu/}).
Finally, we excluded those galaxies that show ring features in their \galex{} FUV images or fall on the edge of the \galex{} field-of-view.
 
In this study, the existence of extraplanar dust is revealed by the galactic off-plane lights which are much weaker than the galactic in-plane lights.
This means that the higher S/N data give us the more reliable information about the extraplanar dust.
In this sense, we winnowed once again the 119 targets to the final 23 targets that show S/N per pixel $\geq3$ \replaced{at all pixels the galaxy covers}{continuously along the galactic disk}.
The S/N was solely calculated from the photon count itself.
\added{We checked if any artifacts from the instrumental scattered light mentioned in \cite{Hodges-Kluck_2014_ApJ_789_131} affect our target galaxy images, and found none. It seems to be due to the lower sensitivity of the \galex{} FUV band than the NUV \citep{Morrissey_2007_ApJS_173_682} and the rarity of the stars bright in the FUV band.}
Fig.~\ref{fig-tg} shows the \galex{} FUV and \sdss{} images of 23 target galaxies, and Table \ref{tbl-tg} lists the relevant target information.

\section{Analysis and Results \label{ana-res}}
The purpose of this study is to search for the galaxies that are likely to have substantial extraplanar dust.
To find these galaxies, we fitted the \galex{} images of the target galaxies (Fig.~\ref{fig-tg} and Table \ref{tbl-tg}) with a three dimensional radiative transfer galaxy model whose dust component is described by a \emph{single} disk.
Under this scheme, the galaxies with \emph{little} extraplanar dust would return typical scale-heights of light-source and dust.
On the other hand, those galaxies with \emph{substantial} extraplanar dust would return distinct scale-heights or systematic (not random) fitting residuals, or both, because the extraplanar dust would affect the radiative transfer of the stellar light.
This point is more elaborated in section \ref{ana-res-test} while mentioning Test C.
In the following, we describe the radiative transfer galaxy model and fitting procedure (section \ref{ana-res-model}); test fitting results (section \ref{ana-res-test}); and target fitting results (section \ref{ana-res-run}).

\subsection{Radiative Transfer Galaxy Model and Fitting Procedure \label{ana-res-model}}
We adopted a galaxy description for the fitting as below, which is equal to the one used in \cite{Shinn_2015_ApJ_815_133} except the additional thick dust disk we excluded here.
\begin{eqnarray}
\kappa(r,z)&=& 
\left\{ 
\begin{array}{ll} \label{eq-kappa}
\kappa_0\,\mathrm{exp}\left(-\frac{r}{h_d}-\frac{|z|}{z_d} \right),\,\mathrm{for}\, r\le R_d \\
0,\,\mathrm{for}\, r>R_d
\end{array} 
\right. \\
I(r,z)&=&
\left\{
\begin{array}{ll} \label{eq-src}
I_0\,\mathrm{exp}\left(-\frac{r}{h_s}-\frac{|z|}{z_s} \right),\,\mathrm{for}\, r\le R_s \\
0,\,\mathrm{for}\, r>R_s 
\end{array}
\right.
\end{eqnarray}
Here, $\kappa(r,z)$ and $I(r,z)$ are the distributions of extinction coefficient and light source, respectively.
$h_d$, $h_s$, $z_d$, and $z_s$ are the corresponding radial scale-length and vertical scale-height of dust and light source.
$\kappa_0$ is the extinction coefficient at the center of the galaxy ($r=0$, $z=0$), where the optical depth along the symmetric axis ($\tau_{FUV}$) is $2\,\kappa_0\,z_d$.
$I_0$ is the light-source density at the center of the galaxy ($r=0$, $z=0$)\deleted{, which is proportional to the galactic luminosity ($L_{FUV}$)}.
\added{The four parameters ($I_0$, $h_s$, $z_s$, and $R_s$) determine the galactic luminosity ($L_{FUV}$).}
$R_d$ and $R_s$ are the truncation radii of dust and light source, respectively.
The radiative transfer calculations were performed at the wavelength of 1538.6 \AA{}, i.e.~the effective wavelength of \galex{} FUV band \citep{Morrissey_2007_ApJS_173_682} using the three dimensional Monte-Carlo method \citep[see][]{Steinacker_2013_ARA&A_51_63}.
Basically, the model produces a two dimensional synthetic image toward the observer, by calculating how many photons arrive at the observer after traveling the three dimensional model galaxy described by eqs.~(\ref{eq-kappa}) and (\ref{eq-src}), while being scattered and absorbed by dust.
Readers are referred to \cite{Seon_2014_ApJ_785_L18} and \cite{Shinn_2015_ApJ_815_133} for more details on the radiative transfer calculations.

In contrast to our previous work \citep{Shinn_2015_ApJ_815_133}, we neither rotated the \galex{} image to align the major axis of the target galaxy to horizontal direction, nor subtracted the background from the \galex{} image before the fitting.
We instead enabled the galaxy model to rotate during the fitting process and treated the background as free parameters.
The background was also set to have linear gradients along the horizontal and vertical directions of the \galex{} image.
The model galaxy was free to move in both directions during the fitting.
In summary, six parameters were newly added: the position angle of the galaxy ($\theta_{pos}$), horizontal and vertical shifts of the galaxy center at the image plane ($\Delta x,\,\Delta y$), the background level at the center of the image ($bg_{ctr}$), and the horizontal and vertical gradients of the background at the center of the image ($\partial bg/\partial x,\,\partial bg/\partial y$).
Now we have 15 free parameters to fit---eight parameters are for the model galaxy itself ($L_{FUV},\,\tau_{FUV},\,h_d,\,z_d,\,R_d,\,h_s,\,z_s,\,R_s$) and the other seven parameters are for the viewpoint and the background ($\theta_{incl},\,\theta_{pos},\,\Delta x,\,\Delta y,\,bg_{ctr},\,\partial bg/\partial x,\,\partial bg/\partial y$); $\theta_{incl}$ is the inclination angle of model galaxy.

When generating the model image, the spatial convolution procedure was included at the last stage to match the spatial resolutions of the model image and the \galex{} image.
We used a new custom point-spread-function (PSF).
This PSF is based on the \galex{} PSF, but extended up to the PSF image size of $600''$ in the same way as \cite{Shinn_2015_ApJ_815_133} did\added{: the part where the radial distance $\gtrsim50''$ was extrapolated from a functional fit}.
We adopted this extended PSF to include the instrumental effects that might vertically broaden the galactic plane emission (see \citealt{Sandin_2014_A&A_567_A97,Sandin_2015_A&A_577_A106}).
The convolution was carried out through the Fourier transformation for speed.

We prepared the images for the fitting as follows.
The fitting area was manually set not to be overwhelmed by the background area, since, if not, the goodness of fit would be primarily determined by the background area.
Bright point sources and uninterested extended features were also masked.
The white areas around the galaxy images (Fig.~\ref{fig-test_a}-\ref{fig-fitdata_en}) indicate the excluded regions during the model fitting.

We found the best-fit model by maximizing the likelihood $\mathcal{L}(M|D)$, where $M$ and $D$ stand for the 15 model parameters and the \galex{} FUV image, respectively.
Poisson distribution was adopted as the parent probability distribution for each image pixel, because the photon counts per pixel of \galex{} image is too small to be treated as Gaussian (see the \galex{} count image in Fig.~\ref{fig-fitdata_st}-\ref{fig-fitdata_en}).
Therefore, the likelihood was calculated from the images in \emph{count} units (not \emph{count per sec}): the \galex{} FUV \emph{count} image and the model image which is converted into \emph{count} units using the \galex{} relative response image (see \citealt{Morrissey_2007_ApJS_173_682}).
In this case, maximizing the likelihood corresponds to minimizing the $C$-statistic \citep{Cash_1979_ApJ_228_939}, not the $\chi^2$-statistic: $C=-2\,\Sigma_{i}(d_i \ln m_i - m_i - \ln d_i !)$, where $m_i$ and $d_i$ are the pixel values of the model and data images, respectively.

Since it is highly uncertain if the likelihood has local maxima or not, partly due to the complex nature of radiative transfer and the large number of free parameters, we employed a global optimization method called Differential Evolution \citep[DE;][]{Storn_1997_J.GlobalOptim._11_341}.
This technique finds the global maximum by making numerous parameter vectors search the likelihood surface through iterations as follows.
Suppose we have $D$ number of model parameters, and they consist of the parameter vector $\mathbf{x}$ = ($p_1$, $p_2$, ... , $p_D$). 
(1) $N\!P$ number of parameter vectors are generated over some reasonable parameter ranges\added{: each parameter of a vector $\mathbf{x}$ is \emph{randomly-and-independently} generated}. That is, we have the vector population [$\mathbf{x}_1$, $\mathbf{x}_2$, ... , $\mathbf{x}_{N\!P}$].
This population is called the 1st generation, say $\mathbb{X}^{(1)}$ = [$\mathbf{x}^{(1)}_1$, $\mathbf{x}^{(1)}_2$, ... , $\mathbf{x}^{(1)}_{N\!P}$].
(2) Among the vector population, one vector is set as the target vector.
(3) Three vectors are \emph{randomly} chosen from the rest of the parameter vectors.
These three are mutated \emph{parameter-wise} to be the mutant vector\added{: that is, $\mathbf{x_{mt}}=\mathbf{x_{1st}}+F\times(\mathbf{x_{2nd}-x_{3rd}})$ where $F$ is an arbitrary constant}.
(4) The mutant vector becomes the trial vector through the ``crossover'' process, where some of the parameters being replaced with the target vector's.\added{ Whether the parameter is replaced or not is determined by a random number which ranges from 0 to 1. When the random number for each parameter is greater than $C\!R$, an arbitrary constant, the corresponding parameter is replaced with the target vector's.}
(5) The two likelihoods are respectively calculated with the target vector and the trial vector, and then compared each other.
If the trial vector's likelihood is greater than that of the target vector, the target vector is replaced with the trial vector in the next generation; if not, the target vector remains in the next generation.
(6) A new target vector is set. It is chosen from the vectors in the current generation that has not been the target vector before.
(7) The processes (3)-(6) are repeated until all vectors have been the target vector.
(8) Then, we have the 2nd generation of the vector population $\mathbb{X}^{(2)}$.
(9) With the 2nd generation, repeat the processes (2)-(7), and then we have the 3rd generation $\mathbb{X}^{(3)}$, and so on.
(10) When each parameter of the vector population converges to a certain range, then the best-fit vector is selected among the population as the one that shows the highest likelihood value.

During the fitting, we impose some constraints as follows.
In the sequence (1), we randomly sampled the scale-length and scale-height parameters so that they are uniformly distributed in log scale, while sampling the other parameters to be uniform in linear scale.
In the sequence (4), we put the parameters back to the initial parameter range limits, when the trial vector exceeds the limits.
Also, we set a constraint of $R_s \leq R_d$, in order to avoid the case that the light source is situated on dust-free space, which seems unlikely for massive stars (the dominant UV sources). 

We show the results of DE fitting in the next two sub-sections.
In section \ref{ana-res-test}, we produce mock observation images and demonstrate how well the DE fitting method reproduces the model input values that are used for making the mock images.
The fitting results for the \galex{} images follow in section \ref{ana-res-run}.

\subsection{Test: Fitting the Mock Observation Images \label{ana-res-test}}
We employed the DE method to find the best-fit parameters of the galaxy model.
The DE method has three control variables ($N\!P, F, C\!R$) that should be adjusted by trial and error.
$N\!P$ is the number of parameter vectors, $F$ is a factor that scales the parameter difference when the mutant vector is created, and $C\!R$ is the crossover constant which determines how many parameters of the mutant vector will be replaced with the ones of the target vector when the trial vector is created (\citealt{Storn_1997_J.GlobalOptim._11_341}); these three variables are related with the sequences (1), (3), and (4) mentioned in section \ref{ana-res-model}, respectively.
To determine these variables and see if the DE method works properly for our study, we ran several test fittings to the mock observation images and checked how much the obtained parameters are close to the model input values.
The mock images were produced from the model described in eqs.~(\ref{eq-kappa}) and (\ref{eq-src}) with $5 \times10^8$ photons, and the background emission was set to zero for simplicity.
The same three dimensional model-grid was used for both producing and fitting the mock observation images.
The number of grid cells were (radial, azimuthal, vertical)=(30, 1, 61) in cylindrical coordinates, and the cell size was set to scale exponentially from the galaxy center to cover a larger space as being further away from the center.

Three different tests were carried out and the results are shown in Fig.~\ref{fig-test_a}-\ref{fig-test_c}.
We adopted $N\!P=90$, $F=0.5$, and $C\!R=0.9$, and obtained a converged population within 500 iterations.
Iterations of 250 took about $1.5-2.0$ hrs with 88 CPUs; the CPU model is the Intel(R) Xeon(R) CPU E5-2699 v4 @ 2.20GHz.
During the fitting, each model image was produced with $5\times10^6$ photons, much less than used for producing the mock image, for speed.

Fig.~\ref{fig-test_a} displays the results of Test A.
Test A represents an edge-on galaxy that shows an absorption feature along the galactic plane.
To show the goodness of fit more intuitively, we show the $\chi^2$ image rather than the $C$-statistic image.
The uncertainty of each pixel was set as $1+\sqrt[]{count+0.75}$, the 1-$\sigma$ upper limit for small number events \citep{Gehrels_1986_ApJ_303_336}, rather than the conventional $\sqrt[]{count}$ in order to reflect the small counts of \galex{} image.
As the results show, the DE method well reproduces the model input values.
$R_d$ is the least converging parameter as its relatively broader population distribution shows at the end of the iteration.

Fig.~\ref{fig-test_b} shows the results of Test B, which represents an edge-on galaxy \emph{without} an absorption feature along the galactic plane.
In this test, again, the DE method well reproduce the model input values, although the inclination angle had been confused at the beginning of the iteration.
The confusion might be related with the absence of the absorption feature at the galactic plane, which hints the direction of galaxy inclination.
$R_d$ is again the least converging parameter.

Fig.~\ref{fig-test_c} displays the results of Test C.
We performed this test in order to see how the fitting results might look like when the target has substantial extraplanar dust.
To simulate the existence of substantial extraplanar dust, we added a second dust disk as adopted in \cite{Seon_2014_ApJ_785_L18} and \cite{Shinn_2015_ApJ_815_133}.
Note that a spherical dust halo was adopted for the extraplanar dust in \cite{Hodges-Kluck_2014_ApJ_789_131}.
In summary, Test C demonstrates how the best-fit parameters turn out when \emph{two dust disks (thin and thick)} exists but the data is fitted with the conventional \emph{one-component dust disk} like our approach in this work.

The $\chi^2$ image of Fig.~\ref{fig-test_c} shows that there is a systematic variation of $\chi^2$ along the galactic plane with higher $\chi^2$ values than those of Test A and Test B.
Note that $z_d$ converges to the value between the two input values (dust scale-heights for thin and thick disks) and that $z_s$ converges to the value higher than the input value (light-source scale-height).
This means that the galaxies with substantial extraplanar dust would return higher $z_d$ and $z_s$ values than the typical ones.
Test C results also show that other parameters such as $R_d$, $\theta_{incl}$, and $\Delta y$ might also be affected by the existence of substantial extraplanar dust.

\subsection{Fitting the \galex{} Images \label{ana-res-run}}
Keeping the results of Test C (section \ref{ana-res-test}) in mind, we performed the model fitting to the \galex{} FUV images of the 23 target galaxies (Fig.~\ref{fig-tg} and Table \ref{tbl-tg}) in order to select the candidate galaxies with substantial extraplanar dust.
The control variables of DE method were adopted as $N\!P=113$, $F=0.5$, and $C\!R=0.9$.
Only $N\!P$ was changed from the test ones, from 90 to 113, to properly handle the increased number of fitting parameters; the three background parameters ($bg_{ctr}$, $\partial bg/\partial x$, and $\partial bg/\partial y$) were additionally set as free parameters.
The configuration of three dimensional model grid was the same with the test one (section \ref{ana-res-test}), and its physical size was set as that of the \galex{} image's diagonal (say, tens of kpc).

We \replaced{performed three independent fitting runs}{ran the fitting three times with different 1st generation parameter vectors,} and found three \deleted{independent }best-fit parameter sets.
Figures \ref{fig-fitdata_st}-\ref{fig-fitdata_en} show one set of the results.
On average, the iteration of 250 took about $2-3$ hrs, a little longer than the test cases (section \ref{ana-res-test}); when the photon has to travel through a denser dust medium (e.g.~higher $\tau_{FUV}$, $h_d$, and $z_d$), the run time was doubled or more.
For most of the target galaxies, the fitting parameters were converged within 500 iterations, but some targets needed $>1000$ iterations to converge.
The $\chi^2$ images were made in the same way with the test cases (Fig.~\ref{fig-test_a}-\ref{fig-test_c}), and the region of $\chi^2<3$ is dominant in most of the targets.
However, some targets (e.g.~EON\_24.788\_-10.504, EON\_163.623\_17.344) show several regions with high residuals, which seems to be due to bright (or dark) local features.
Table \ref{tbl-bestfit} shows the best model parameters with errors.
The mean and standard deviation of the three best-fit parameter sets are quoted as the best model parameters and their errors, respectively, in a similar way to \cite{DeGeyter_2013_A&A_550_A74} and \cite{DeGeyter_2014_MNRAS_441_869}.

Fig.~\ref{fig-zh} shows how light sources and dust distribute in terms of scale-heights and scale-lengths.
It shows that $z_s$ is smaller than $z_d$ for all targets except EON\_20.349\_-1.863 and EON\_189.100\_40.005, while $h_s$ and $h_d$ are almost randomly distributed on either sides of the one-to-one line.
\added{Large $h_s$ and $h_d$ ($>100$ kpc) mean the relatively uniform distribution of light source and dust along the radial direction.}

To find the candidate galaxies with substantial extraplanar dust, we plot the $z_s/D_{25,ph}$ vs $z_d/D_{25,ph}$ (Fig.~\ref{fig-zz}a).
If the galactic disk is the manifestation of the equilibrium between internal kinetic pressure and gravitational force (see \citealt{Spitzer_1942_ApJ_95_329}), the scale-height would be proportional to the galactic mass (and roughly to the galactic diameter).
Therefore, the spiral galaxies in dynamical equilibrium would fall into a certain region in the plot of $z_s/D_{25,ph}$ vs $z_d/D_{25,ph}$.
Under these circumstances, we expect that the galaxies with substantial extraplanar dust would fall on a distinct region from the stable spiral galaxies, as Test C (section \ref{ana-res-test}) shows that the existence of substantial extraplanar dust make $z_s$ and $z_d$ overestimated.

Our fitting results are plotted as circles in Fig.~\ref{fig-zz}a, and the circles show two distinct distributions.
One group is on the region $(z_s/D_{25,ph}\times100)>0.2$ (\emph{open circles}), while the other is on the region $(z_s/D_{25,ph}\times100)<0.2$ (\emph{filled circles}).
We name the former and the latter as ``high-group'' and ``low-group,'' respectively.
``high-group'' contains 17 galaxies, while ``low-group'' contains 6 galaxies.
To help the identification of these two groups, we plot together the values of other edge-on galaxies (\emph{gray sqaures}) obtained from optical radiative transfer studies (Table \ref{tbl-zD_ref}).
These \emph{gray squares} represent the galactic thin disks \citep[see][]{Freeman_2002_ARA&A_40_487}, and occupy the lower-right corner of Fig.~\ref{fig-zz}a.

In Fig.~\ref{fig-zz}b, we compare the $L_{FUV}/D^2_{25,ph}$ and $z_d/D_{25,ph}$ to see if the FUV luminosity plays any role in separating ``high-group'' and ``low-group,'' since the luminosity is one important physical parameter of the host galaxy.
The factor $D^2_{25,ph}$, proportional to the galactic disk area, is used for scaling the luminosity to compensate the size difference among galaxies.
As Fig.~\ref{fig-zz}b shows, no prominent difference of $L_{FUV}/D^2_{25,ph}$ exists between ``high-group'' and ``low-group,'' although ``high-group'' seems to have somewhat lower values.
This means that $L_{FUV}/D^2_{25,ph}$ is not sensitive in separating ``high-group'' and ``low-group.''

\section{Discussion \label{discu}}
As seen in Fig.~\ref{fig-zz}a, ``high-group'' and ``low-group'' occupy the right and left regions, respectively.
The reference values for comparison (\emph{gray squares}) are located at the lower-right corner.
These reference values are from optical radiative transfer studies (Table \ref{tbl-zD_ref}), and represent the galactic thin disk (see \citealt{Freeman_2002_ARA&A_40_487}); hence, they would help to identify ``high-group'' and ``low-group.''
As in the following two paragraphs, we conclude that ``high-group'' likely represents the galaxies with \emph{substantial} extraplanar dust, while ``low-group'' likely represents the ones with \emph{little} extraplanar dust.

``Low-group'' shows lower $z_s/D_{25,ph}$ than the reference values, while showing similar $z_d/D_{25,ph}$ to the reference values.
Considering that the reference values are from optical studies, the reference $z_s$ values represent the old stellar population.
On the other hand, our $z_s$ represent the young stellar population, since our $z_s$ are from ultraviolet data analyses.
As mentioned in section \ref{data-tg}, the young stellar population has a \emph{smaller} scale-height than the old stellar population \citep{Bahcall_1980_ApJS_44_73,Wainscoat_1992_ApJS_83_111,Martig_2014_MNRAS_442_2474}.
Therefore, smaller $z_s/D_{25,ph}$ of ``low-group'' than the reference values can be understood on the basis of stellar population difference.
Based on this explanation on $z_s/D_{25,ph}$, as well as the similarity of $z_d/D_{25,ph}$ between ``low-group'' and the reference values, we think that ``low-group'' likely represents the galactic thin disks, i.e. galaxies with little extraplanar dust.

``High-group'' shows similar $z_s/D_{25,ph}$ to the reference values, while showing similar or higher $z_d/D_{25,ph}$ than the reference values.
The $z_s$ of ``high-group'' comparable to the reference values is contradictory to the fact that the young stellar population has a lower scale-height than the old stellar population.
It is hard to imagine any dynamical process that can spread the young stellar population more away from the galactic plane as a factor of $\sim10$ than the typical thin disk (``low-group'').
The higher $z_d/D_{25,ph}$ than the reference values which some ``high-group'' galaxies show might simply be interpreted as the existence of substantial dust above the galactic plane.
In section \ref{ana-res-test}, Test C shows that the existence of substantial extraplanar dust make $z_s$ and $z_d$ overestimated in the fitting results.
Considering these test results as well as the contradictory $z_s/D_{25,ph}$ and the higher $z_d/D_{25,ph}$, we think that ``high-group'' is likely to be the galaxies with substantial extraplanar dust.
In other words, ``high-group'' is hard to be explained within the framework of galactic thin disk.

We additionally note that one target from the optical radiative transfer studies, NGC 4302, happens to be on our target list as EON\_185.427\_14.598.
This target belongs to ``high-group,'' and its optical study showed poor fitting results that remains big residual features at the bulge and halo regions \citep{Bianchi_2007_A&A_471_765}.
These large residuals might be caused by the substantial extraplanar dust in NGC 4302, rather than the additional stellar disk as suspected by \cite{Bianchi_2007_A&A_471_765}.

In Fig.~\ref{fig-zz}b, ``high-group'' and ``low-group'' show no stark difference along the axis of $L_{FUV}/D^2_{25,ph}$.
This means that $L_{FUV}/D^2_{25,ph}$ is not a good discriminator for the galaxies with substantial extraplanar dust.
Test C in section \ref{ana-res-test} shows a consistent result that the best-fit $L_{FUV}$ is not much different from the input value for the mock data even when substantial extraplanar dust exists.
For ``low-group'' which likely represents the galactic thin disk, we could examine any relation between $z_d/D_{25,ph}$ and $L_{FUV}/D^2_{25,ph}$.
However, the number of members in ``low-group'' (six) is too small to remark on any relation.
The same kind examination is improper for ``high-group.''
The fitting parameters of ``high-group'' were obtained from an \emph{improper} model (single dust disk) which does not include a description for the extraplanar dust.
Any relation between the properties of the extraplanar dust and the host galaxy must be studied with the model that includes a proper description for the extraplanar dust, such as a geometrically-thick dust disk assumed in \cite{Shinn_2015_ApJ_815_133} or a spherical dust halo assumed in \cite{Hodges-Kluck_2014_ApJ_789_131}.
Model selection (see \citealt{Sharma_2017_ARA&A_55_213}) would be required to determine which model is better at describing the extraplanar dust.
\added{Then, based on the better model, we can obtain the model parameters that characterize the extraplanar dust through the parameter estimation process.}

\section{Conclusions \label{concl}}
To list the candidate galaxies with substantial extraplanar dust, we fitted the \galex{} FUV images of edge-on galaxies using three dimensional radiative transfer model.
The DE method was employed to find a global maximum in the highly complex likelihood surface.
Twenty-three galaxies were selected as target galaxies from the edge-on galaxy catalog of \cite{Bizyaev_2014_ApJ_787_24}.
The best model parameters were analyzed by plotting the ratios of scale-height to galactic diameter (Fig.~\ref{fig-zz}a): $z_s/D_{25,ph}$ for light source and $z_d/D_{25,ph}$ for dust.
We found that 17 galaxies are populated at the region of $(z_s/D_{25,ph}\times100)>0.2$ (``high-group''), while the rest six (``low-group'') are populated at the region of $(z_s/D_{25,ph}\times100)<0.2$.
To help the identification of these two groups, we plotted the corresponding values of other galaxies from optical radiative transfer studies (Table \ref{tbl-zD_ref}).
These optical reference values represent the galactic thin disks and are populated at the lower-right corner of the plot $z_s/D_{25,ph}$ vs $z_d/D_{25,ph}$ (Fig.~\ref{fig-zz}a).

We conclude that ``low-group'' likely represents the galactic thin disk (i.e.~galaxies with little extraplanar dust), while ``high-group'' likely represents the ones with substantial extraplanar dust, based on the followings: (1) the positions of ``low-group'' and ``high-group'' relative to the optical reference values in the plot of $z_s/D_{25,ph}$ versus $z_d/D_{25,ph}$ (Fig.~\ref{fig-zz}a); (2) $z_s/D_{25,ph}$ from ultraviolet data analysis should be lower than $z_s/D_{25,ph}$ from optical data analysis, because the optical light sources are the old stellar population while the ultraviolet light sources are the young stellar population; and (3) Test C in section \ref{ana-res-test} shows that the existence of substantial extraplanar dust makes $z_s$ and $z_d$ overestimated in the fitting results.
The distributions of ``low-group'' and ``high-group'' along the luminosity values ($L_{FUV}/D^2_{25,ph}$) were also examined.
We found that the group separation does not sensitively depend on $L_{FUV}/D^2_{25,ph}$ values, which is also observed in the result of Test C (section \ref{ana-res-test}).

The candidate galaxies with substantial extraplanar dust we found (``high-group'') would be useful for studying the relation between the properties of the extraplanar dust and the host galaxy.
Prior to this study, a proper model well describing the extraplanar dust must be determined through model selection.

\acknowledgments
J.-H.S. appreciate the anonymous referee's comments that improve the manuscript substantially.
J.-H.S. is also grateful to Kwang-Il Seon for providing the galaxy model, to Yujin Yang for his helpful comments on the manuscript, and to Young-Dae Lee for his help in retrieving \galex{} data with a script.

\bibliographystyle{aasjournal}
\bibliography{extraplanar_dust_2017}

\input{ms_fig.tex}

\input{ms_tbl.tex}

\end{document}

%% file: macro.tex
\newcommand{\Ha}{H${\alpha}$}

\newcommand{\galex}{\textit{GALEX}}
\newcommand{\galexfull}{\textit{Galaxy Evolution Explorer}}
\newcommand{\sdss}{\textit{SDSS}}
\newcommand{\sdssfull}{\textit{Sloan Digital Sky Survey}}

%% file: ms_fig.tex

\input{fig/fig_target.tex}

\clearpage
\begin{figure}
\center{
\includegraphics[scale=0.45]{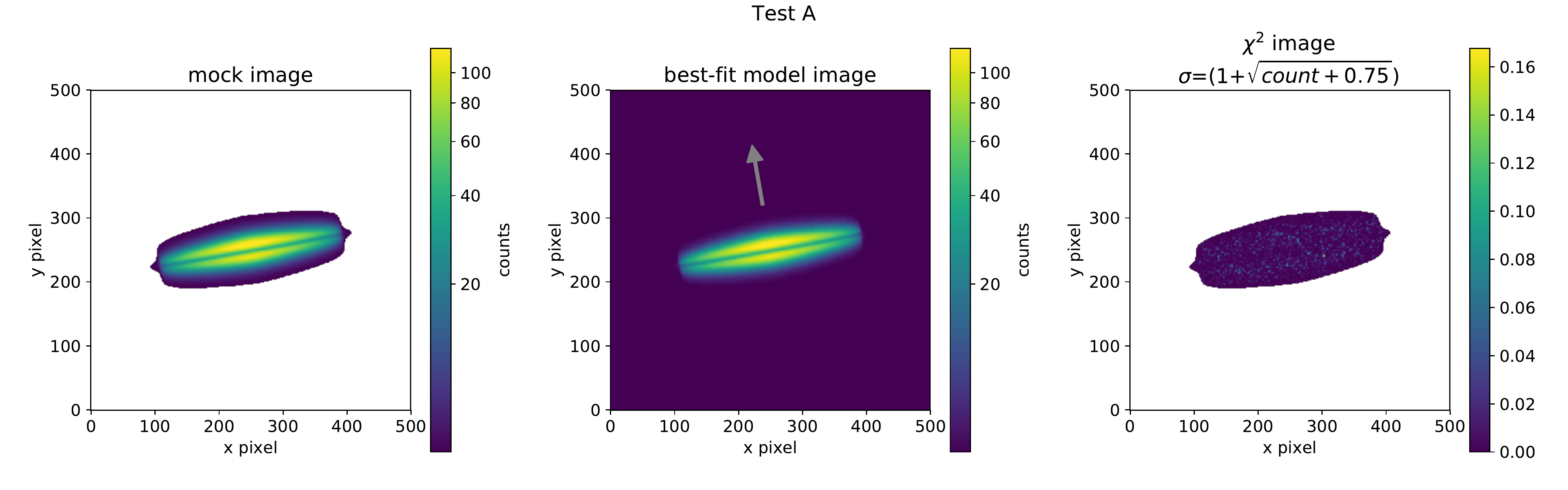}
\includegraphics[scale=0.45]{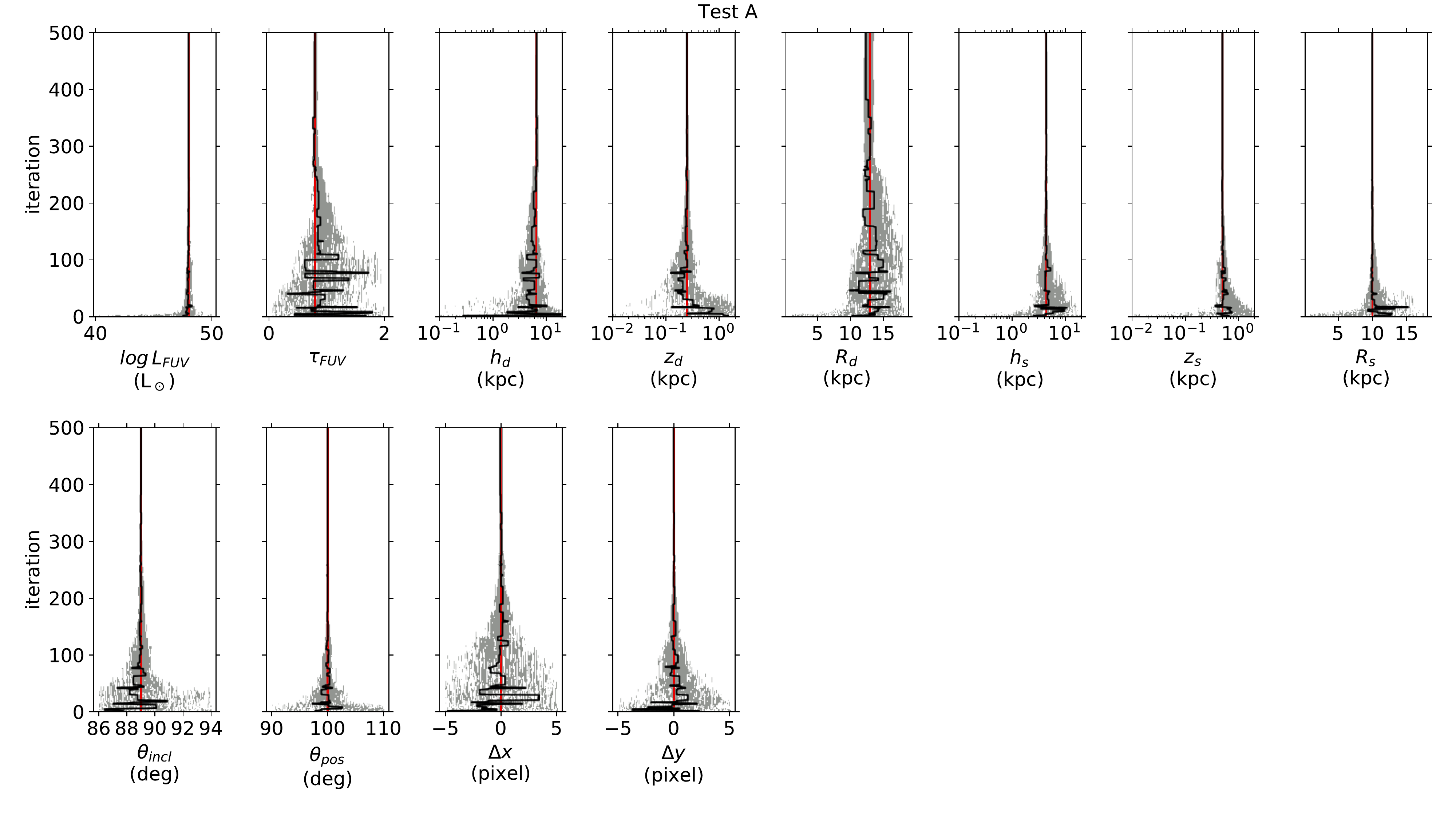}
}
\caption{The results of Test A. The three \emph{top panels} show the mock observation image, best-fit model image, and the goodness of fit image from left to right, respectively. See text for the definition of the goodness of fit image. The white areas are excluded from the model fitting.\added{ The gray arrow on the best-fit model image is to define the inclination angle which increases from the arrow to the direction into the page.} The \emph{middle and bottom panels} show the evolution of fitting parameters over the iteration. The eight \emph{middle panels} are for the intrinsic galactic structural parameters, while the four \emph{bottom panels} are for the viewpoints. The \emph{gray points} indicate the parameter values of the population. The \emph{black line} shows the evolution of the best-fit parameter along the iteration. The \emph{red vertical lines} indicate the model input values. } \label{fig-test_a}
\end{figure}

\clearpage
\begin{figure}
\center{
\includegraphics[scale=0.45]{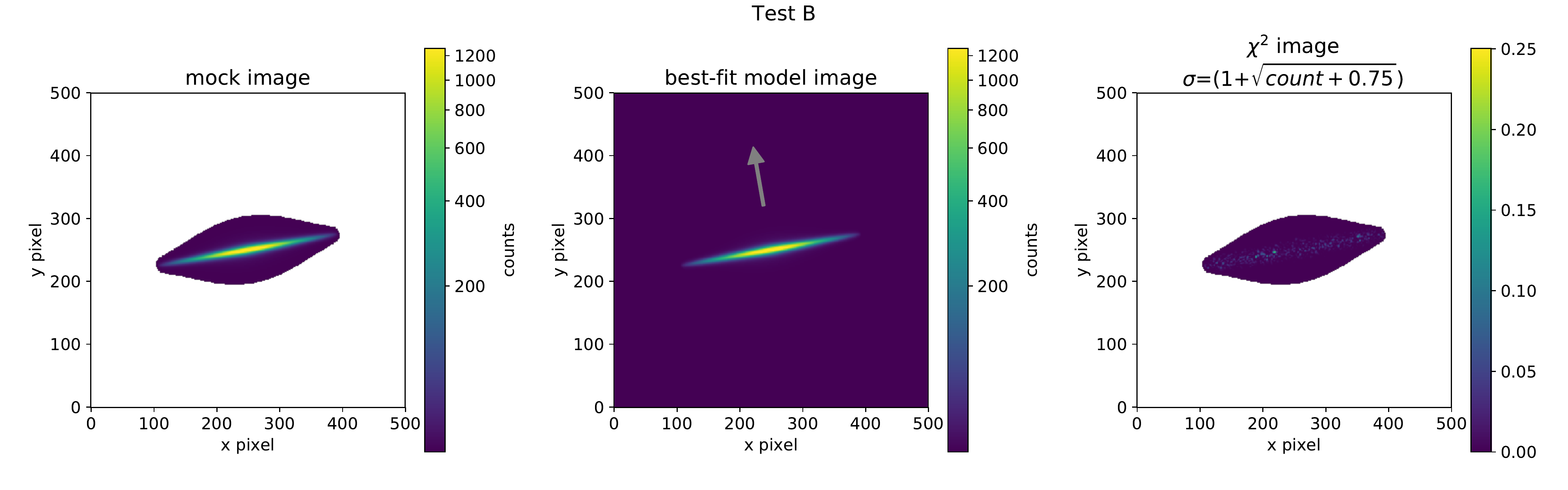}
\includegraphics[scale=0.45]{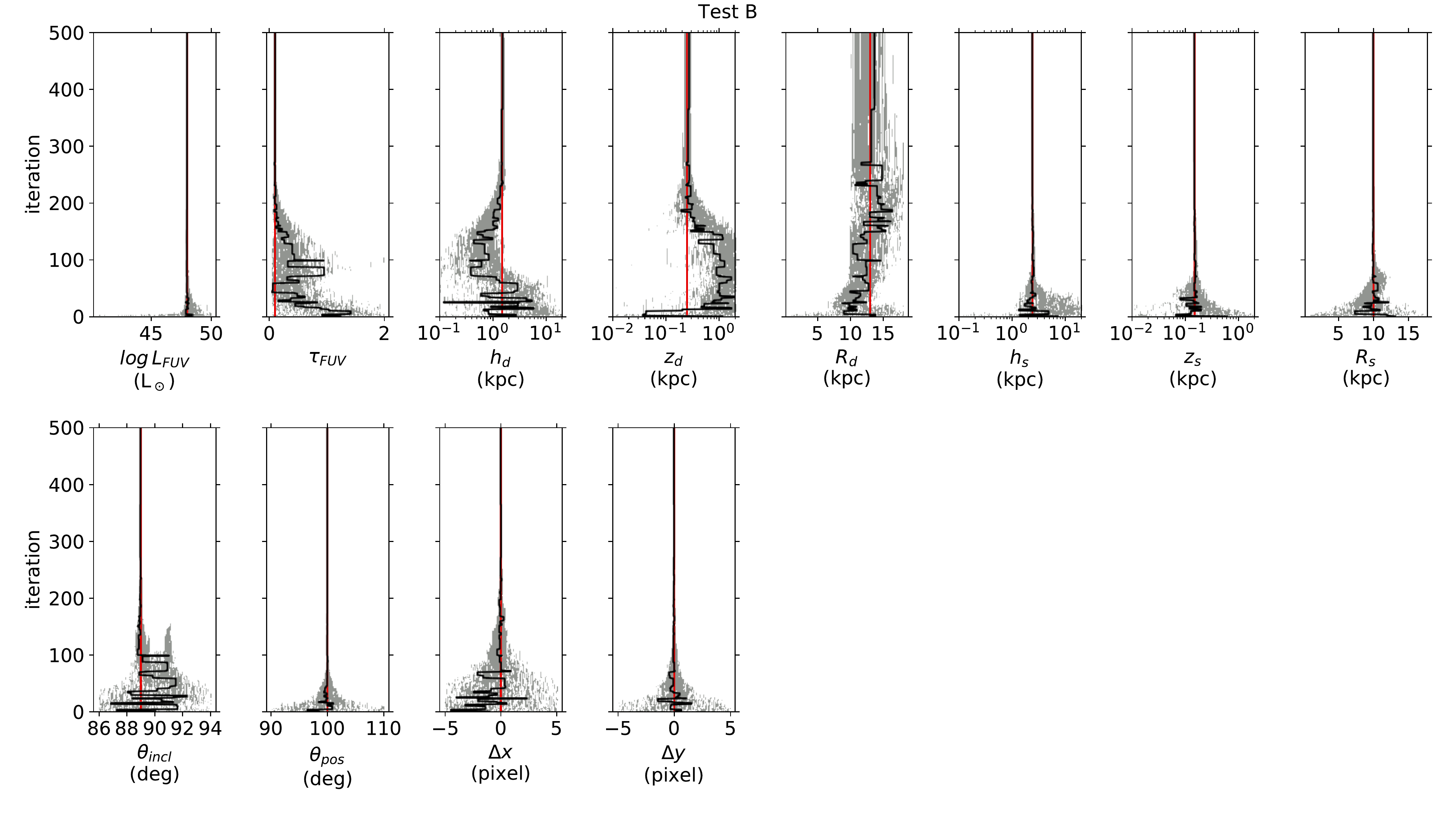}
}
\caption{The results of Test B. The rest are the same with Fig.~\ref{fig-test_a}} \label{fig-test_b}
\end{figure}

\clearpage
\begin{figure}
\center{
\includegraphics[scale=0.45]{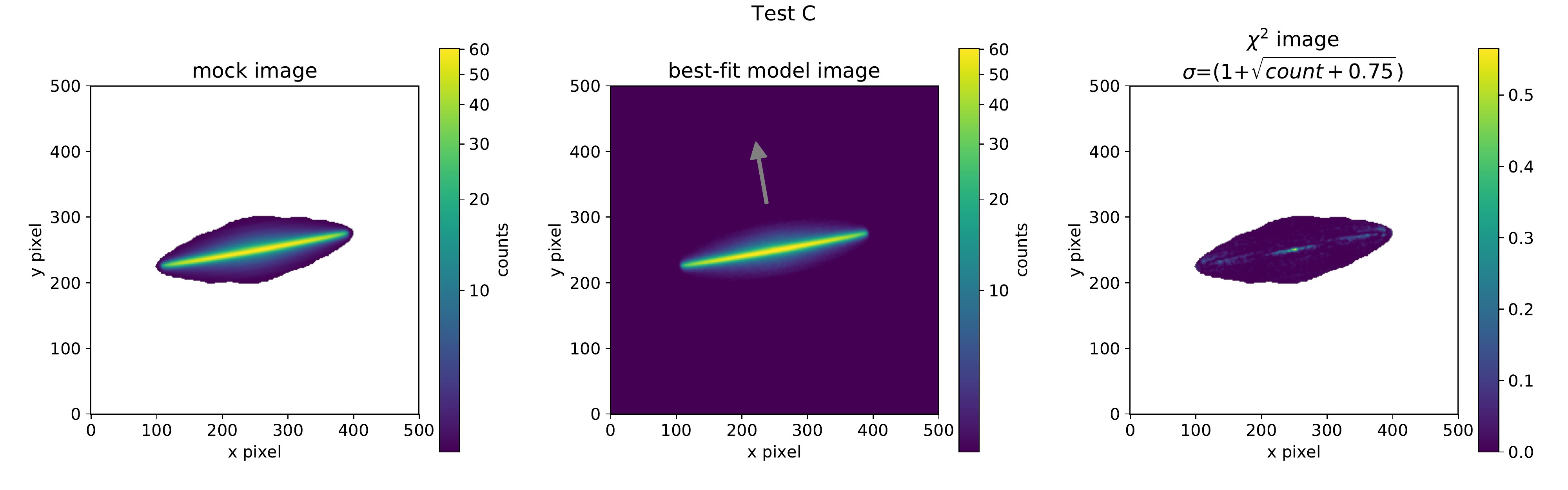}
\includegraphics[scale=0.45]{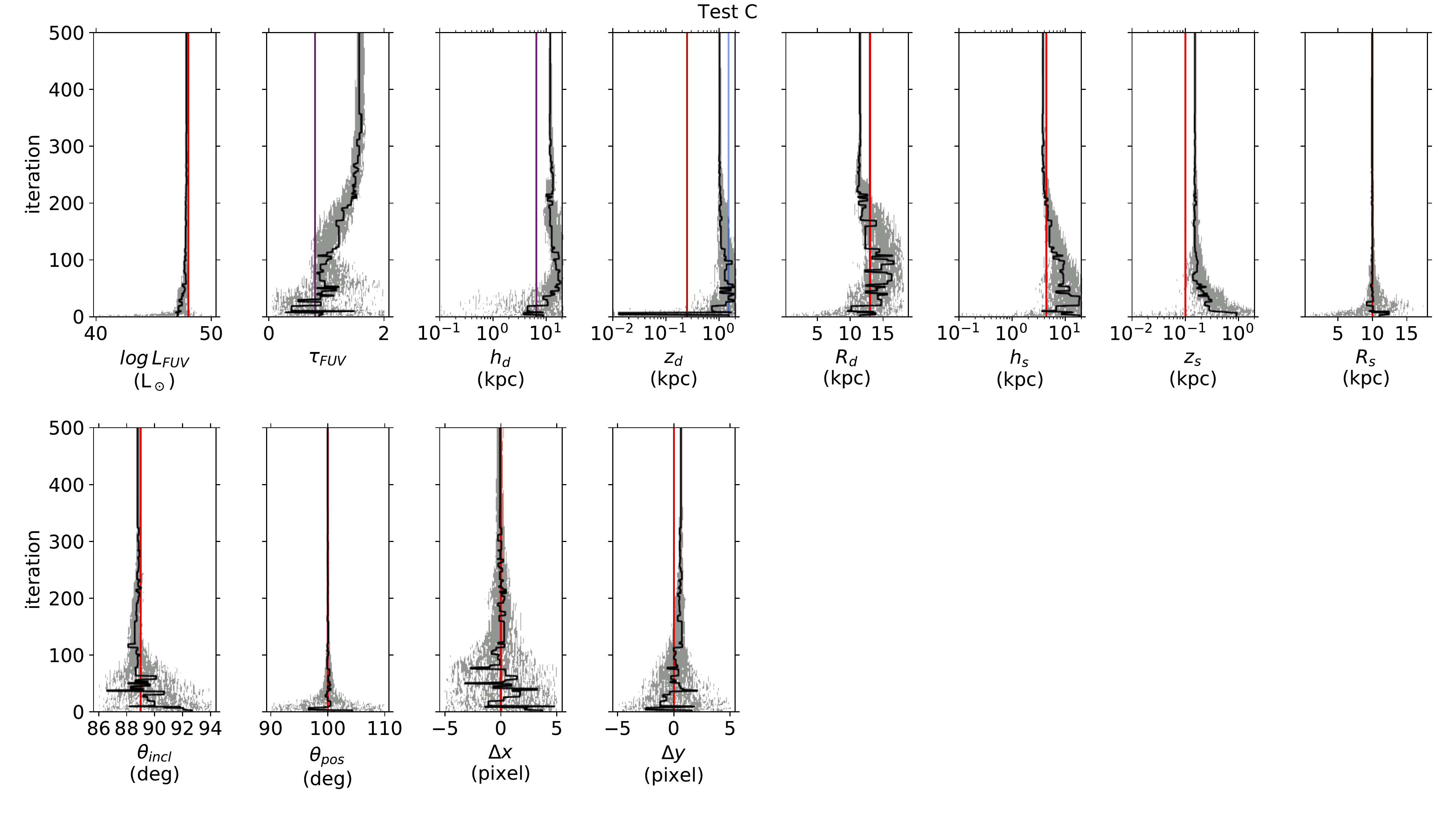}
}
\caption{The results of Test C. The \emph{blue vertical lines} are additionally displayed to indicate the second (thick) dust component (see text). In the case of $\tau_{FUV}$ and $h_d$, the red and blue lines are overlapped because we adopted the same $\tau_{FUV}$ and $h_d$ values for the two dust components. The rest are the same with Fig.~\ref{fig-test_a} } \label{fig-test_c}
\end{figure}

\input{fig/fig_fitdata.tex}

\clearpage
\begin{figure}
\figurenum{28}
\center{
\includegraphics[scale=0.45]{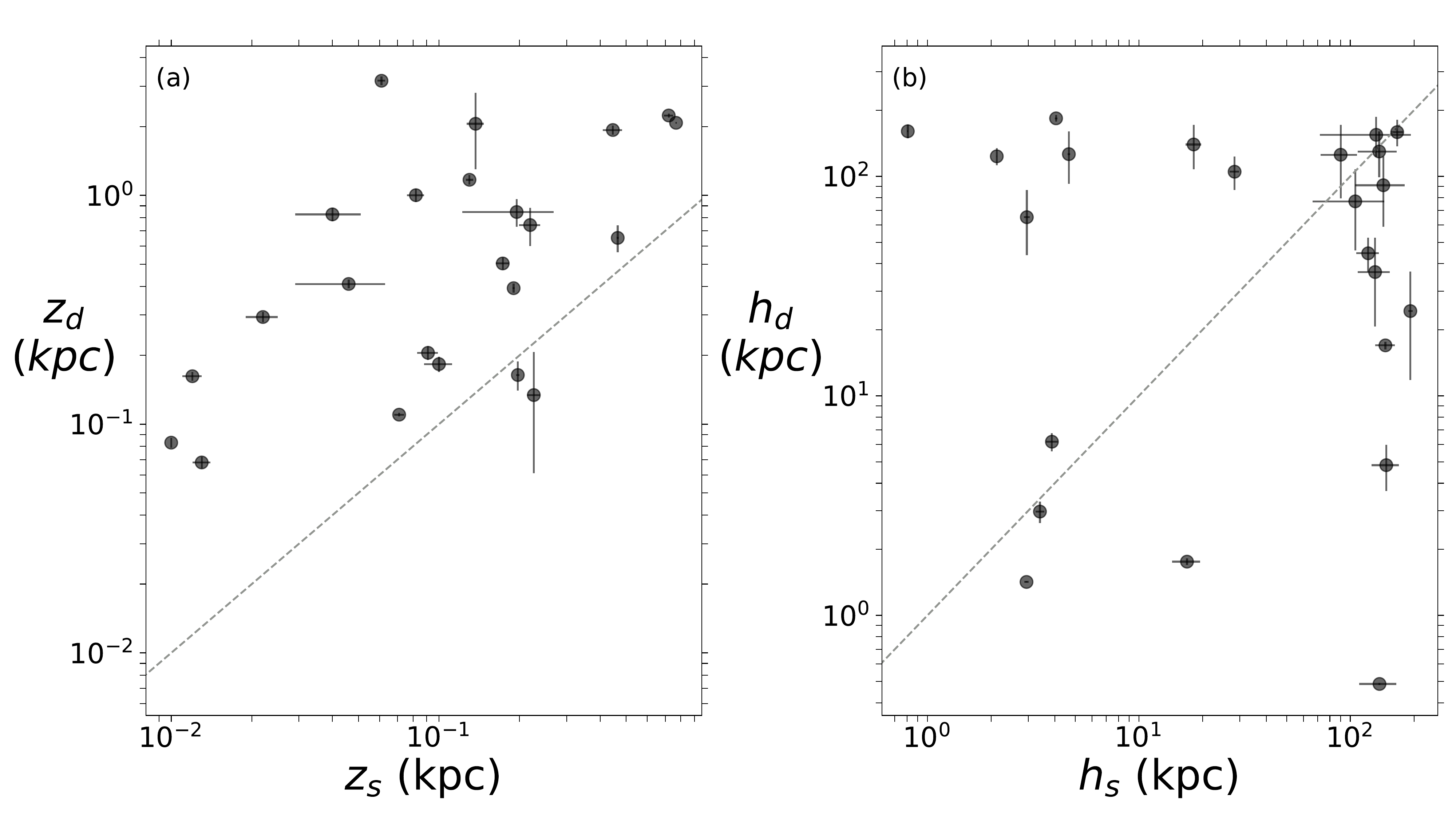}
}
\caption{Plots of scale-heights and scale-lengths. (a) light source scale-height vs dust scale-height. (b) light source scale-length vs dust scale-length. The \emph{grey dashed lines} indicate the one-to-one lines. 
} \label{fig-zh}
\end{figure}

\clearpage
\begin{figure}
\figurenum{29}
\center{
\includegraphics[scale=0.45]{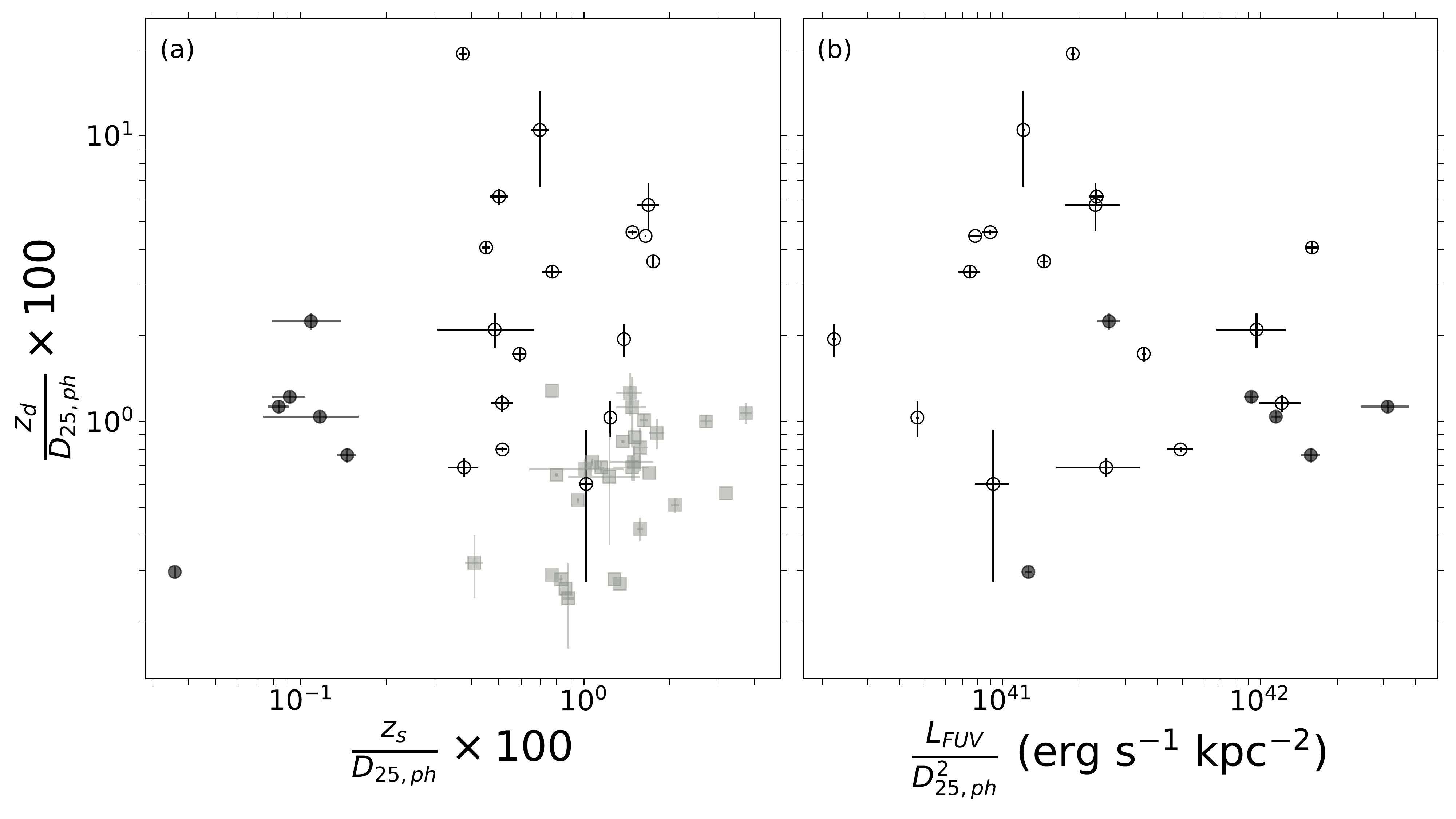}
}
\caption{Plots for galaxy-size independent comparisons. (a) the plot between the ratios of scale-height to galactic diameter for light source and dust. \emph{Open} and \emph{filled circles} indicate ``high-group'' and ``low-group,'' respectively (see text). \emph{Gray squares} indicate the values obtained from the optical radiative transfer studies of other galaxies (Table \ref{tbl-zD_ref}). (b) the plot between the ratio of dust scale-height to galactic diameter and $L_{FUV}/D^2_{25}$ which is proportional to the surface density of FUV luminosity. The symbols are the same as in the panel (a).} \label{fig-zz}
\end{figure}

%% file: fig/fig_target.tex
\clearpage
\begin{figure}
\center{
(a)\includegraphics[scale=0.7]{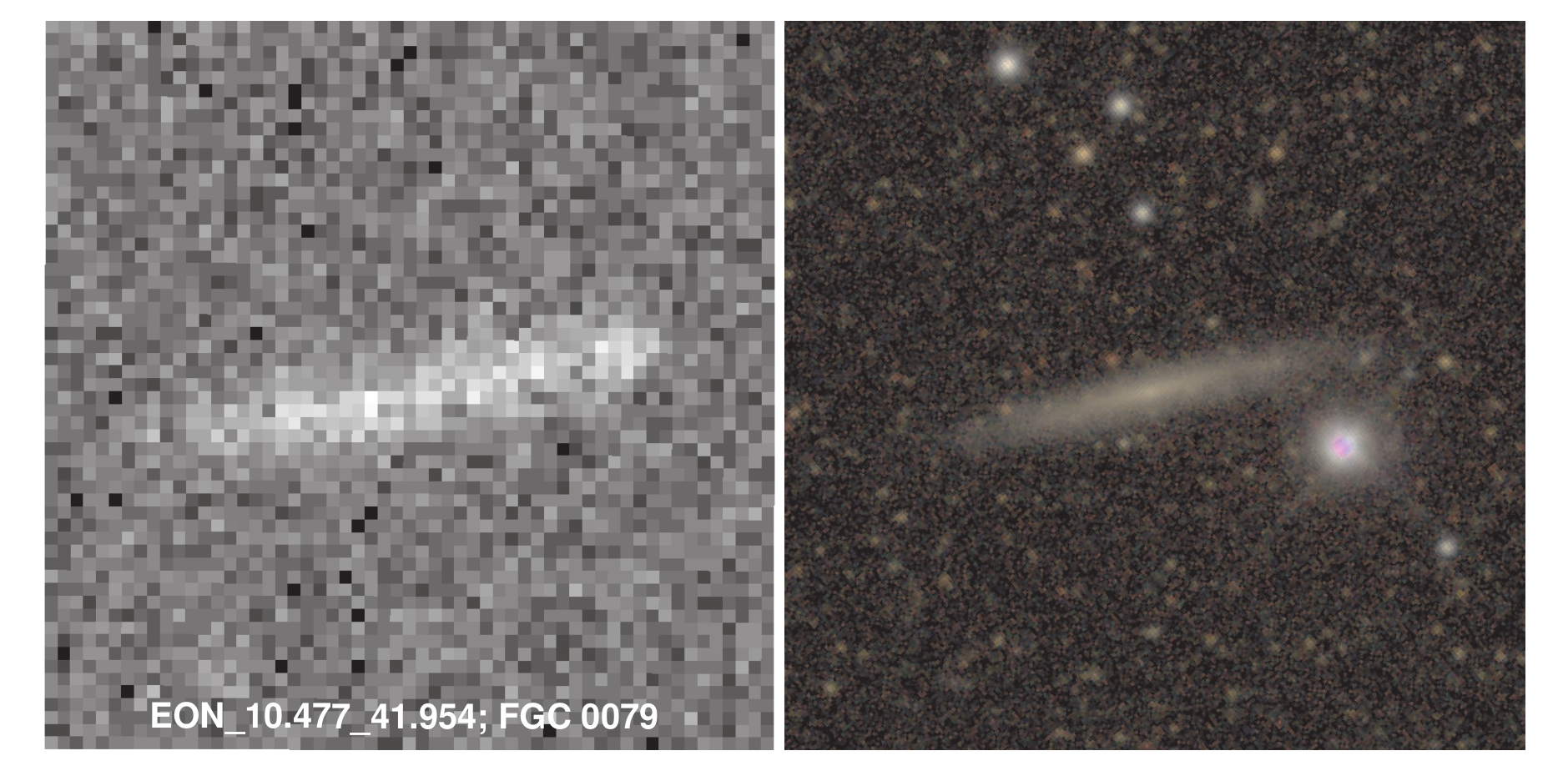}
(b)\includegraphics[scale=0.7]{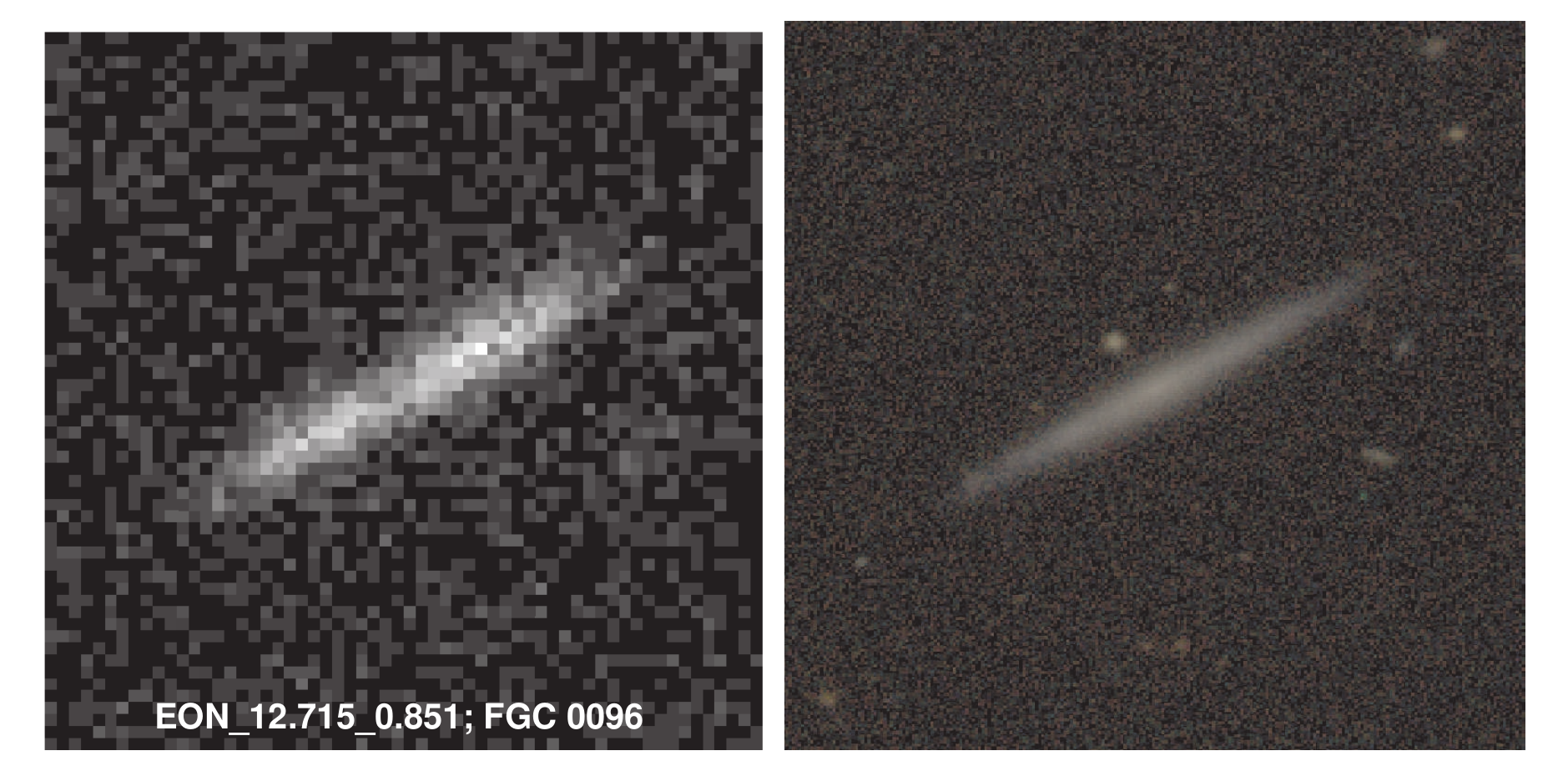}
(c)\includegraphics[scale=0.7]{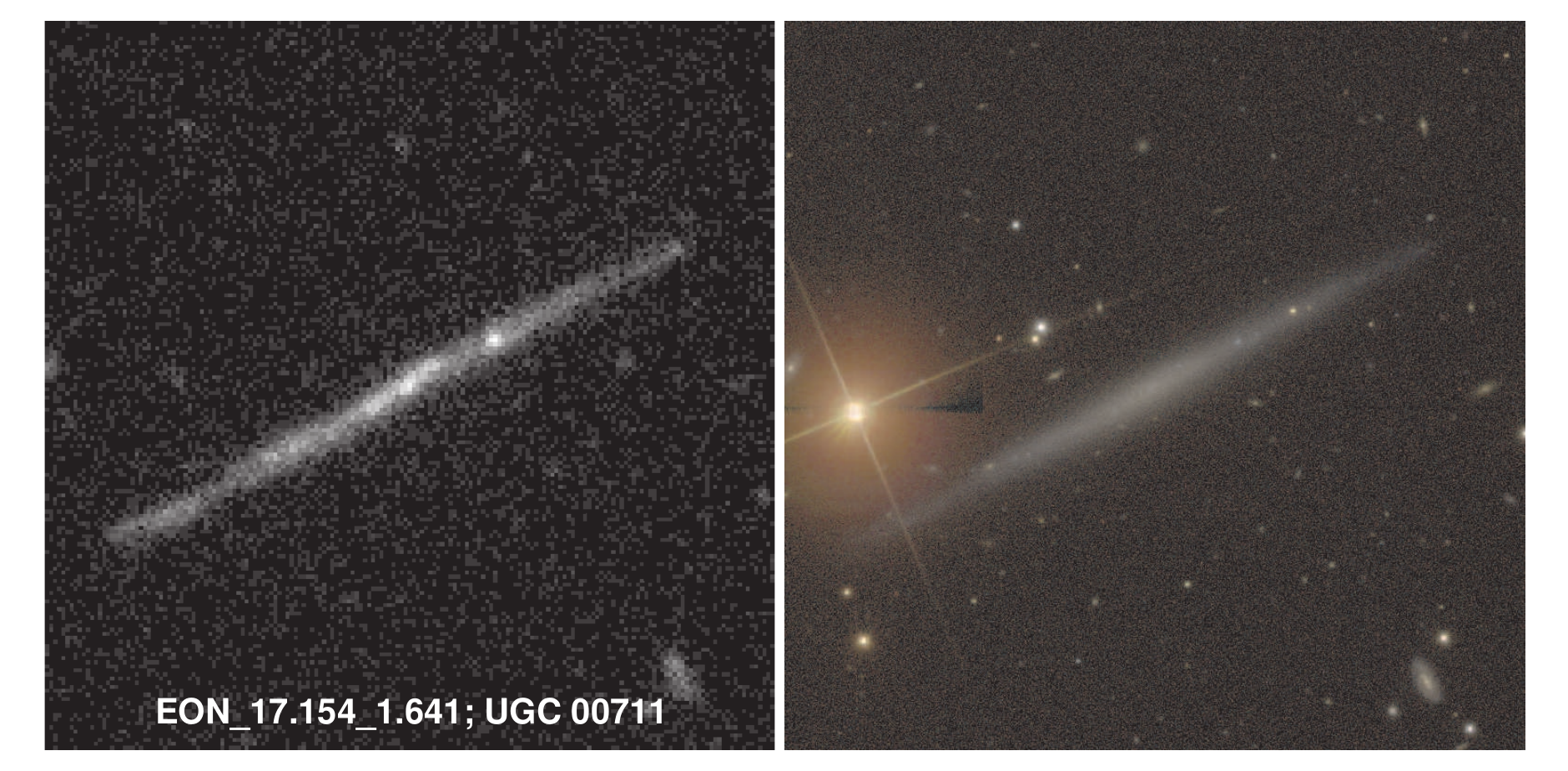}
}
\caption{\input{caption/fig_targets0.tex}} \label{fig-tg}
\end{figure}

\clearpage
\begin{figure}
\figurenum{1}
\center{
(d)\includegraphics[scale=0.7]{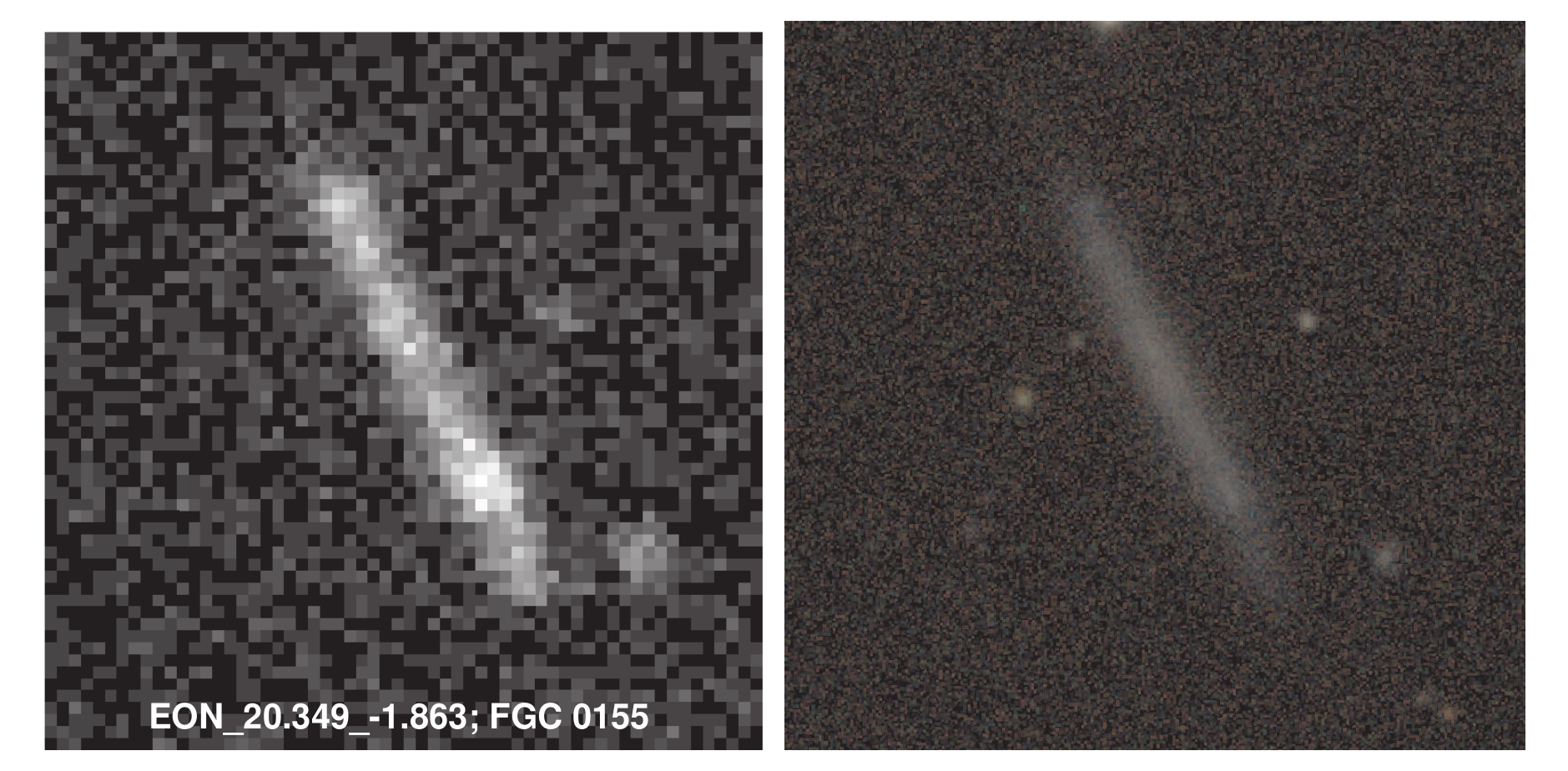}
(e)\includegraphics[scale=0.7]{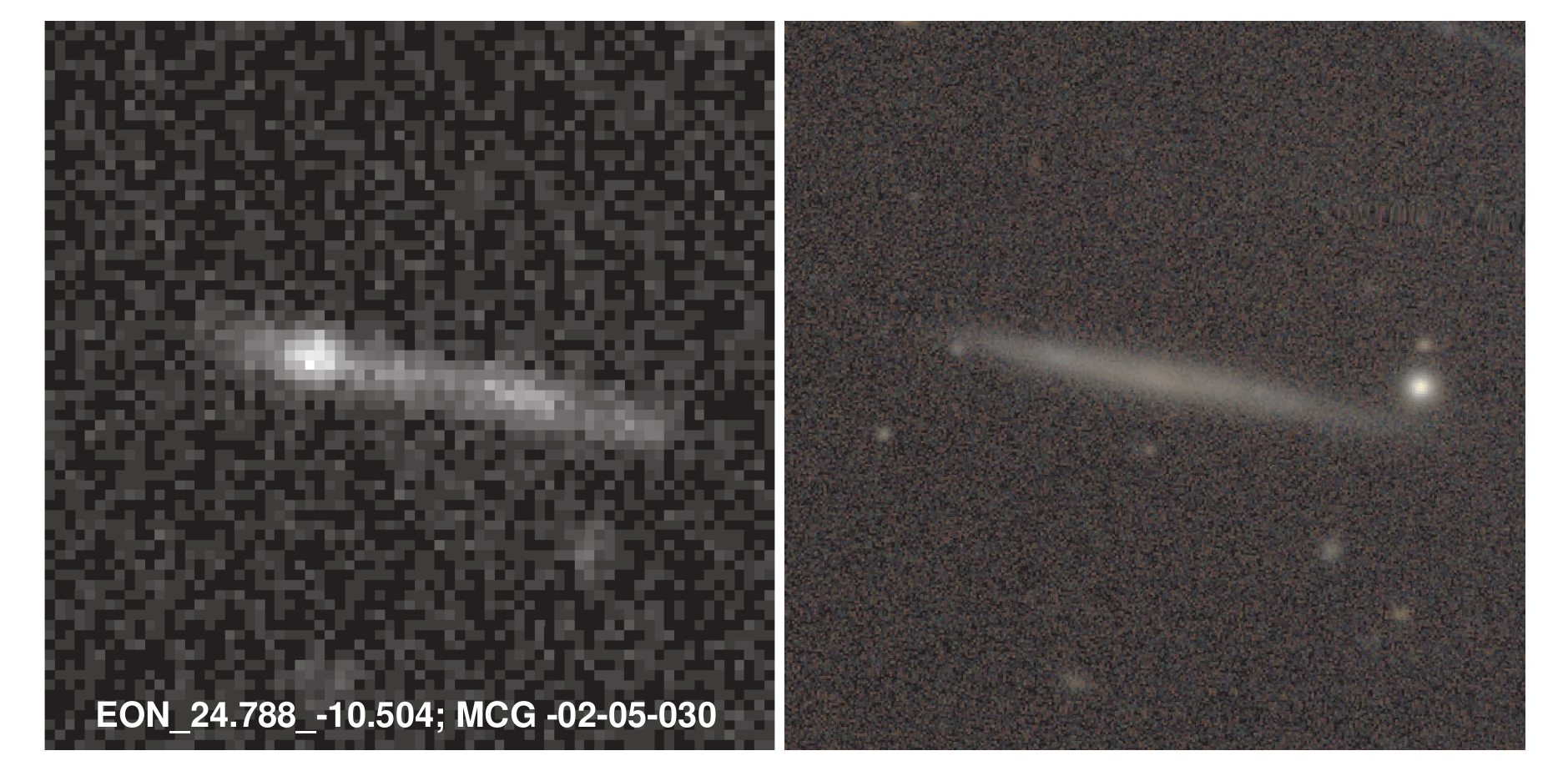}
(f)\includegraphics[scale=0.7]{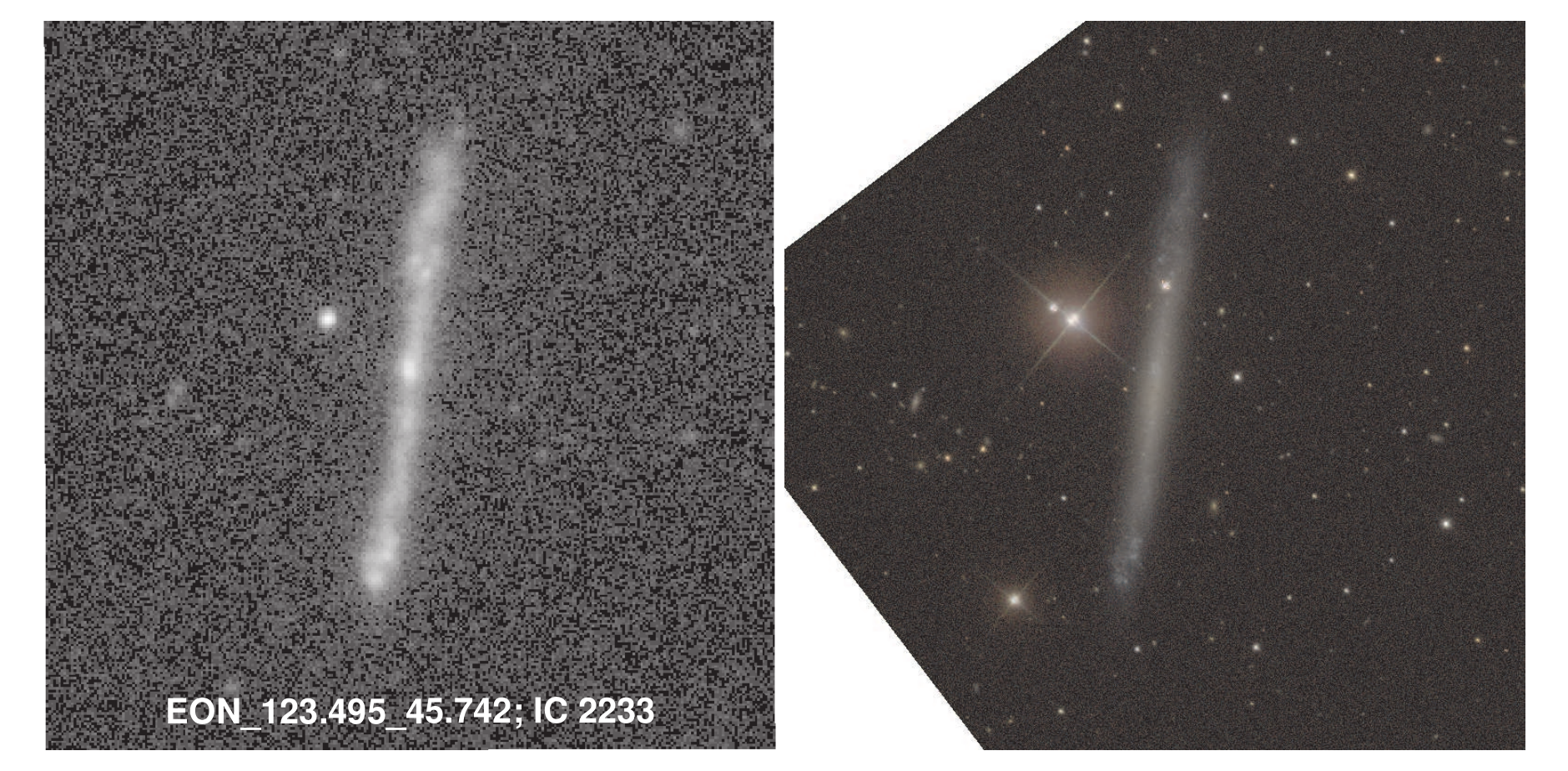}
}
\caption{\input{caption/fig_targets1.tex}} 
\end{figure}

\clearpage
\begin{figure}
\figurenum{1}
\center{
(g)\includegraphics[scale=0.7]{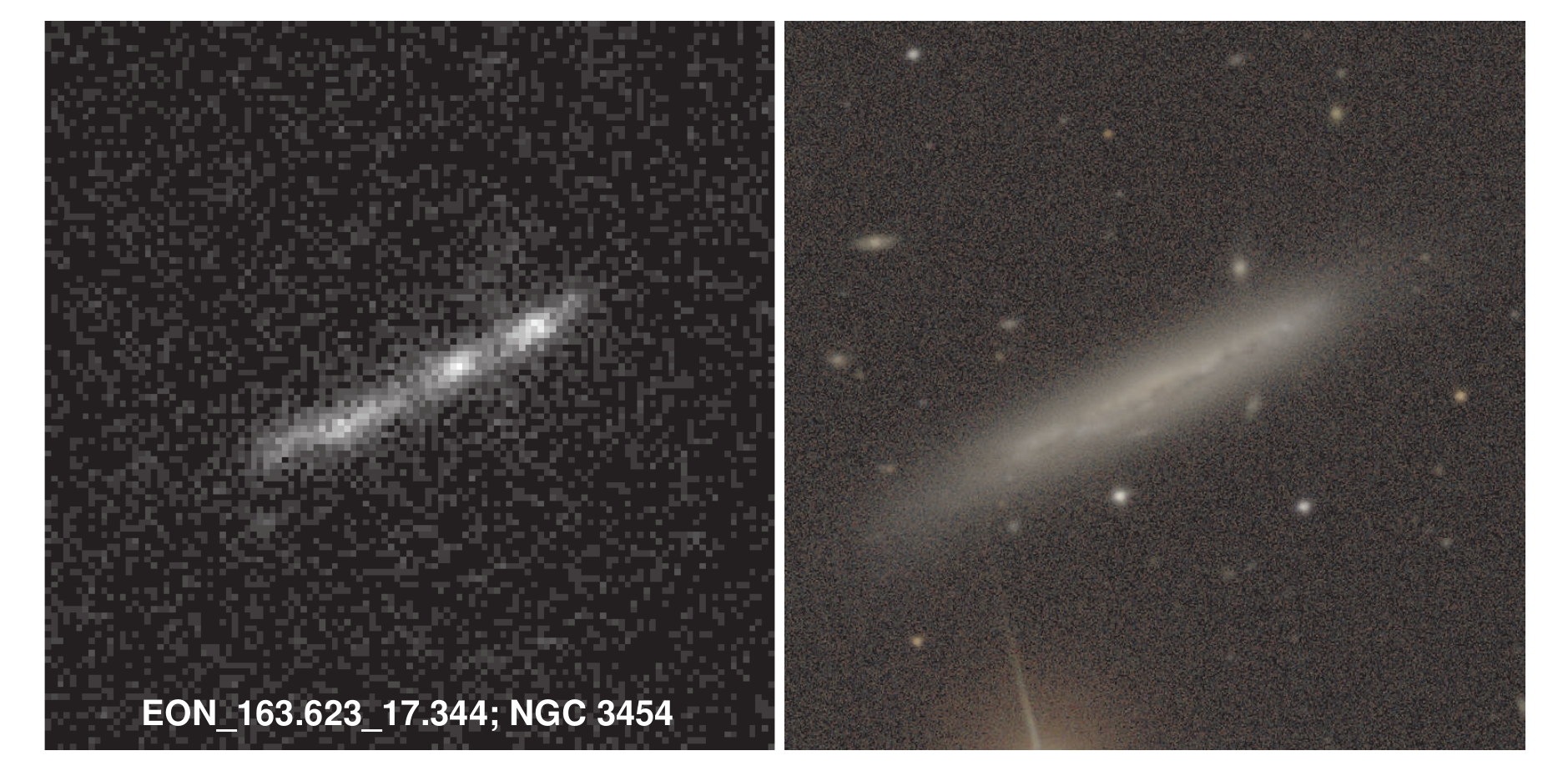}
(h)\includegraphics[scale=0.7]{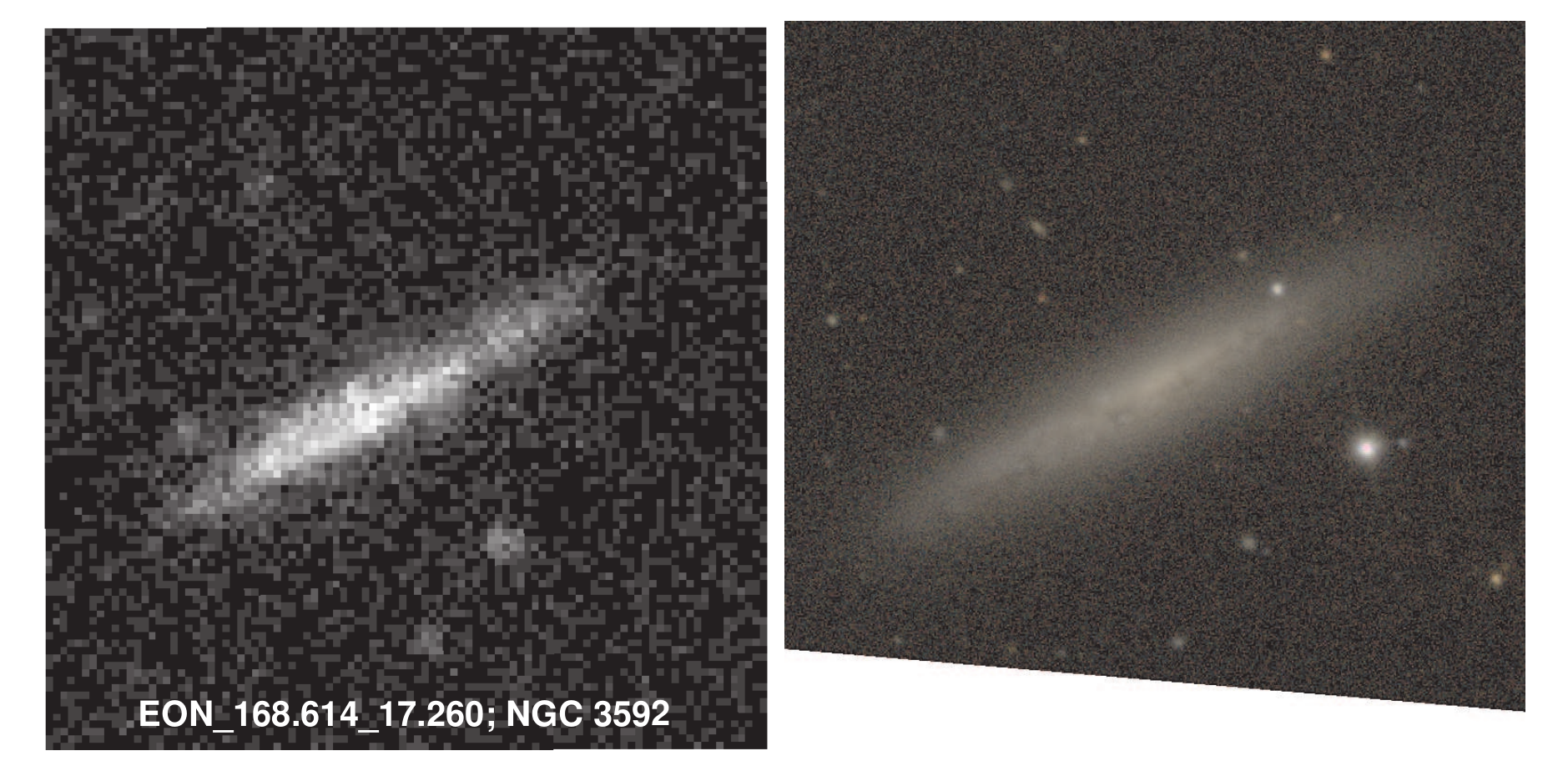}
(i)\includegraphics[scale=0.7]{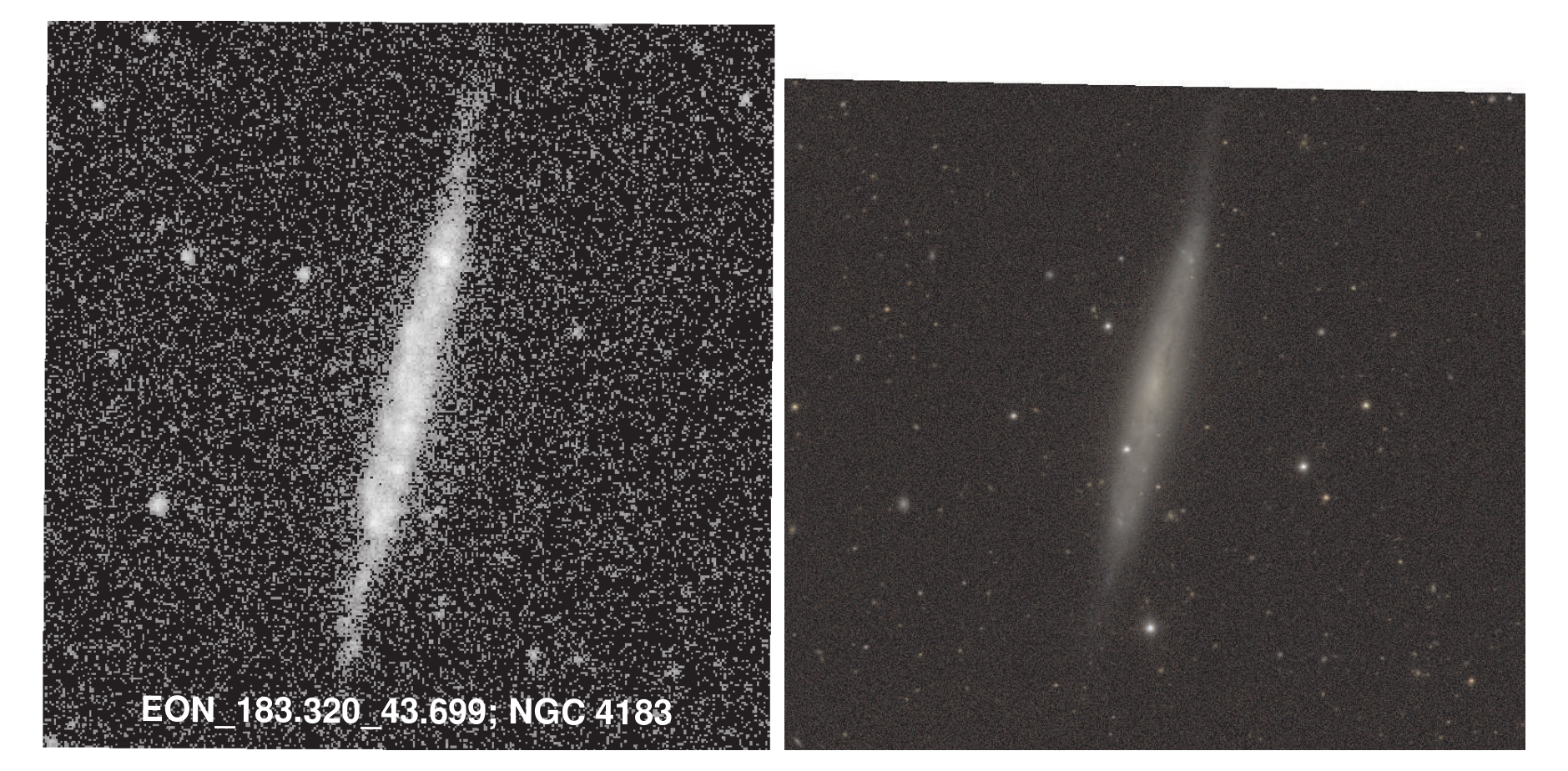}
}
\caption{\input{caption/fig_targets1.tex}} 
\end{figure}

\clearpage
\begin{figure}
\figurenum{1}
\center{
(j)\includegraphics[scale=0.7]{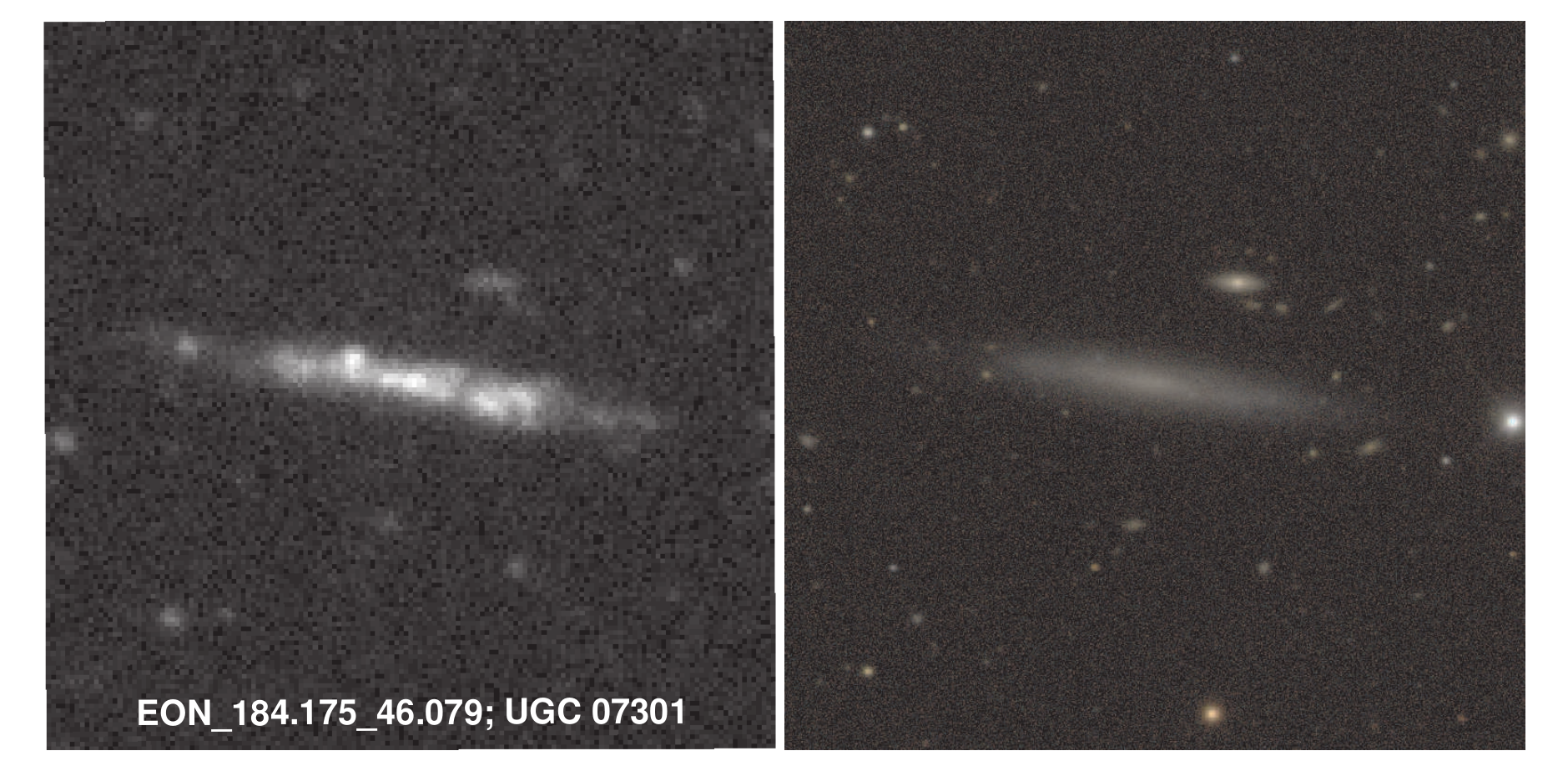}
(k)\includegraphics[scale=0.7]{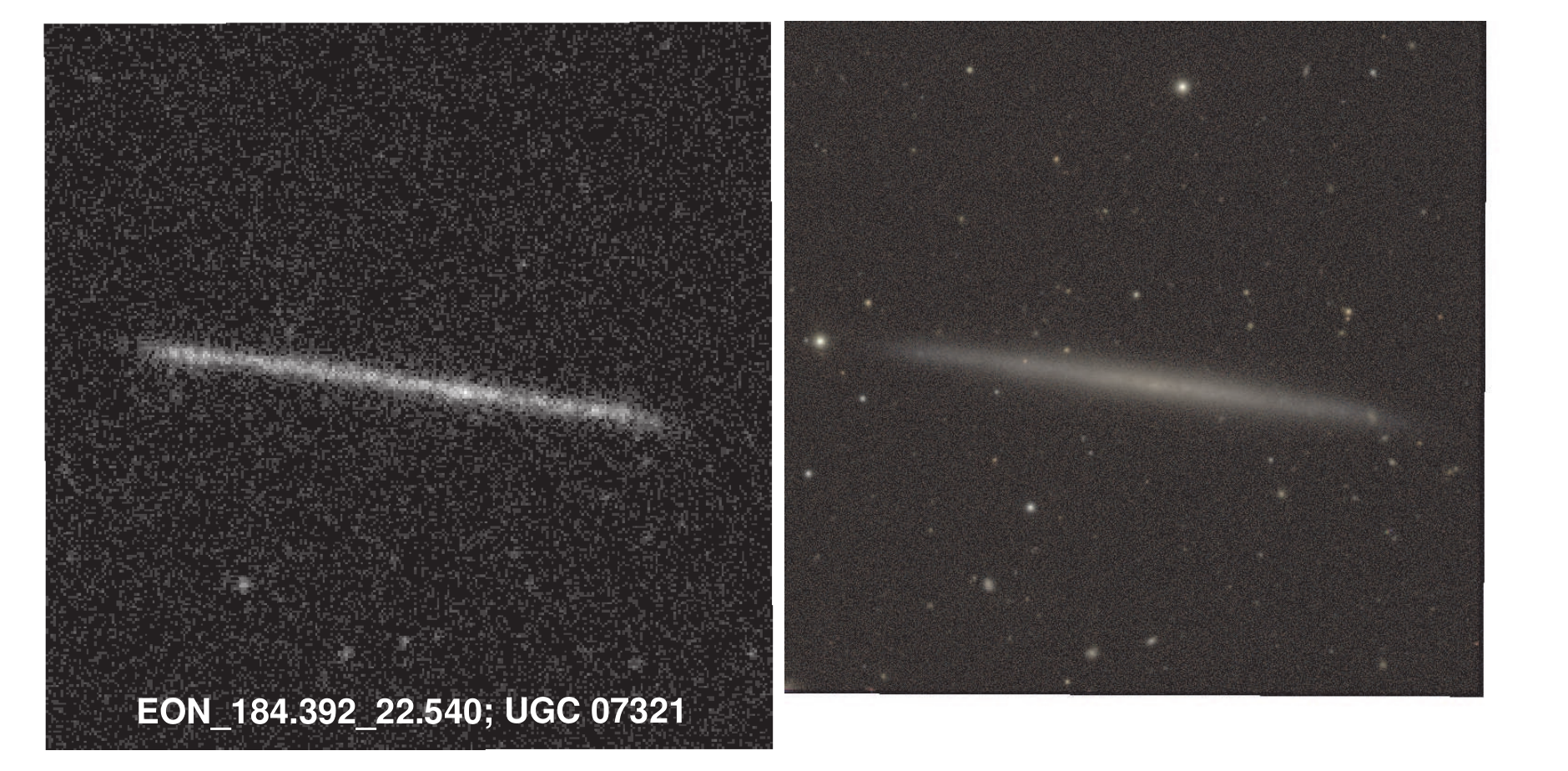}
(l)\includegraphics[scale=0.7]{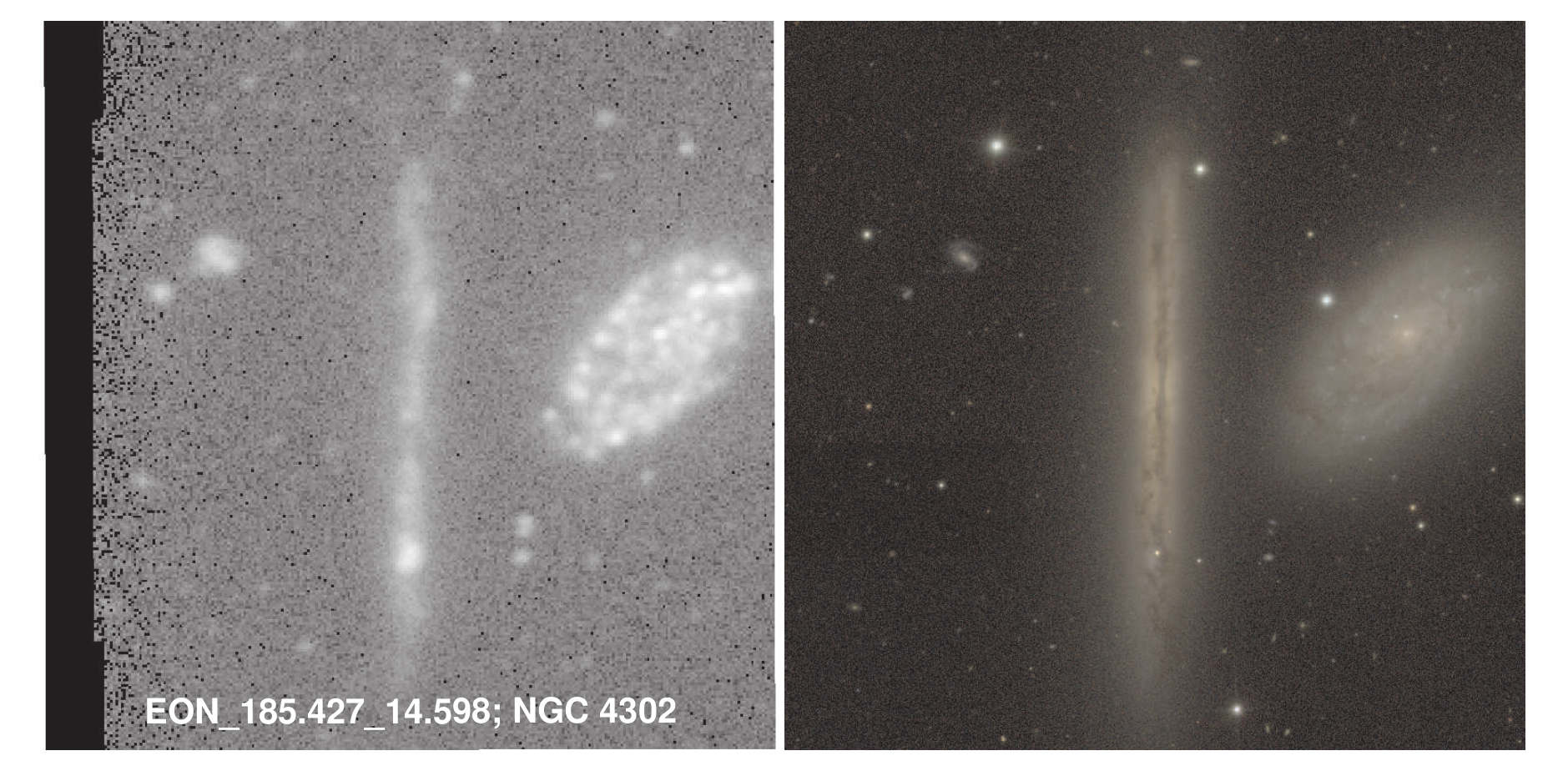}
}
\caption{\input{caption/fig_targets1.tex}} 
\end{figure}

\clearpage
\begin{figure}
\figurenum{1}
\center{
(m)\includegraphics[scale=0.7]{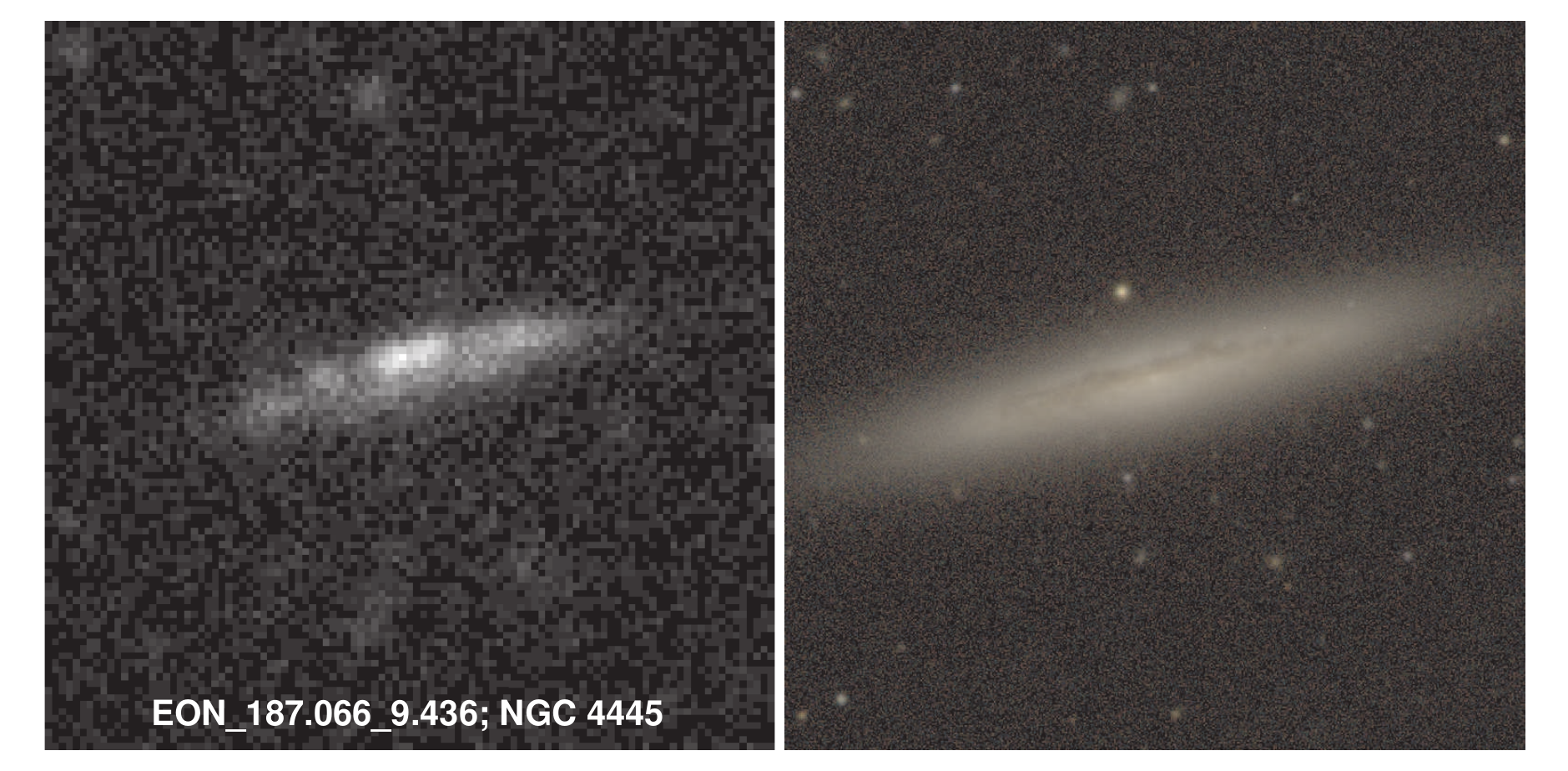}
(n)\includegraphics[scale=0.7]{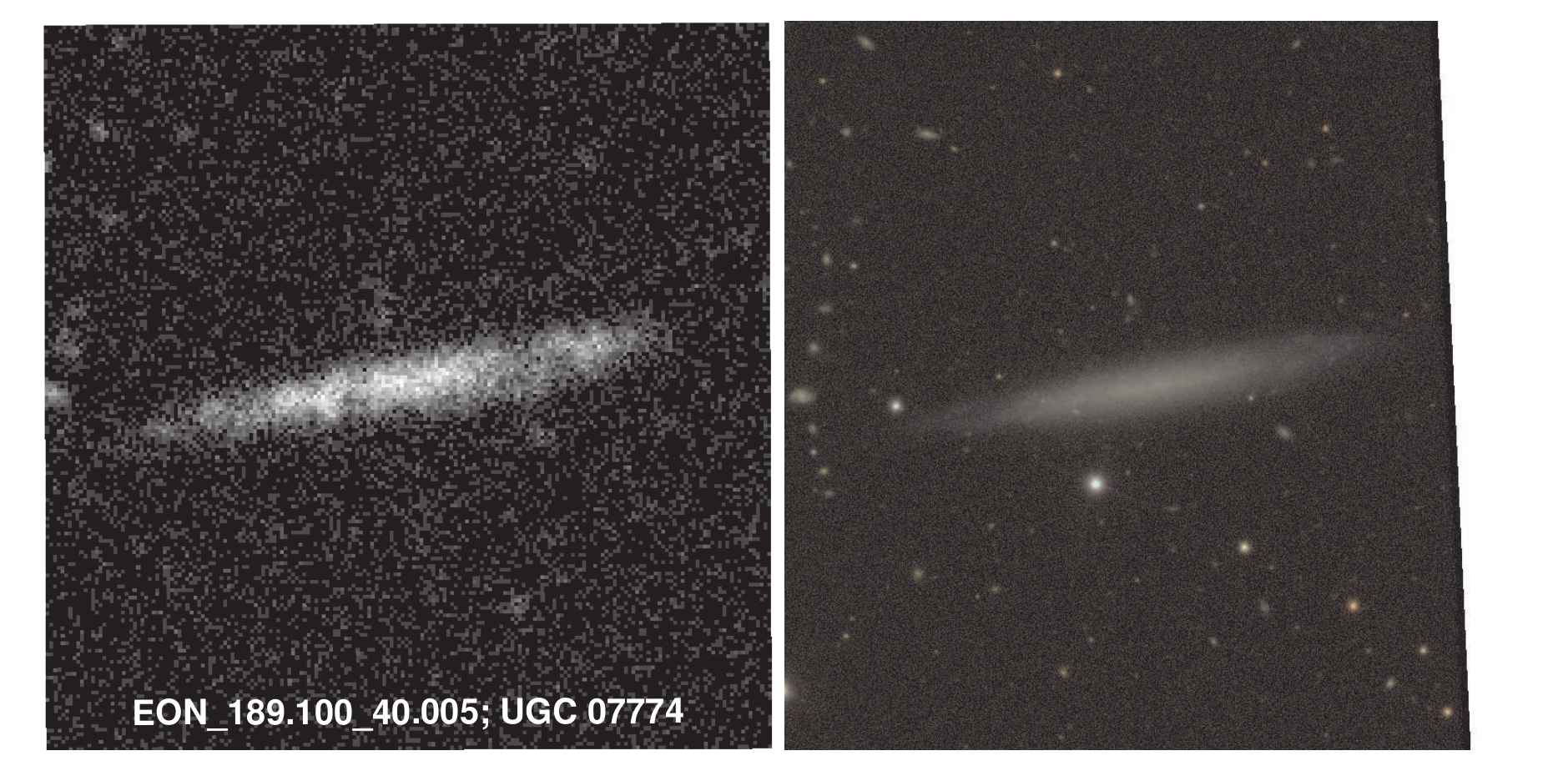}
(o)\includegraphics[scale=0.7]{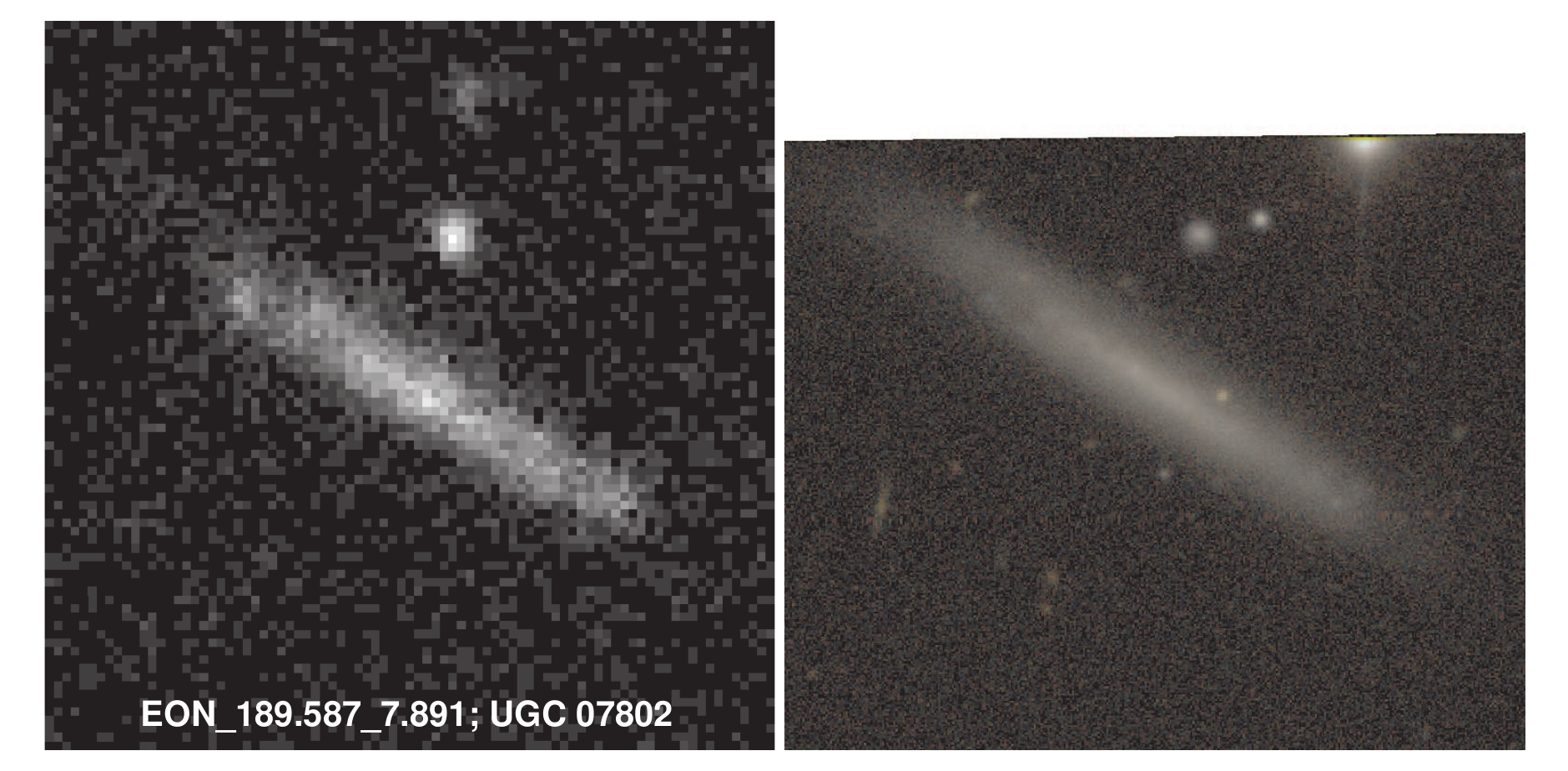}
}
\caption{\input{caption/fig_targets1.tex}} 
\end{figure}

\clearpage
\begin{figure}
\figurenum{1}
\center{
(p)\includegraphics[scale=0.7]{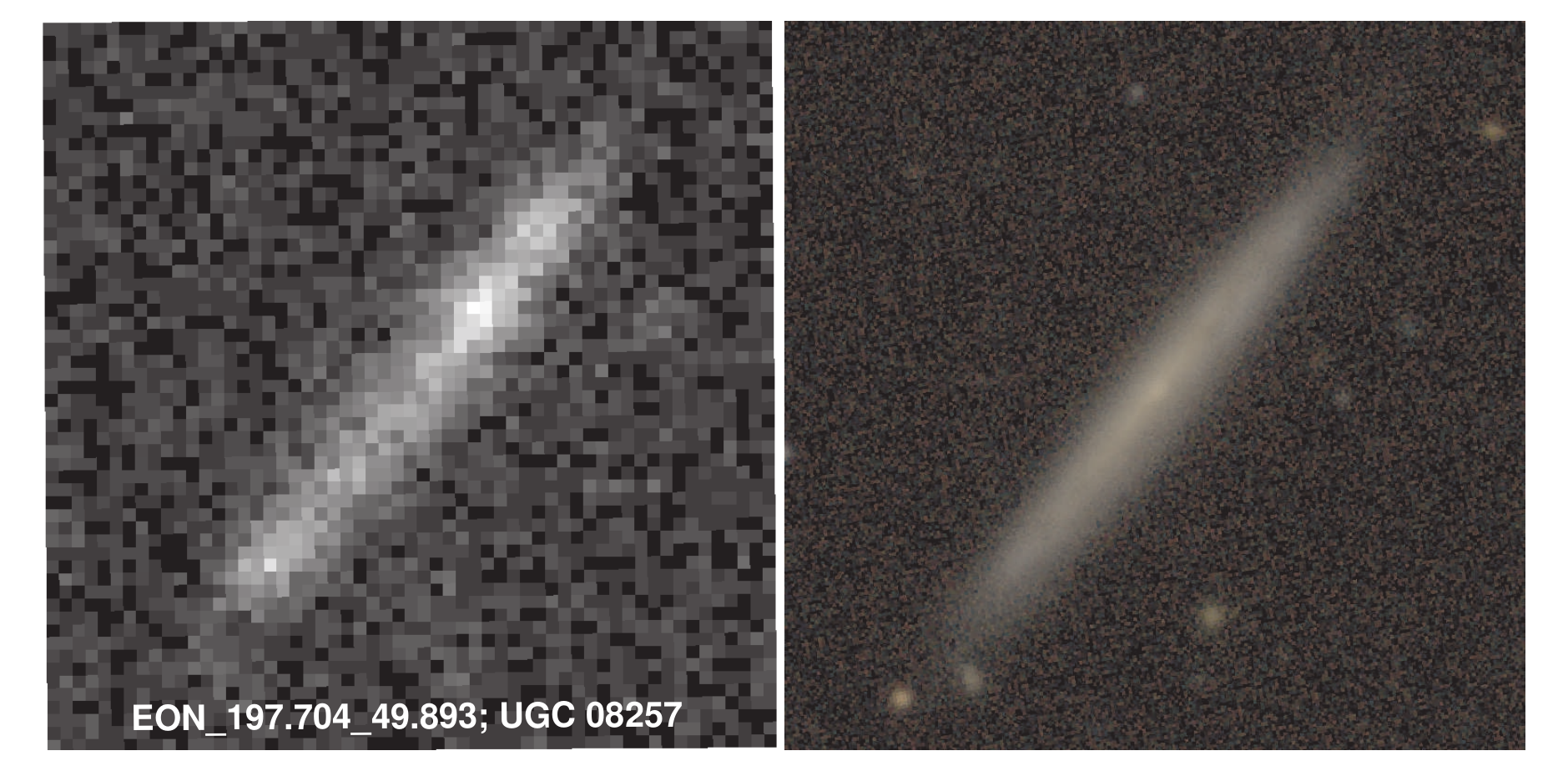}
(q)\includegraphics[scale=0.7]{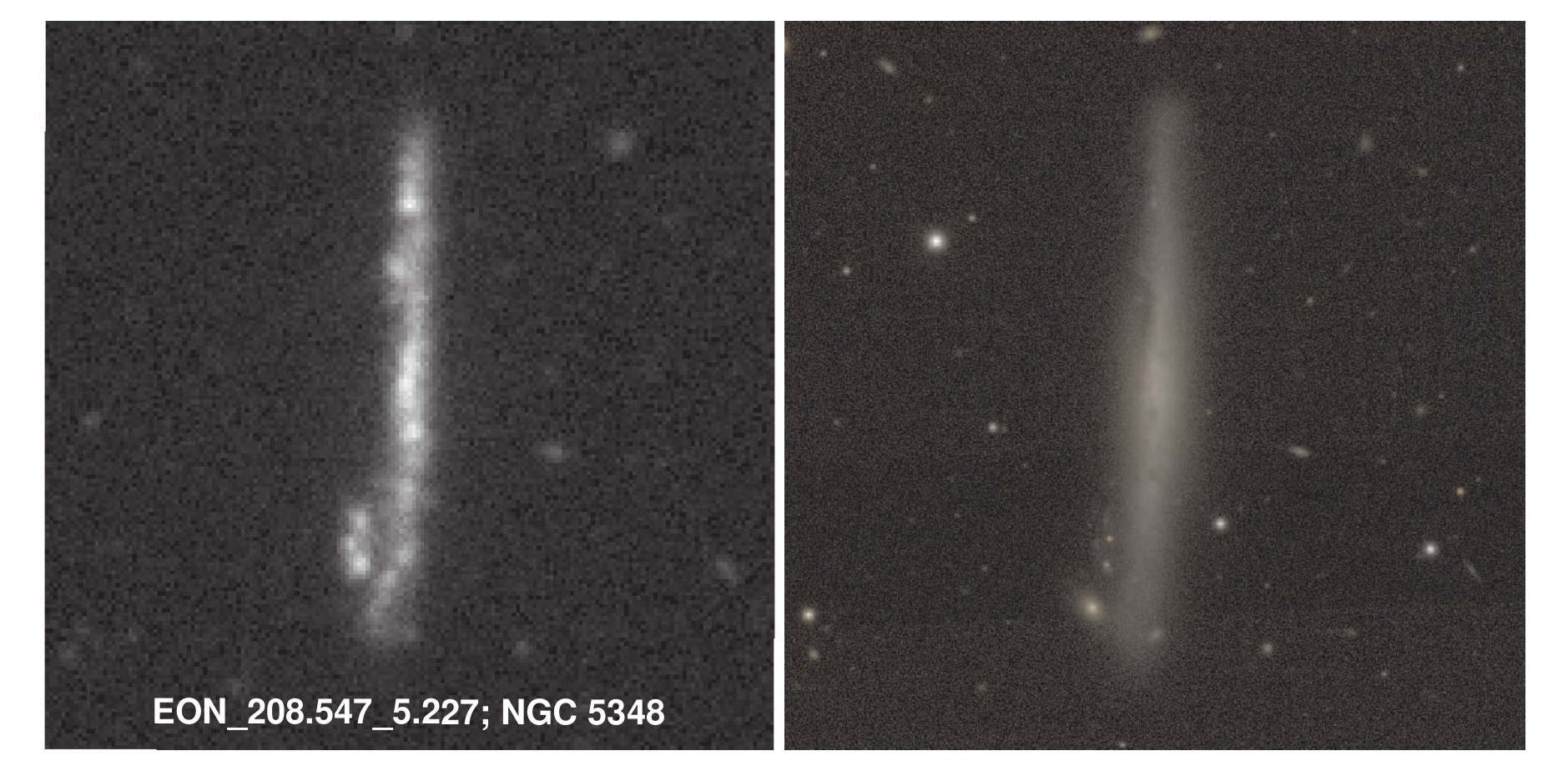}
(r)\includegraphics[scale=0.7]{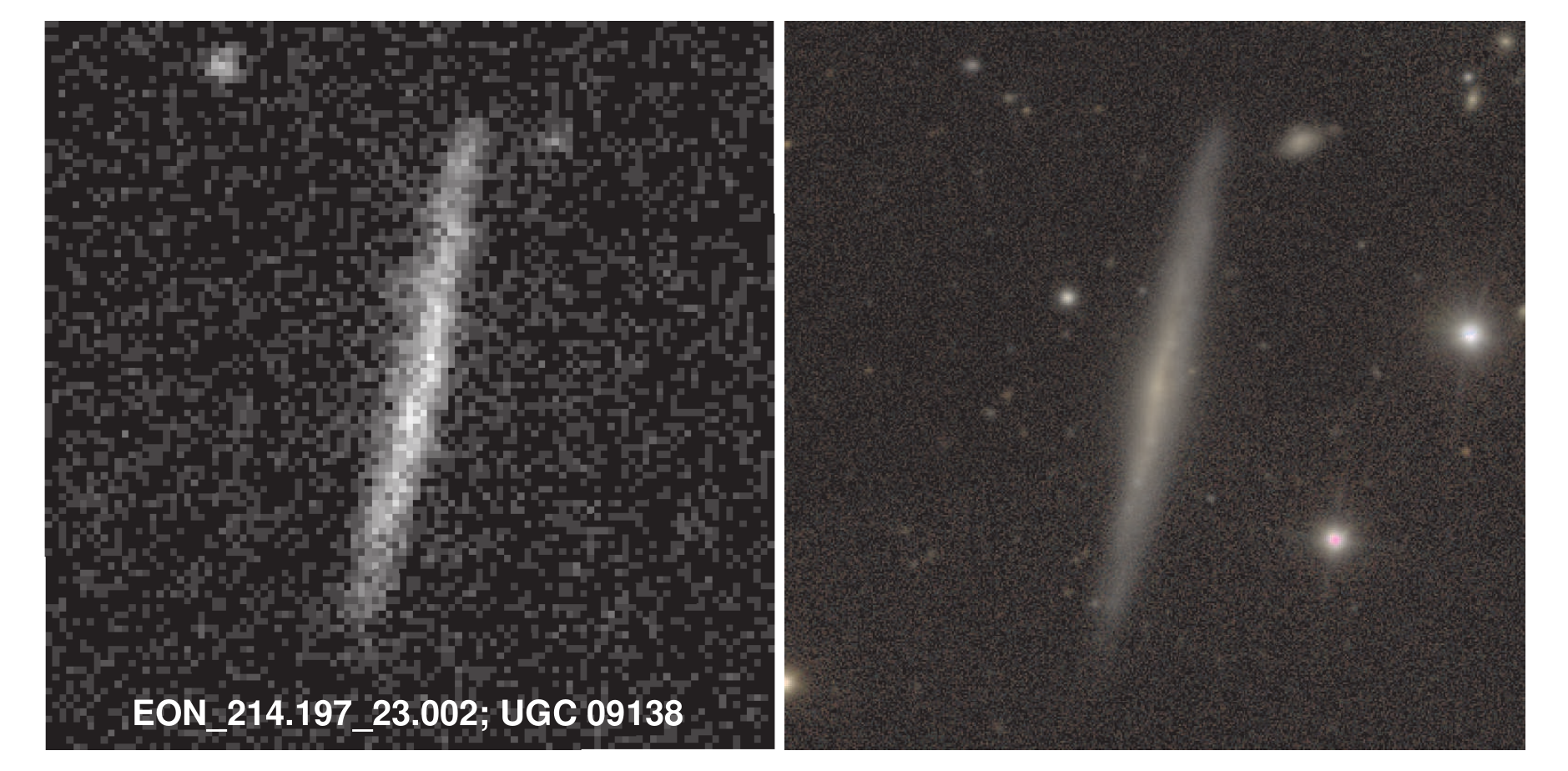}
}
\caption{\input{caption/fig_targets1.tex}} 
\end{figure}

\clearpage
\begin{figure}
\figurenum{1}
\center{
(s)\includegraphics[scale=0.7]{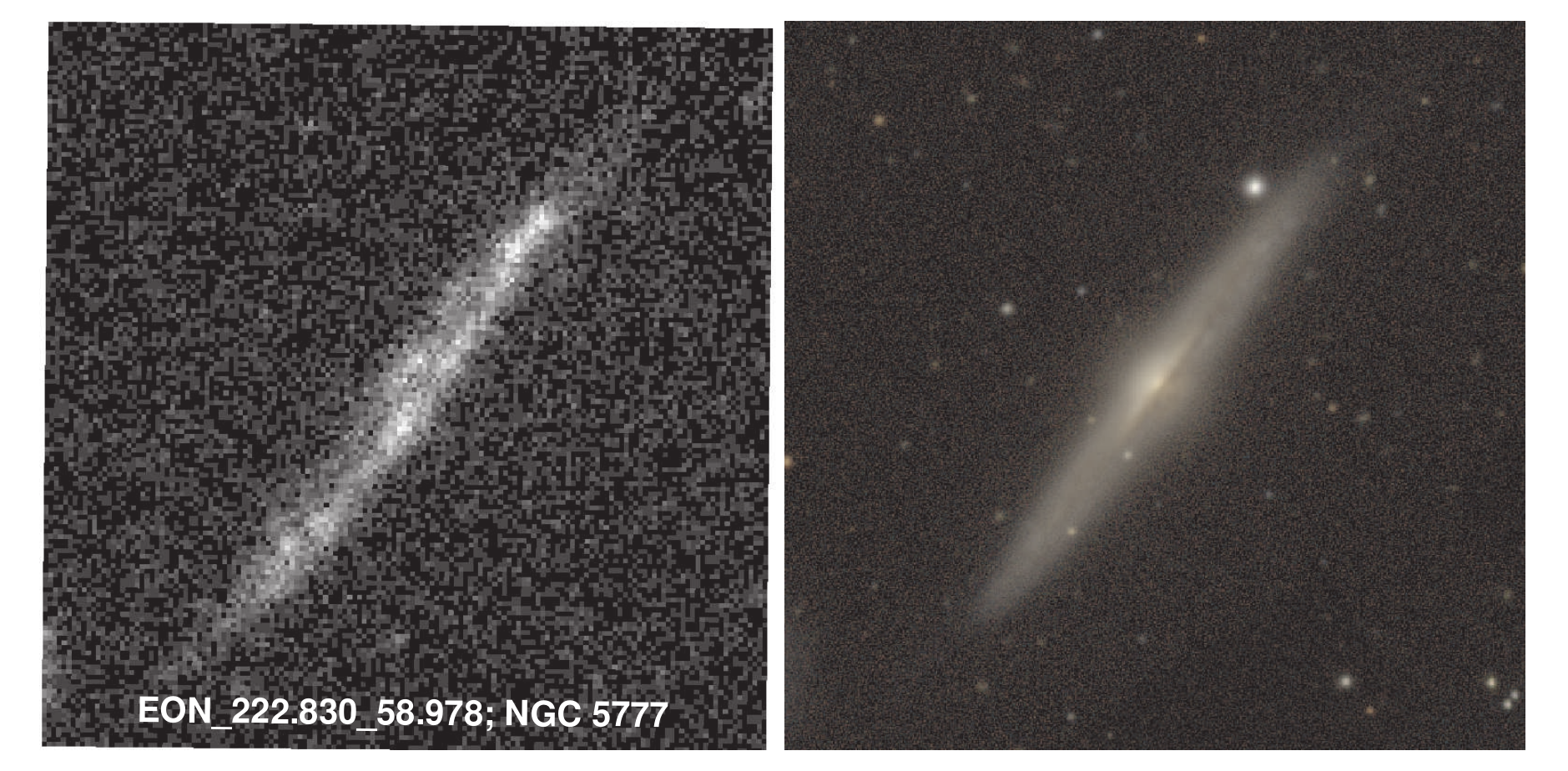}
(t)\includegraphics[scale=0.7]{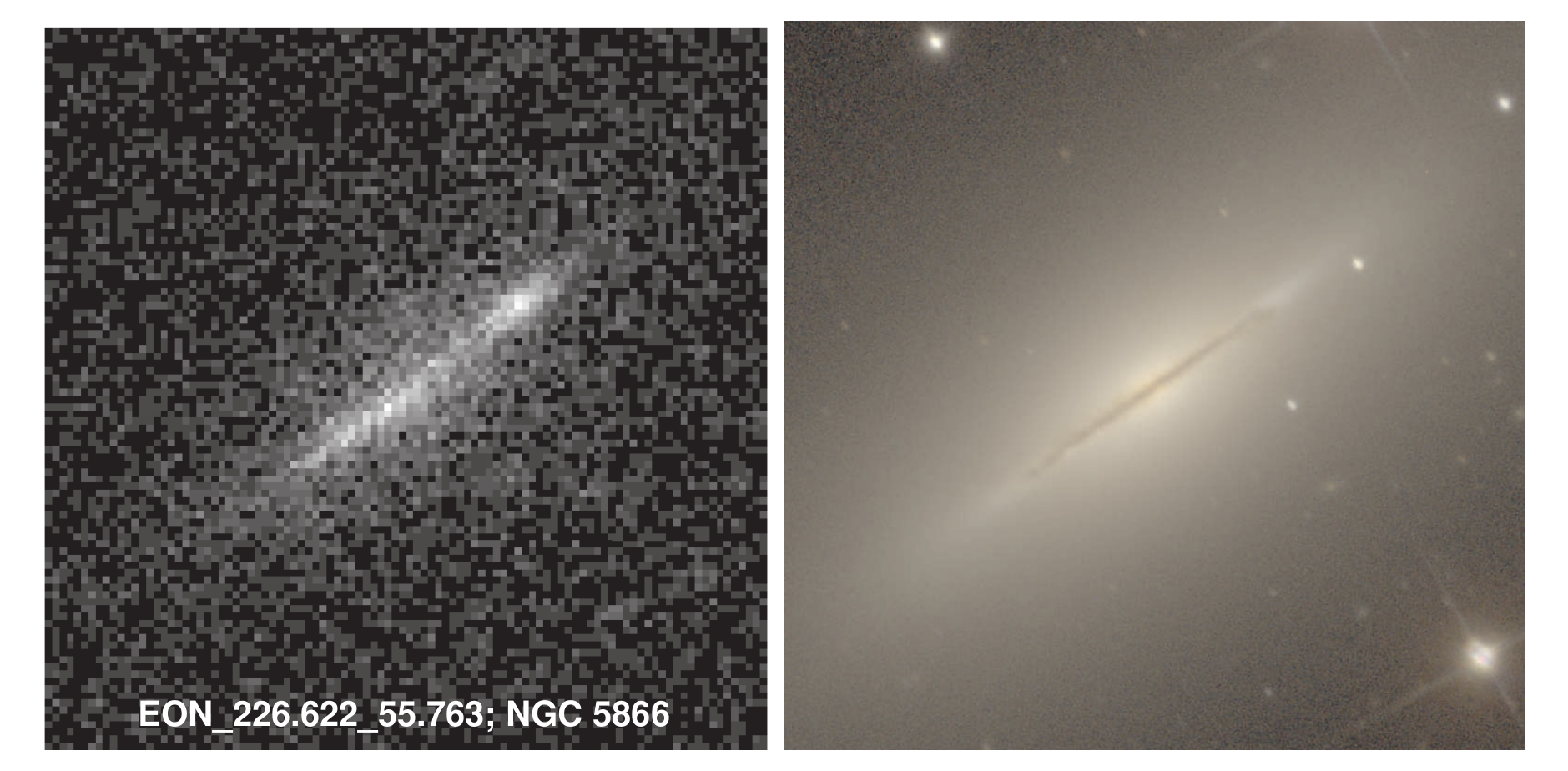}
(u)\includegraphics[scale=0.7]{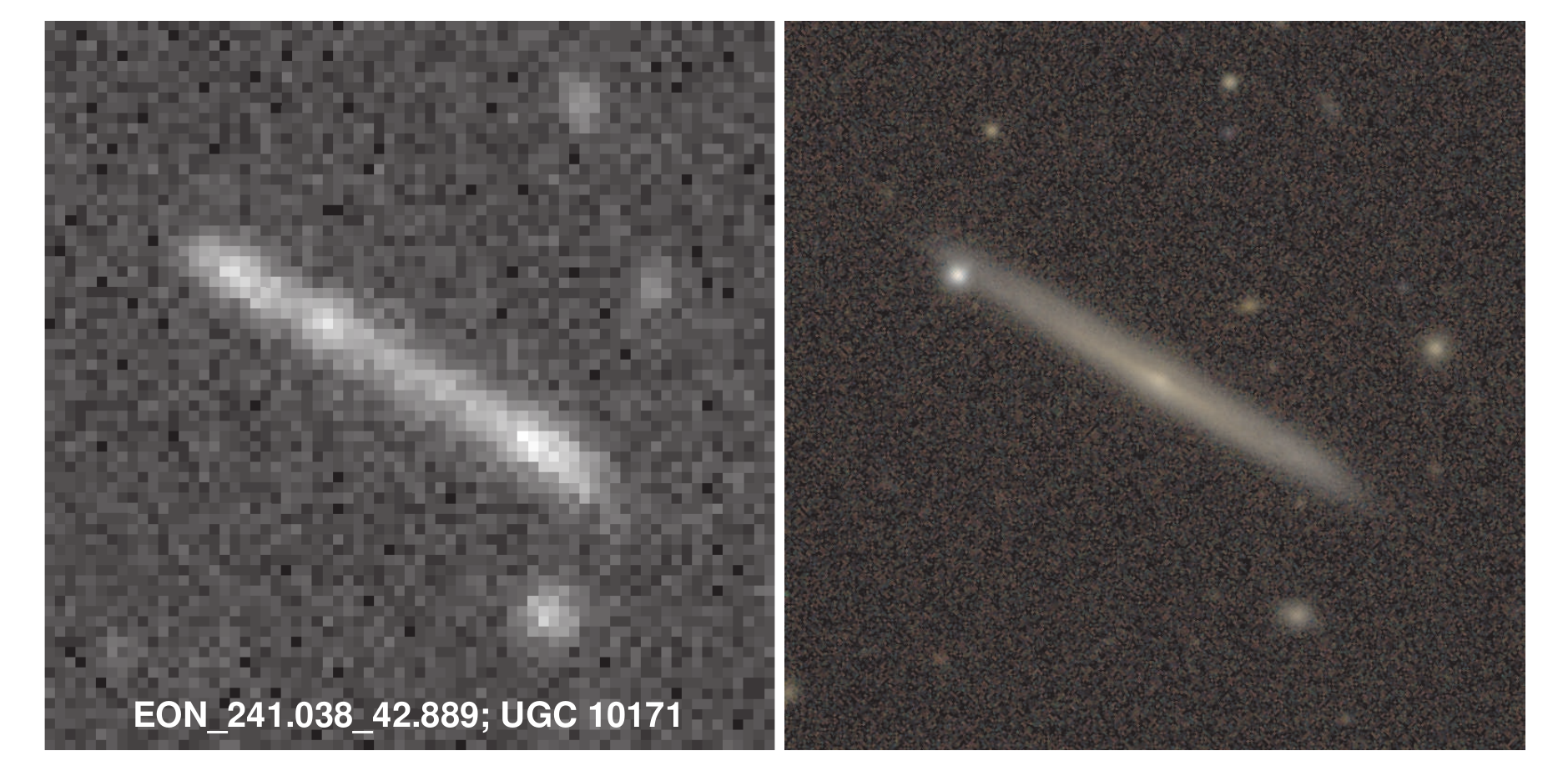}
}
\caption{\input{caption/fig_targets1.tex}} 
\end{figure}

\clearpage
\begin{figure}
\figurenum{1}
\center{
(v)\includegraphics[scale=0.7]{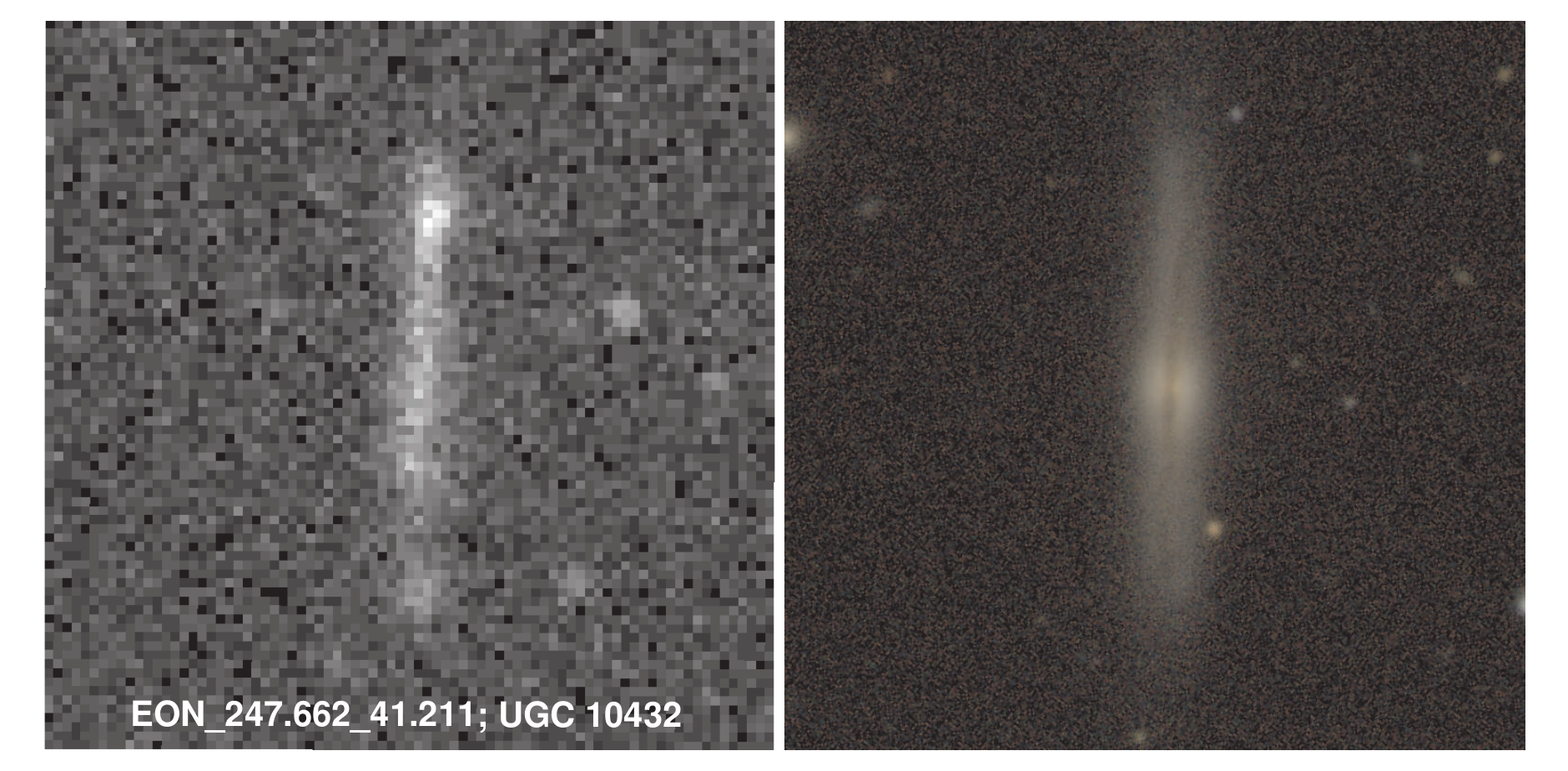}
(w)\includegraphics[scale=0.7]{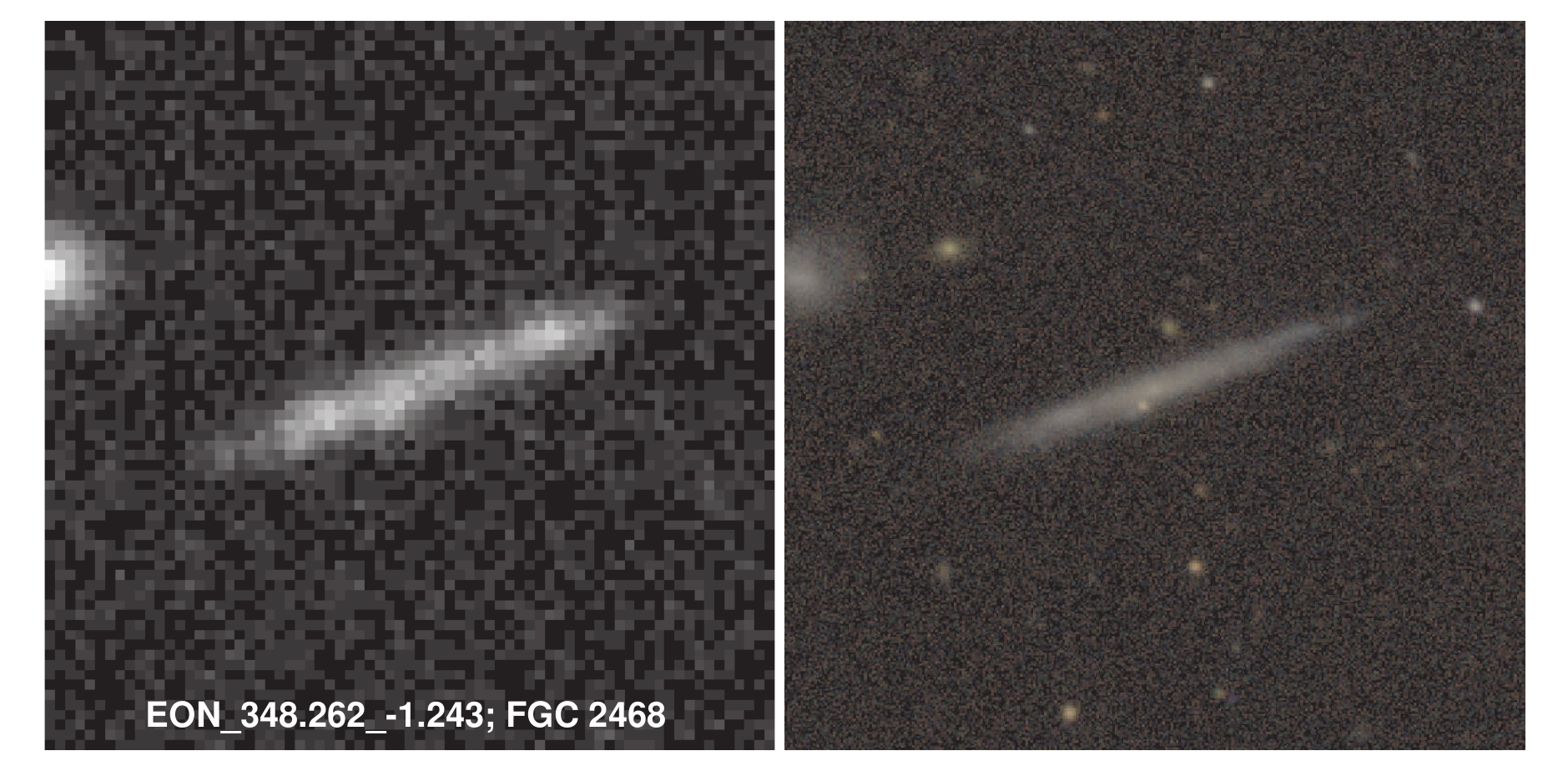}
}
\caption{\input{caption/fig_targets1.tex}} 
\end{figure}

%% file: caption/fig_targets0.tex
\galex{} FUV images (\emph{left}) and \sdss{} RGB images (\emph{right}) of the 23 target edge-on galaxies.
The \sdss{} RGB image is made from $g$=blue, $r$=green, and $i$=red.
The \galex{} image field-of-view is the same with the one in Fig.~\ref{fig-fitdata_st}-\ref{fig-fitdata_en}.

%% file: caption/fig_targets1.tex
Continued.

%% file: fig/fig_fitdata.tex
\clearpage
\begin{figure}
\center{
\includegraphics[scale=0.45]{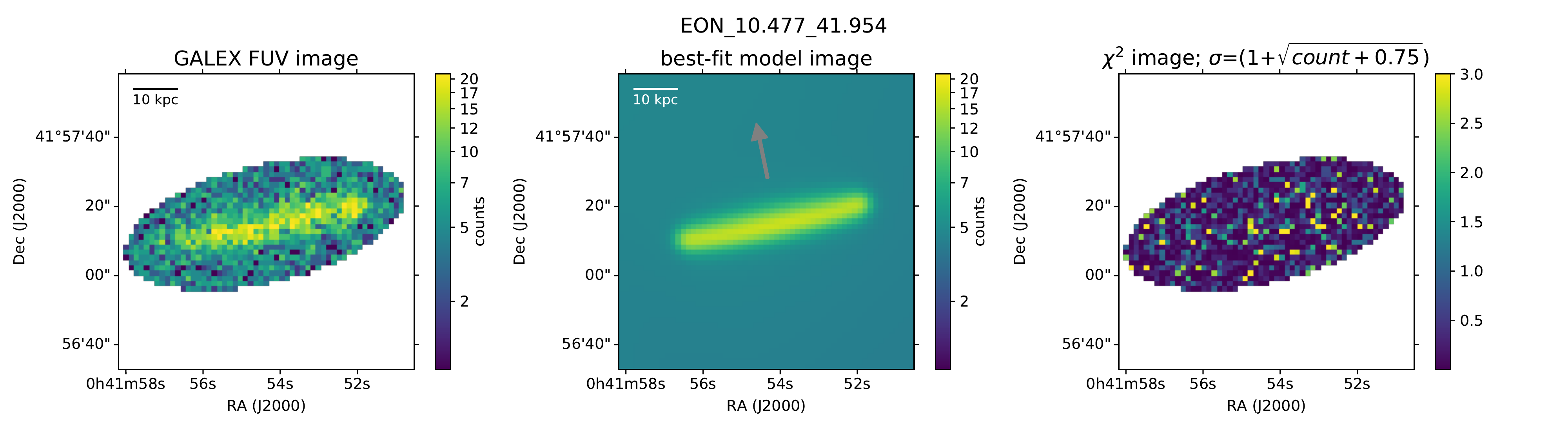}
\includegraphics[scale=0.45]{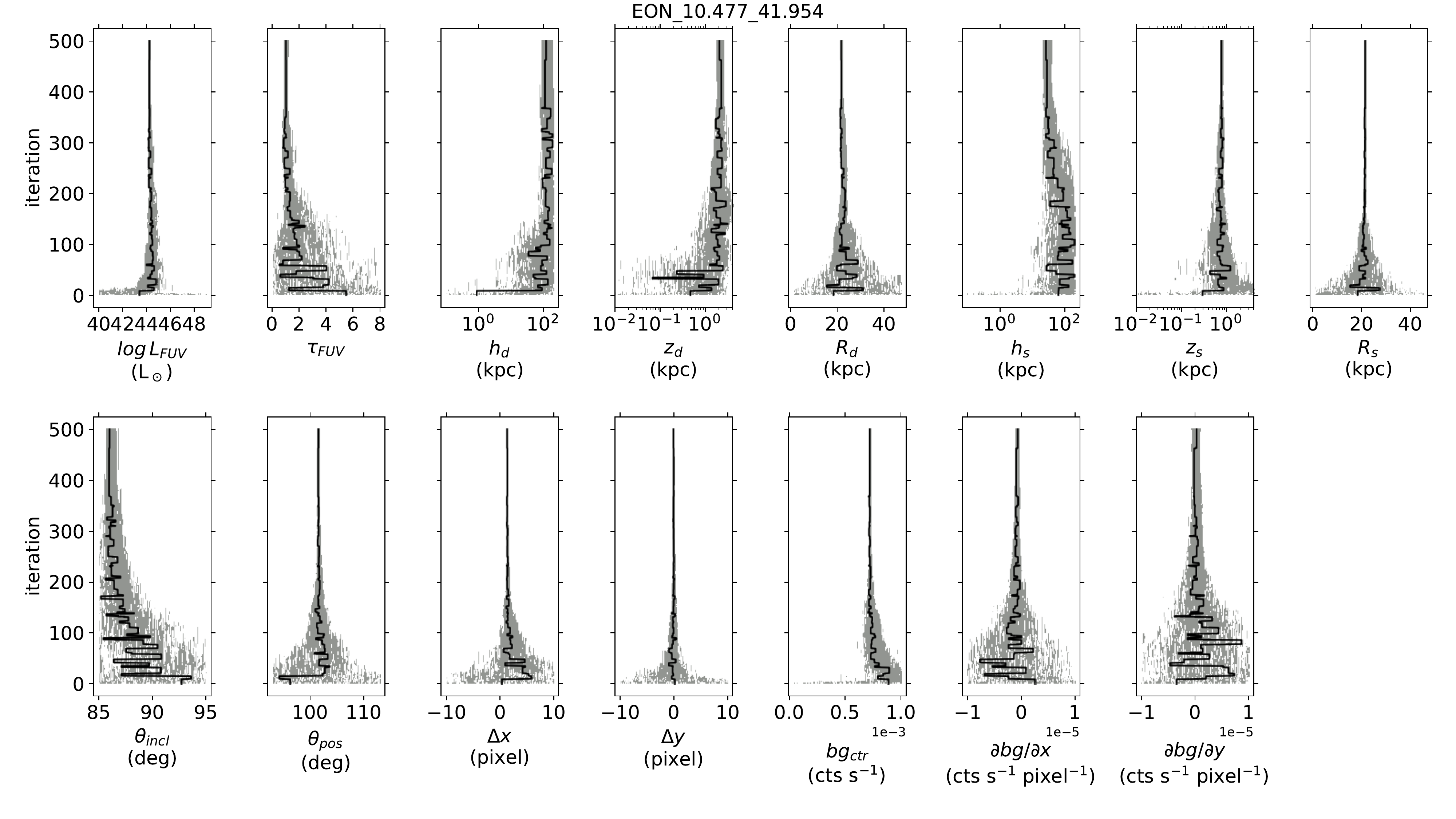}
}
\caption{Fitting results for EON\_10.477\_41.954. \input{caption/fig_fitdata0.tex}} 
\end{figure}

\clearpage
\begin{figure}
\figurenum{6}
\center{
\includegraphics[scale=0.45]{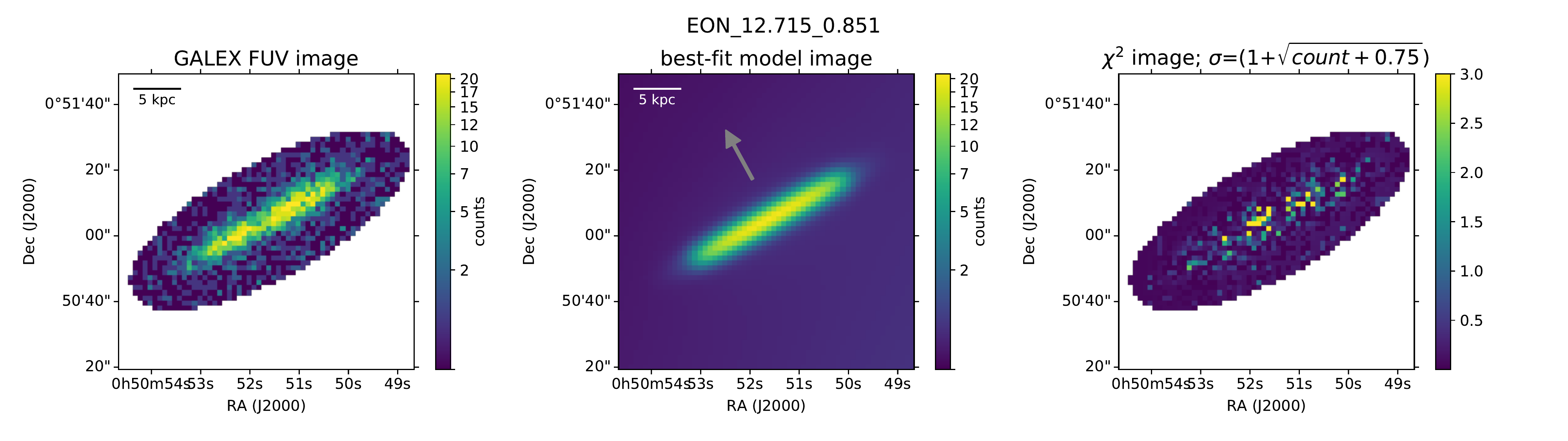}
\includegraphics[scale=0.45]{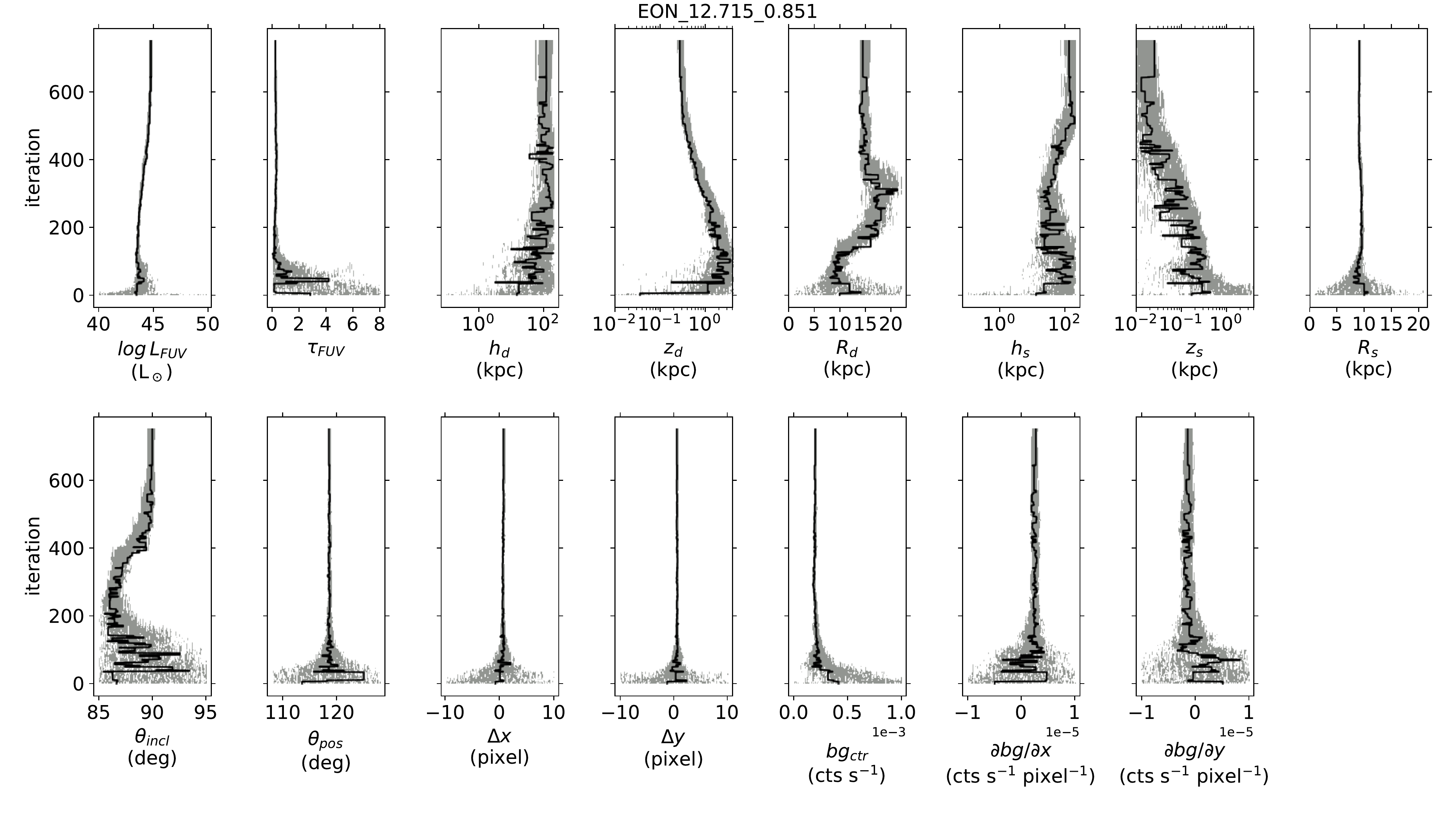}
}
\caption{Fitting results for EON\_12.715\_0.851. \input{caption/fig_fitdata1.tex}} 
\end{figure}

\clearpage
\begin{figure}
\figurenum{7}
\center{
\includegraphics[scale=0.45]{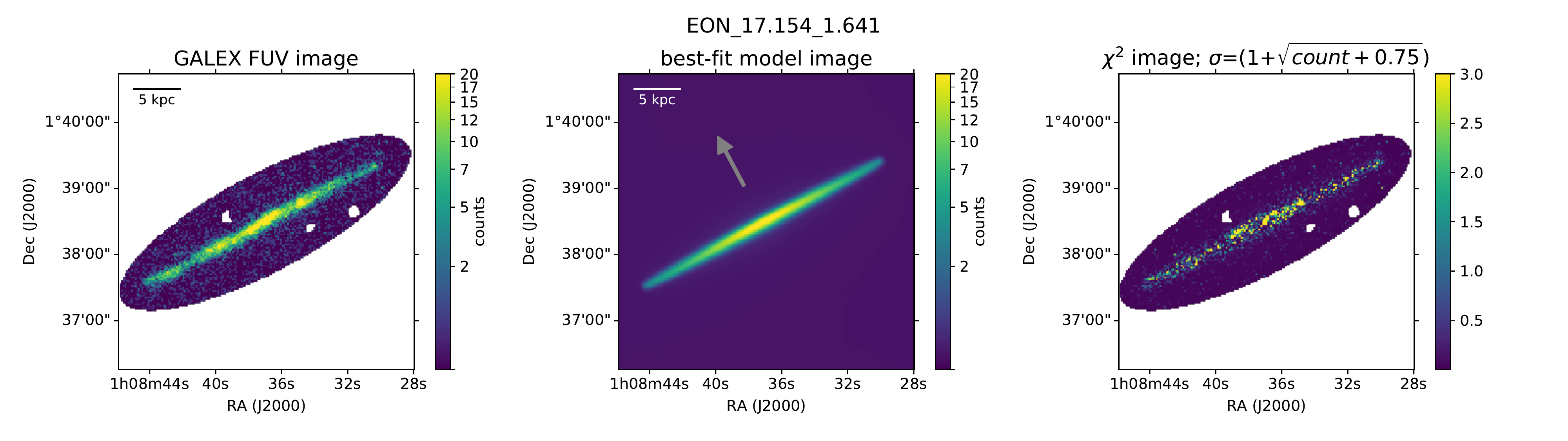}
\includegraphics[scale=0.45]{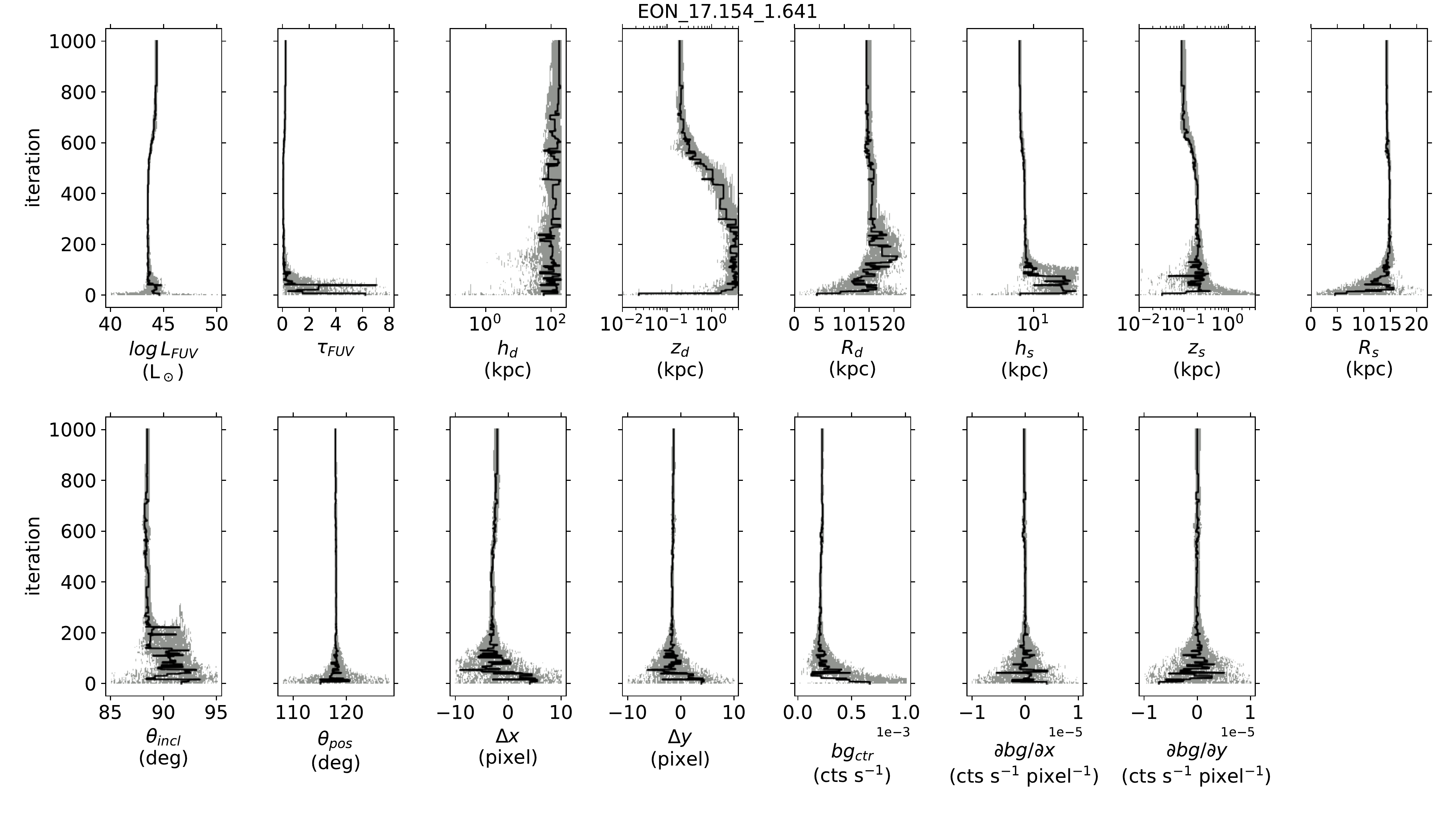}
}
\caption{Fitting results for EON\_17.154\_1.641. \input{caption/fig_fitdata1.tex}} 
\end{figure}

\clearpage
\begin{figure}
\figurenum{8}
\center{
\includegraphics[scale=0.45]{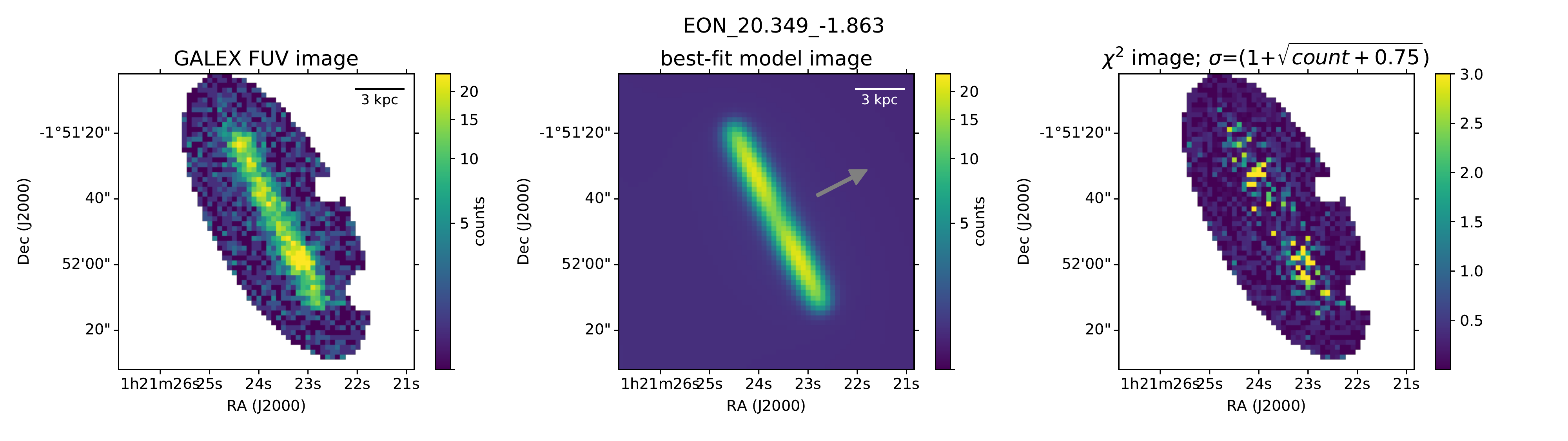}
\includegraphics[scale=0.45]{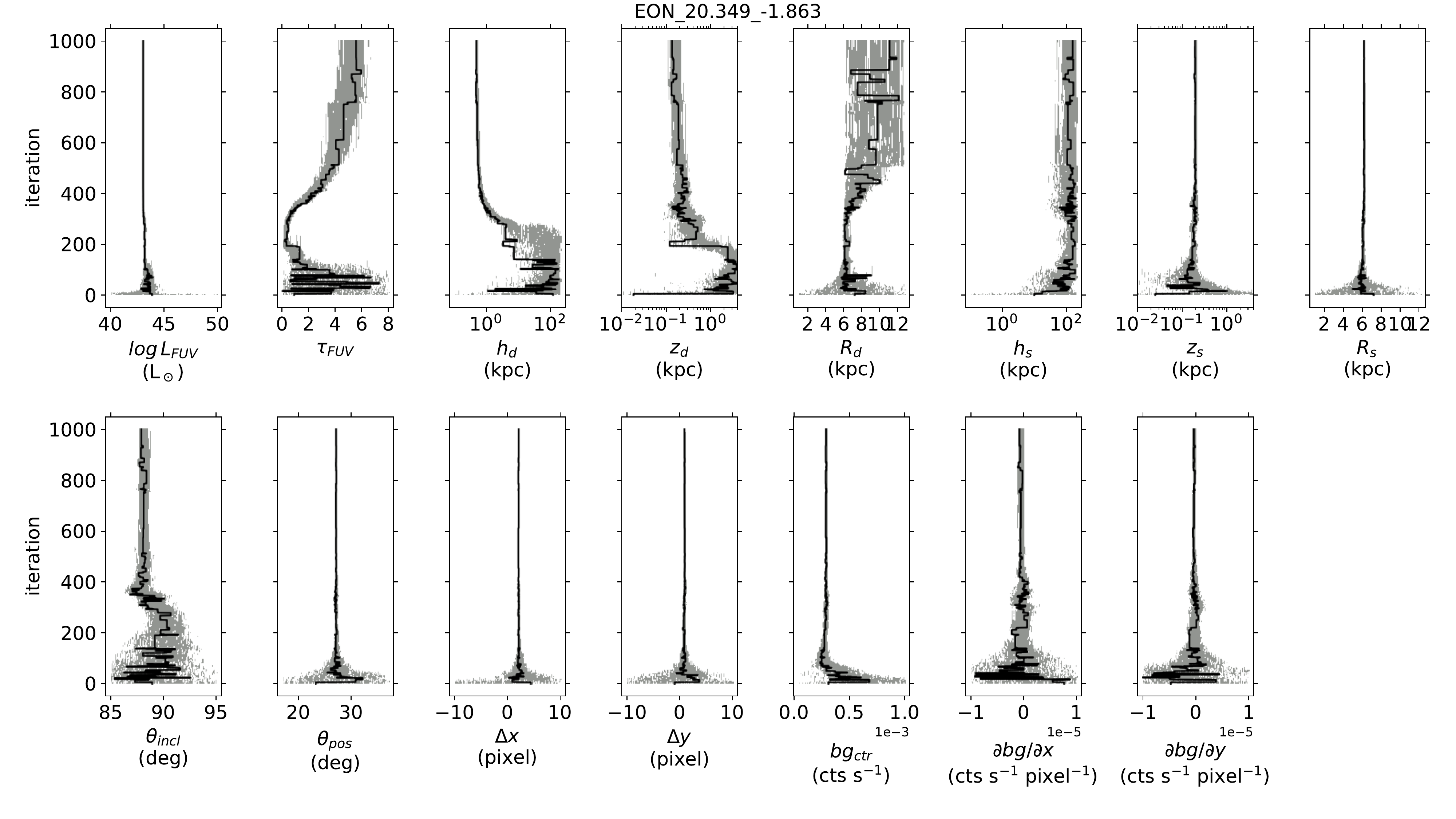}
}
\caption{Fitting results for EON\_20.349\_-1.863. \input{caption/fig_fitdata1.tex}} 
\end{figure}

\clearpage
\begin{figure}
\figurenum{9}
\center{
\includegraphics[scale=0.45]{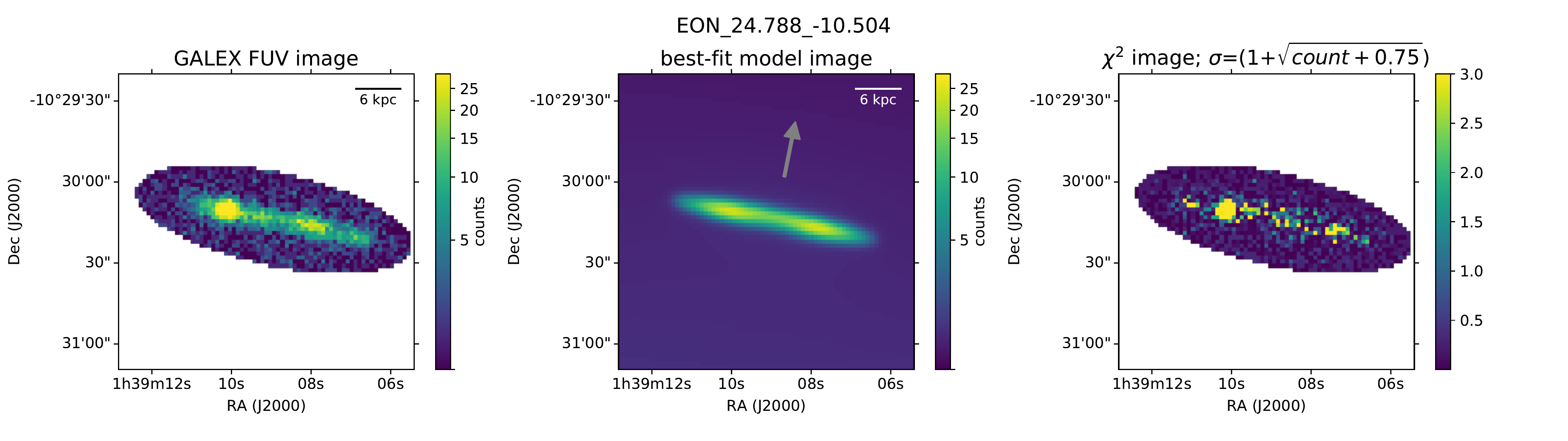}
\includegraphics[scale=0.45]{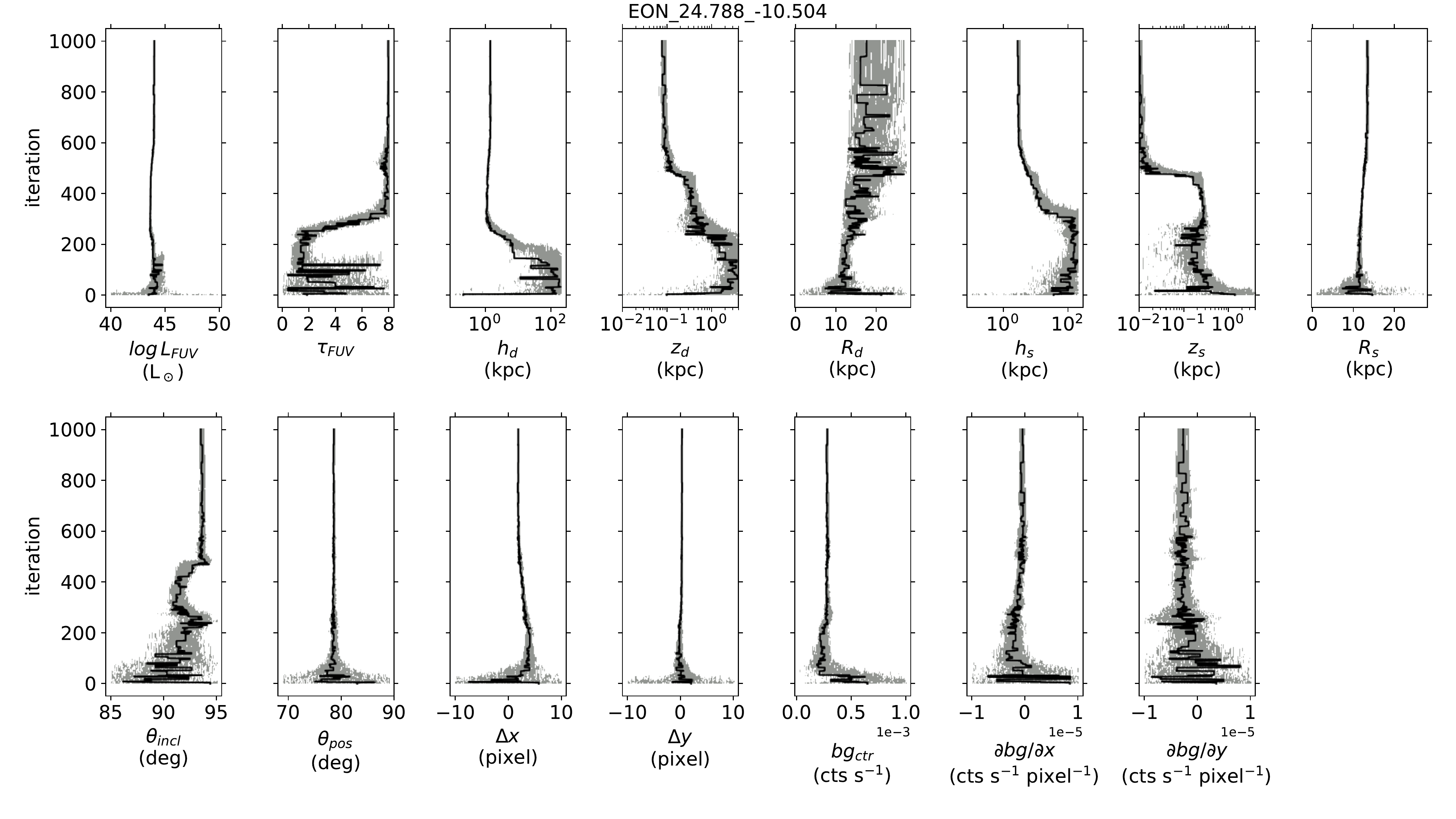}
}
\caption{Fitting results for EON\_24.788\_-10.504. \input{caption/fig_fitdata1.tex}} 
\end{figure}

\clearpage
\begin{figure}
\figurenum{10}
\center{
\includegraphics[scale=0.45]{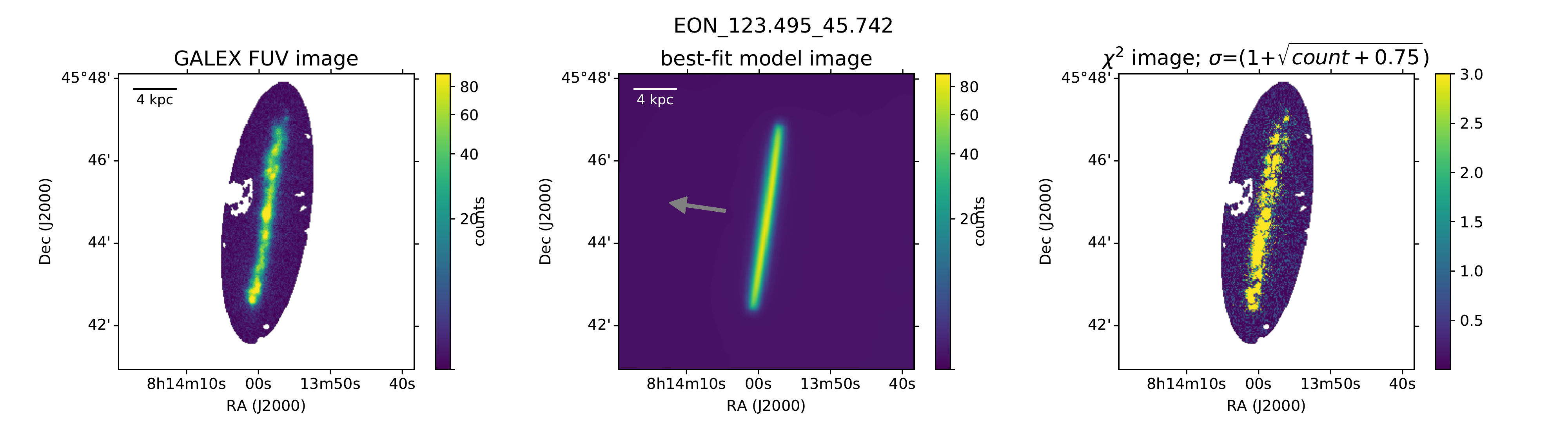}
\includegraphics[scale=0.45]{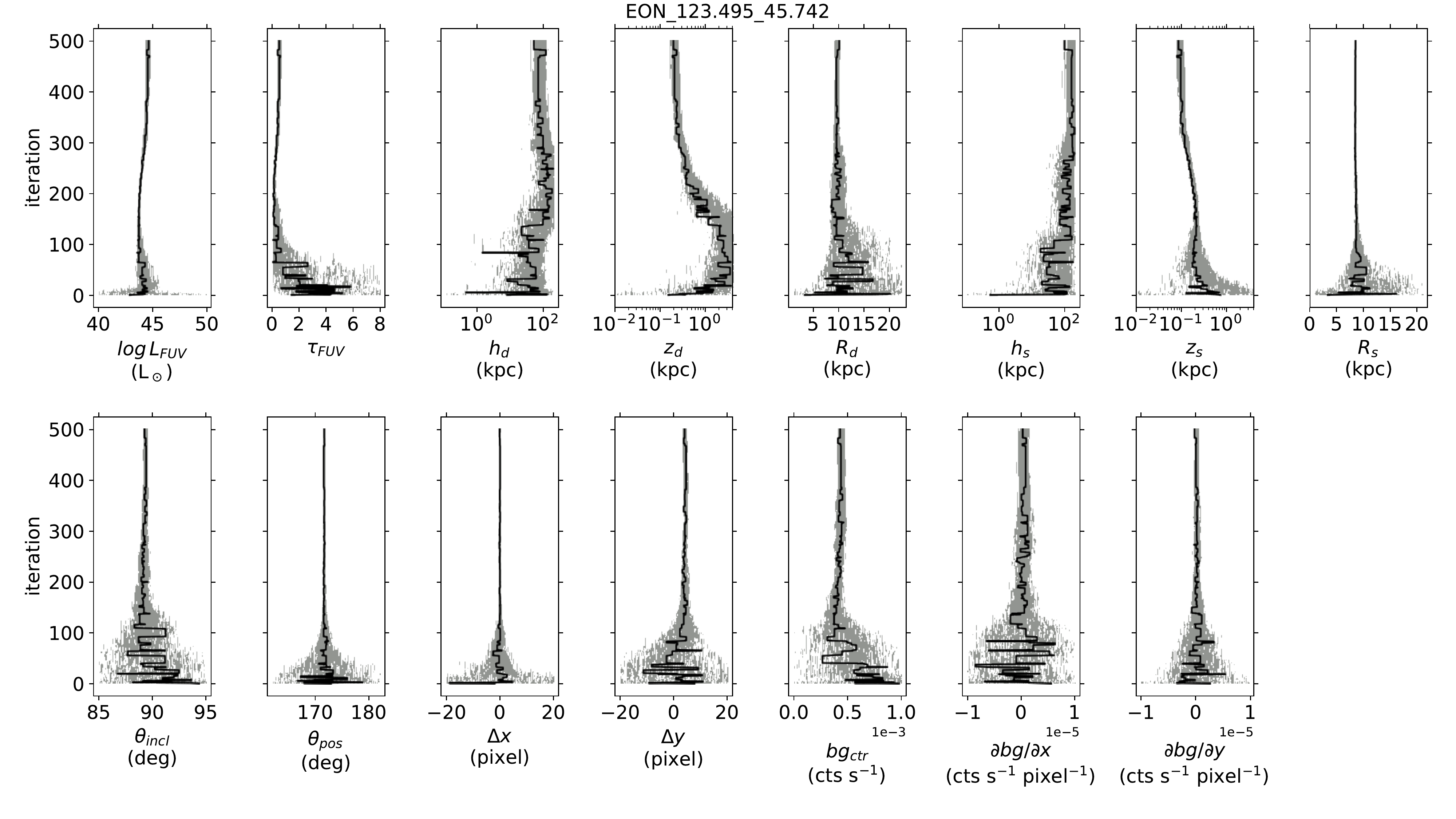}
}
\caption{Fitting results for EON\_123.495\_45.742. \input{caption/fig_fitdata1.tex}} 
\end{figure}

\clearpage
\begin{figure}
\figurenum{11}
\center{
\includegraphics[scale=0.45]{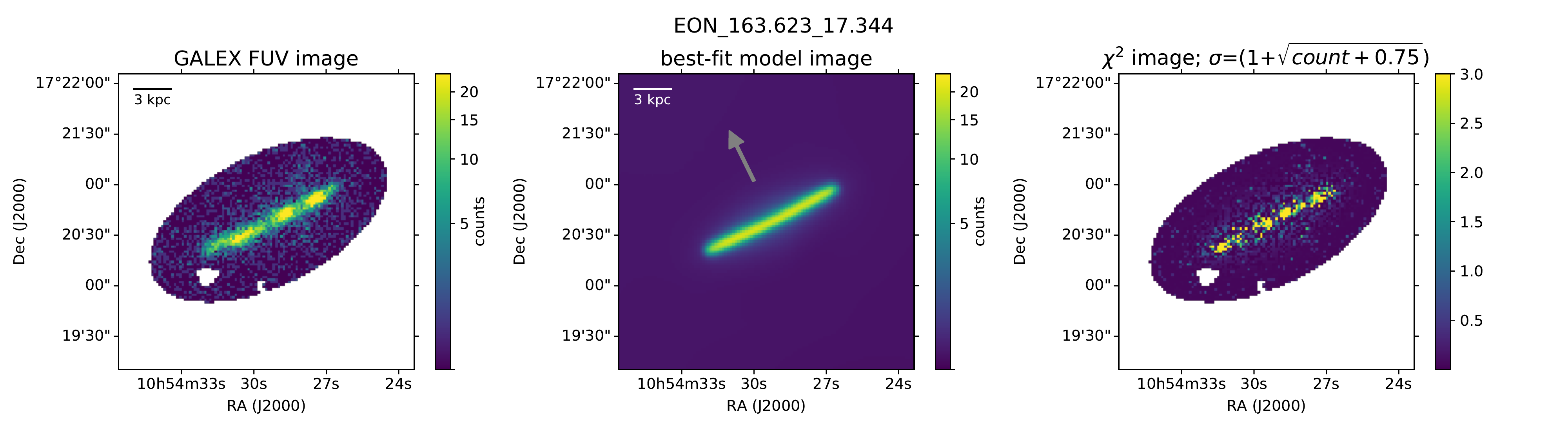}
\includegraphics[scale=0.45]{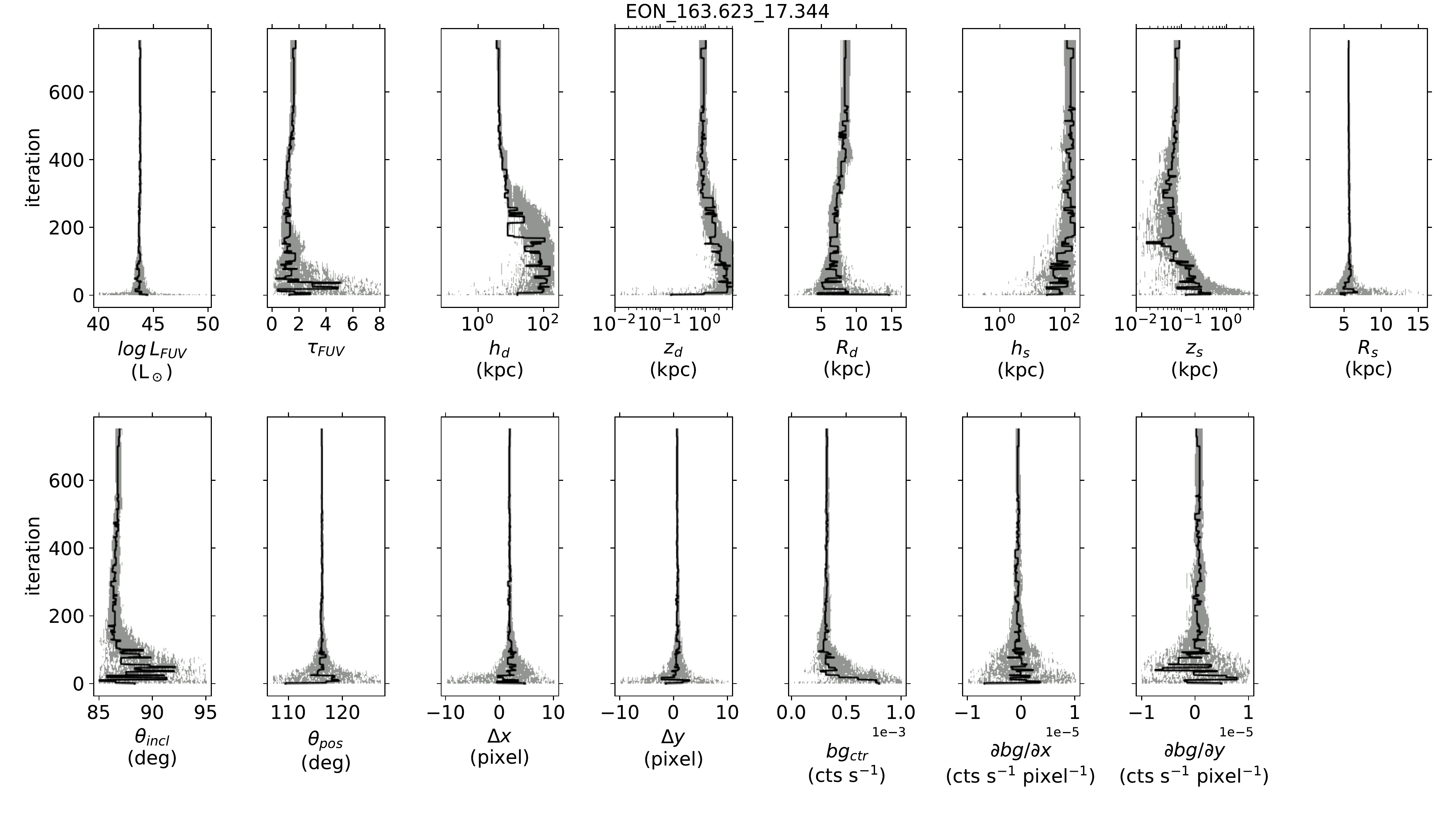}
}
\caption{Fitting results for EON\_163.623\_17.344. \input{caption/fig_fitdata1.tex}} 
\end{figure}

\clearpage
\begin{figure}
\figurenum{12}
\center{
\includegraphics[scale=0.45]{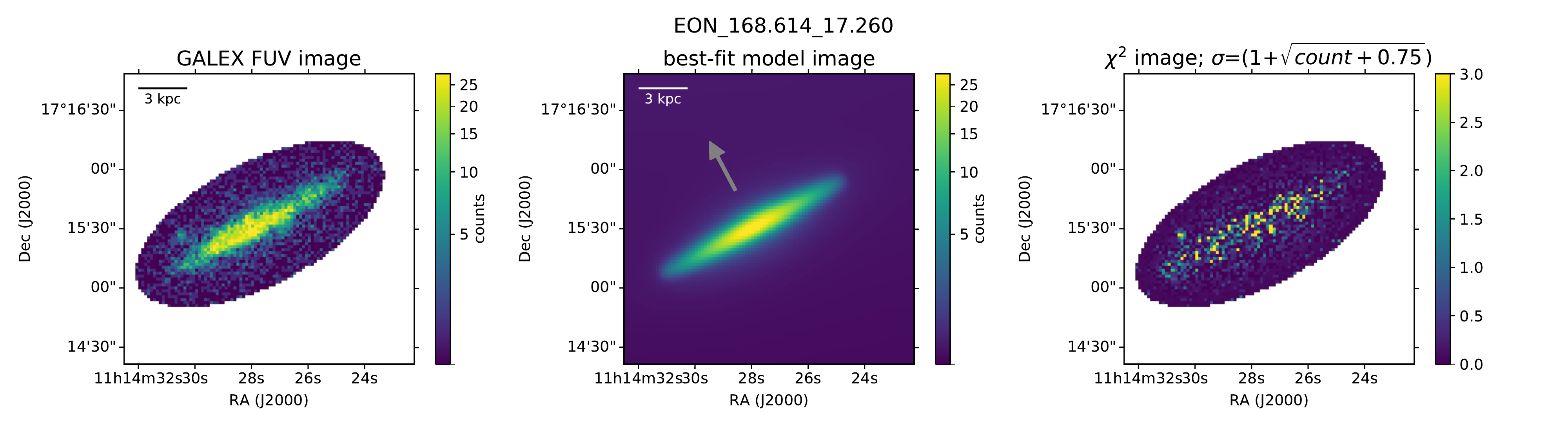}
\includegraphics[scale=0.45]{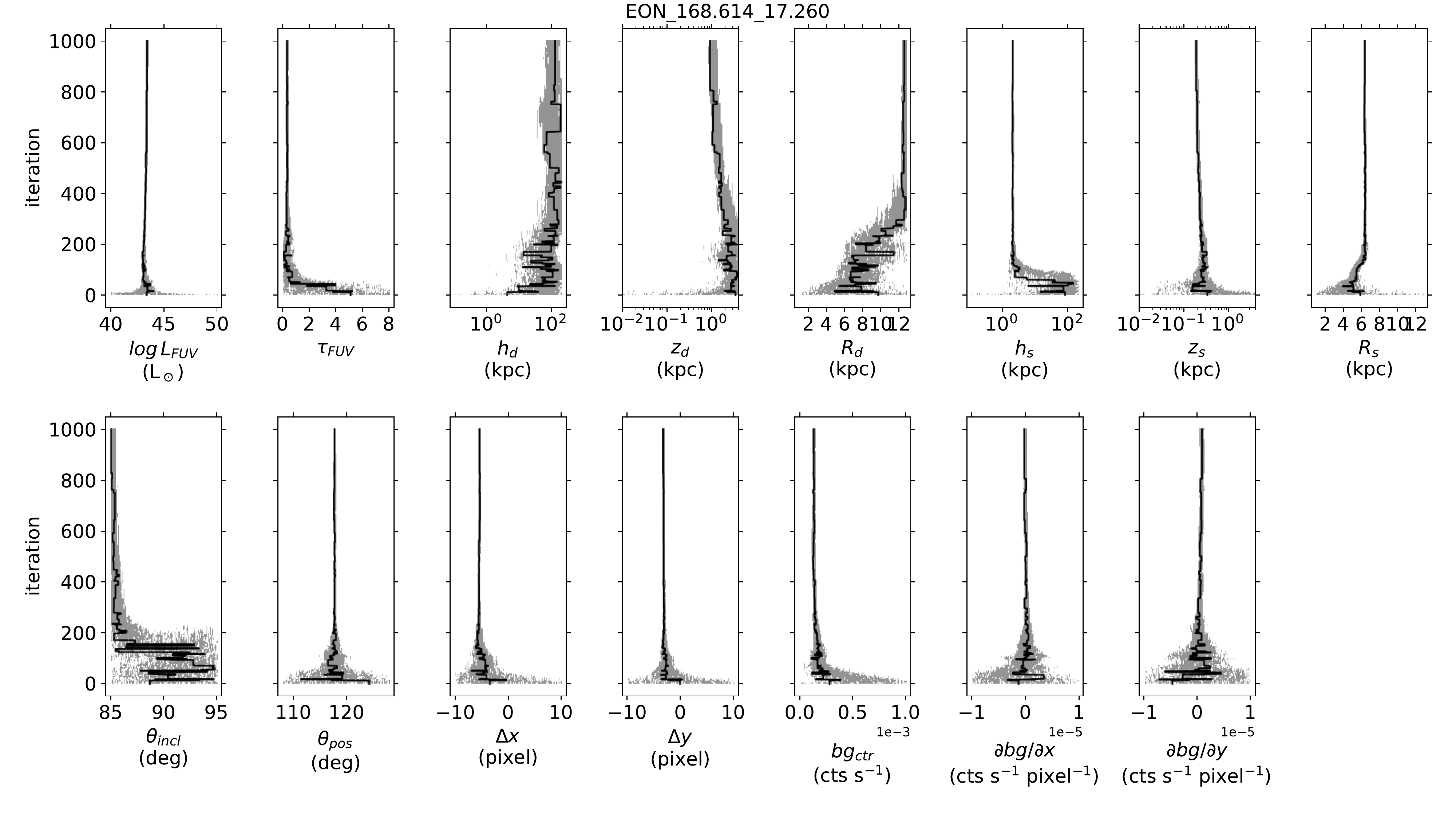}
}
\caption{Fitting results for EON\_168.614\_17.260. \input{caption/fig_fitdata1.tex}} 
\end{figure}

\clearpage
\begin{figure}
\figurenum{13}
\center{
\includegraphics[scale=0.45]{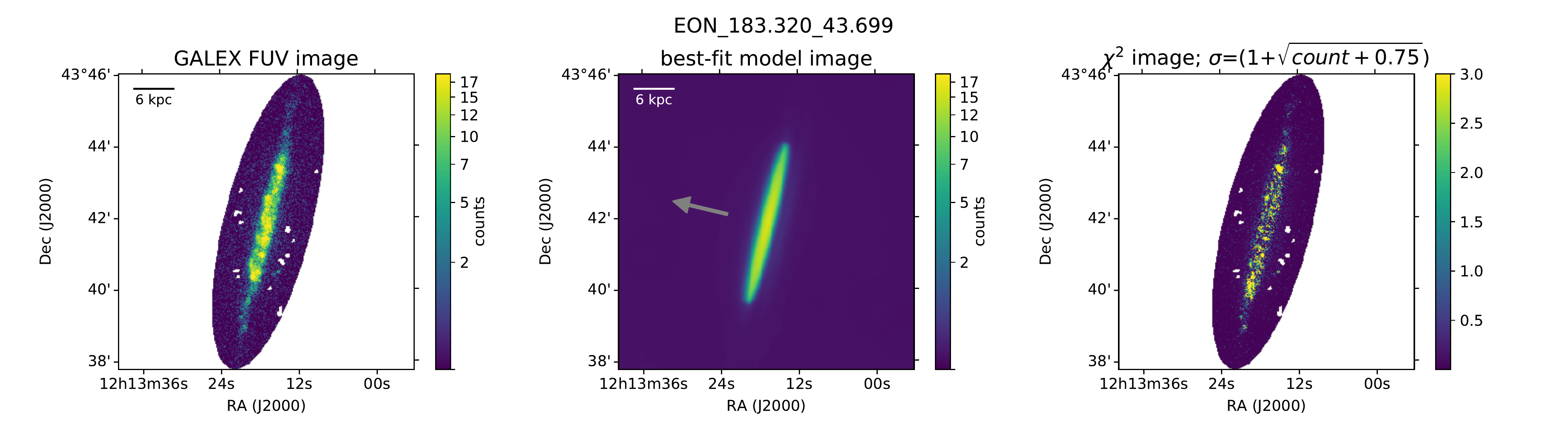}
\includegraphics[scale=0.45]{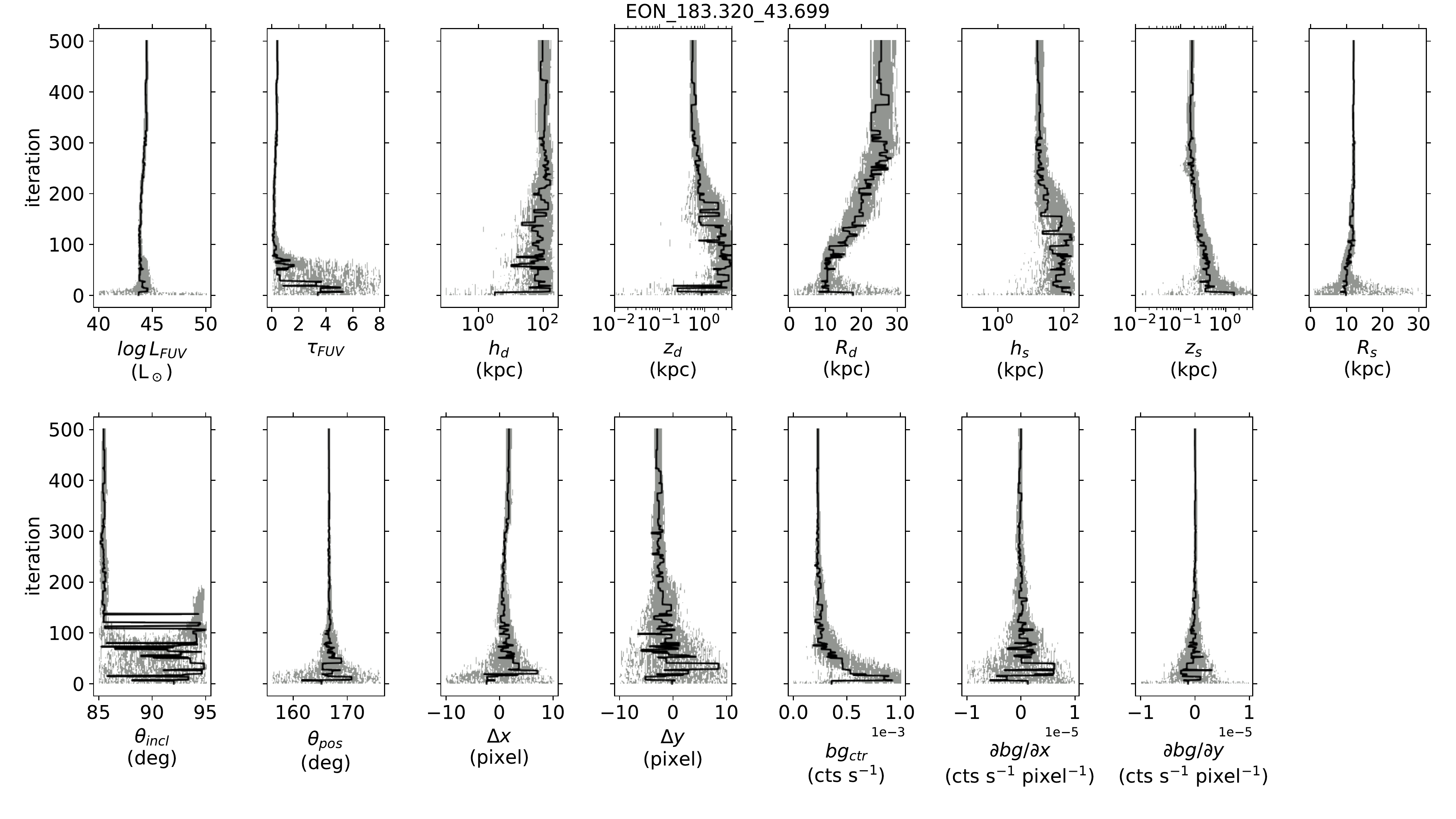}
}
\caption{Fitting results for EON\_183.320\_43.699. \input{caption/fig_fitdata1.tex}} 
\end{figure}

\clearpage
\begin{figure}
\figurenum{14}
\center{
\includegraphics[scale=0.45]{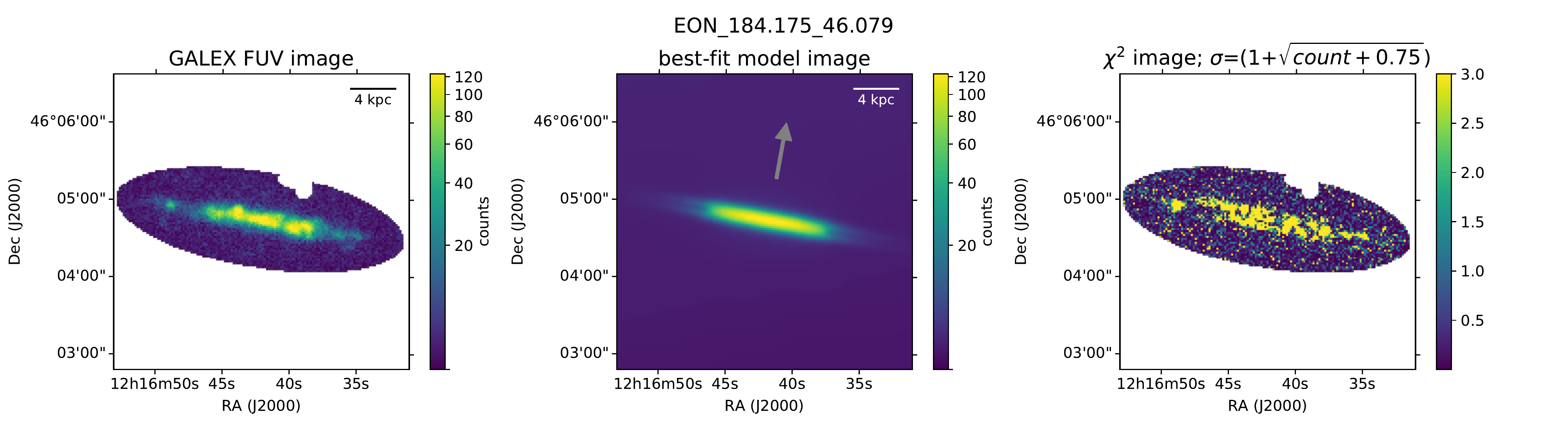}
\includegraphics[scale=0.45]{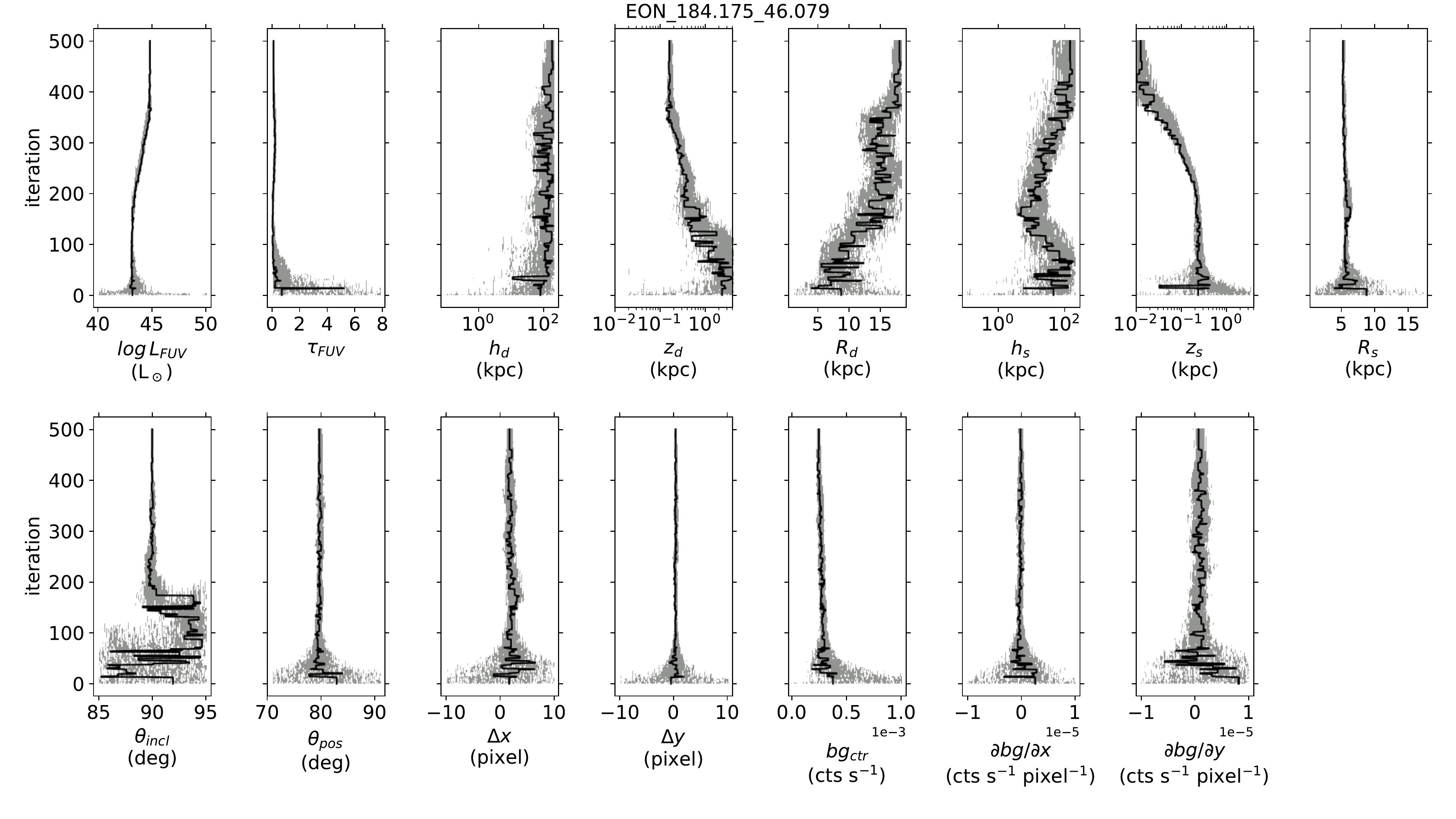}
}
\caption{Fitting results for EON\_184.175\_46.079. \input{caption/fig_fitdata1.tex}} 
\end{figure}

\clearpage
\begin{figure}
\figurenum{15}
\center{
\includegraphics[scale=0.45]{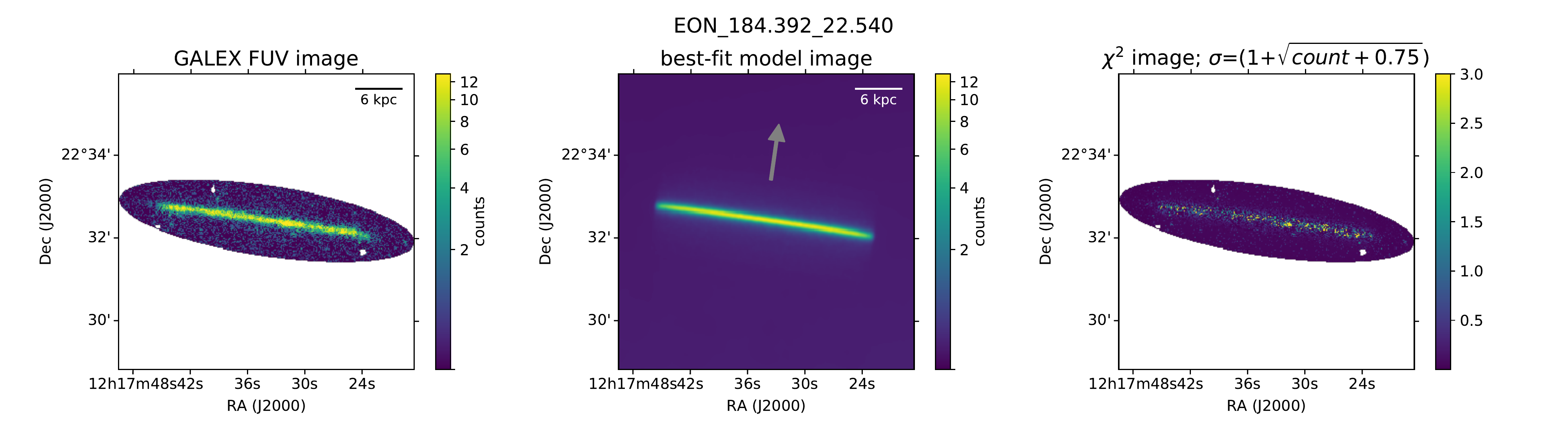}
\includegraphics[scale=0.45]{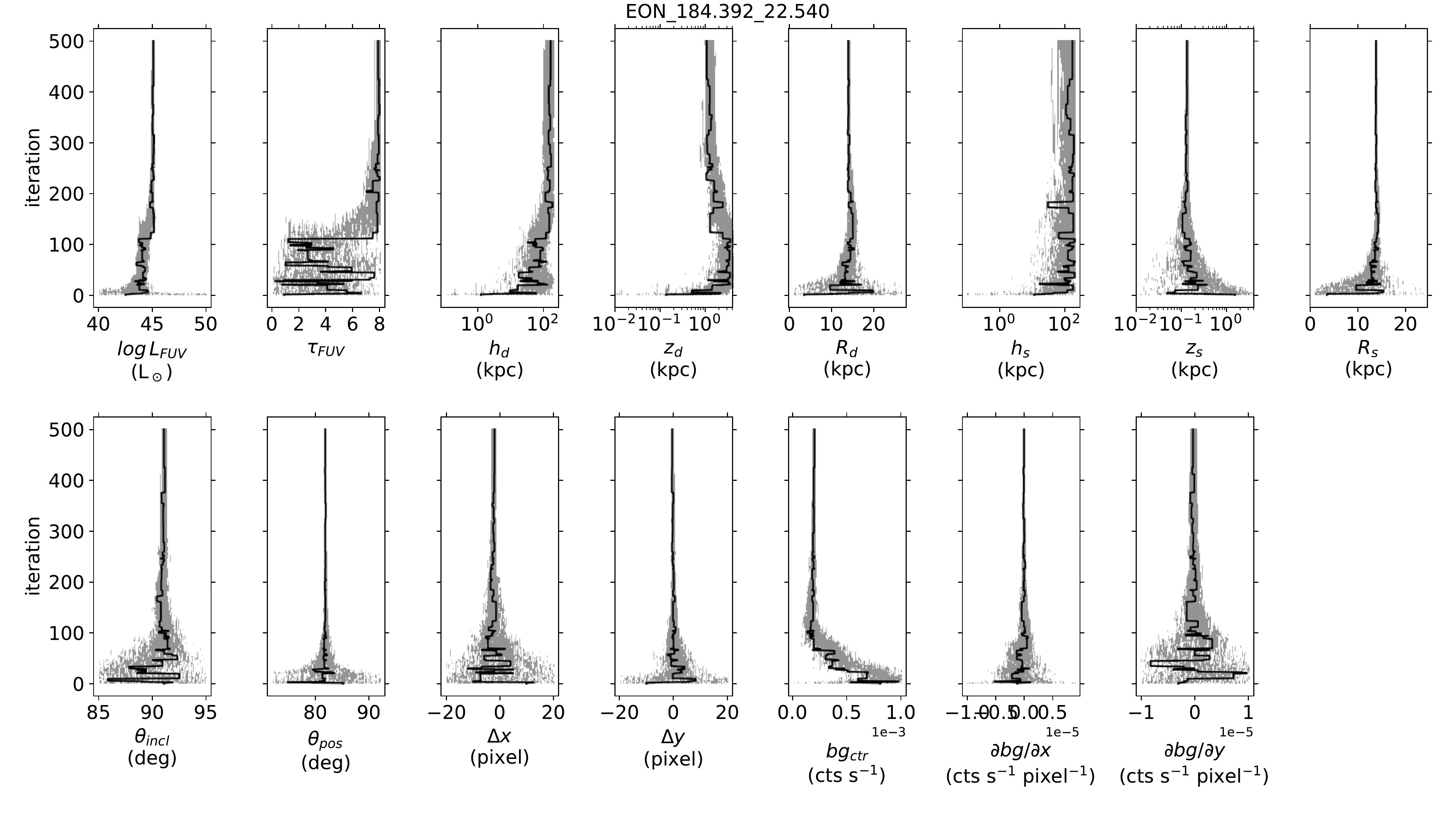}
}
\caption{Fitting results for EON\_184.392\_22.540. \input{caption/fig_fitdata1.tex}} 
\end{figure}

\clearpage
\begin{figure}
\figurenum{16}
\center{
\includegraphics[scale=0.45]{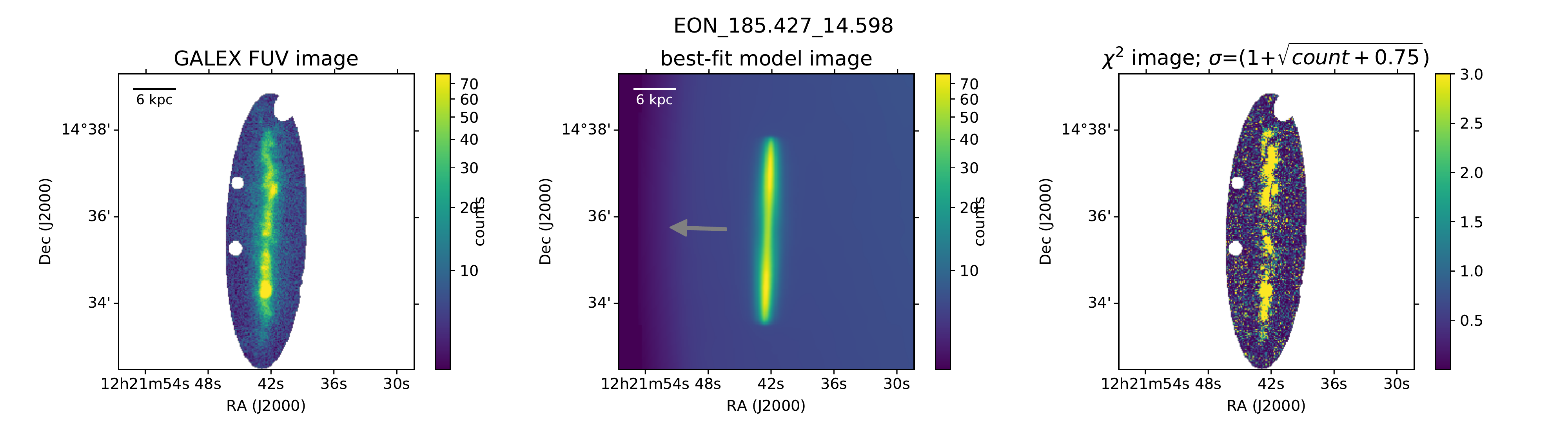}
\includegraphics[scale=0.45]{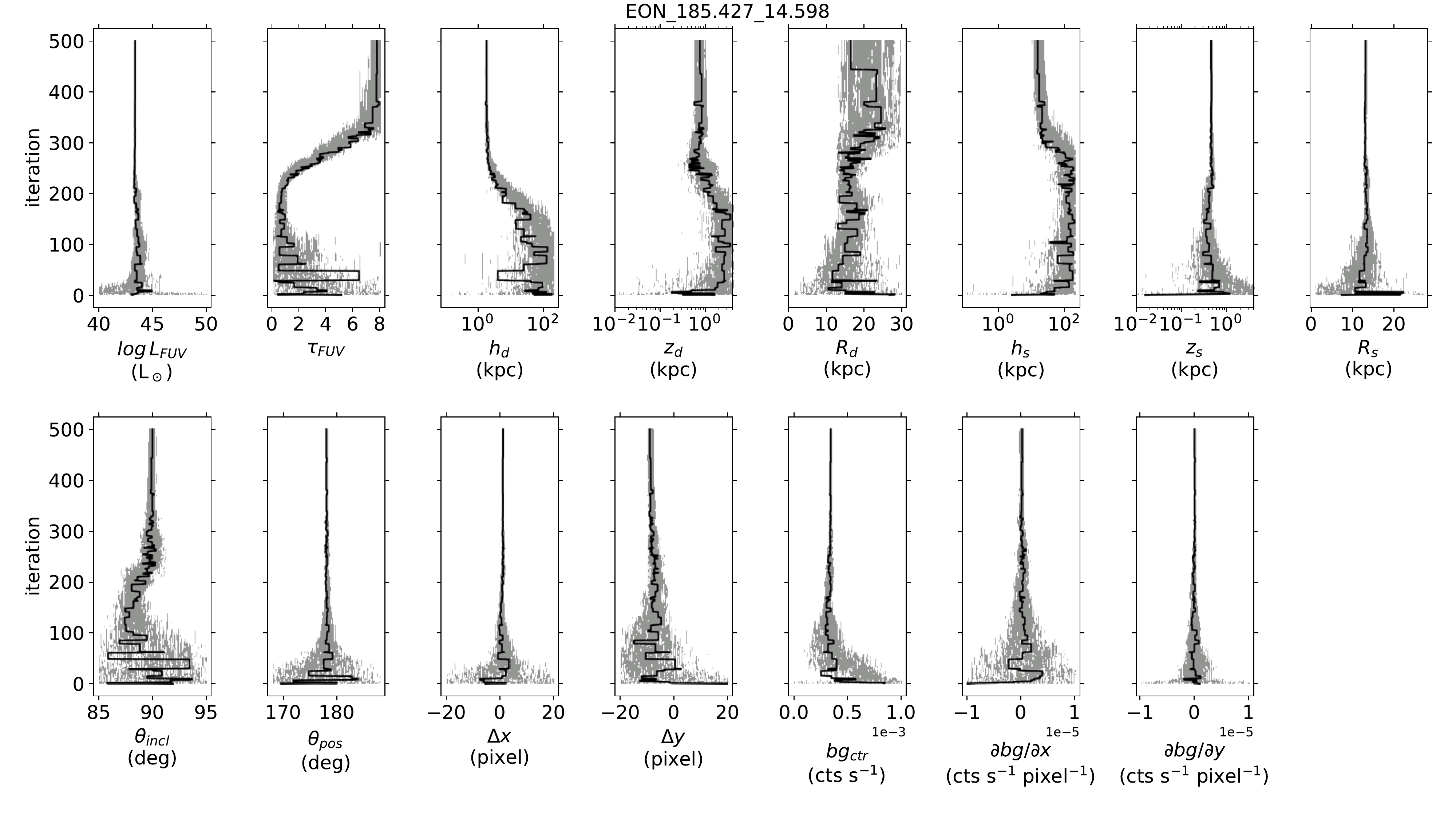}
}
\caption{Fitting results for EON\_185.427\_14.598. \input{caption/fig_fitdata1.tex}} 
\end{figure}

\clearpage
\begin{figure}
\figurenum{17}
\center{
\includegraphics[scale=0.45]{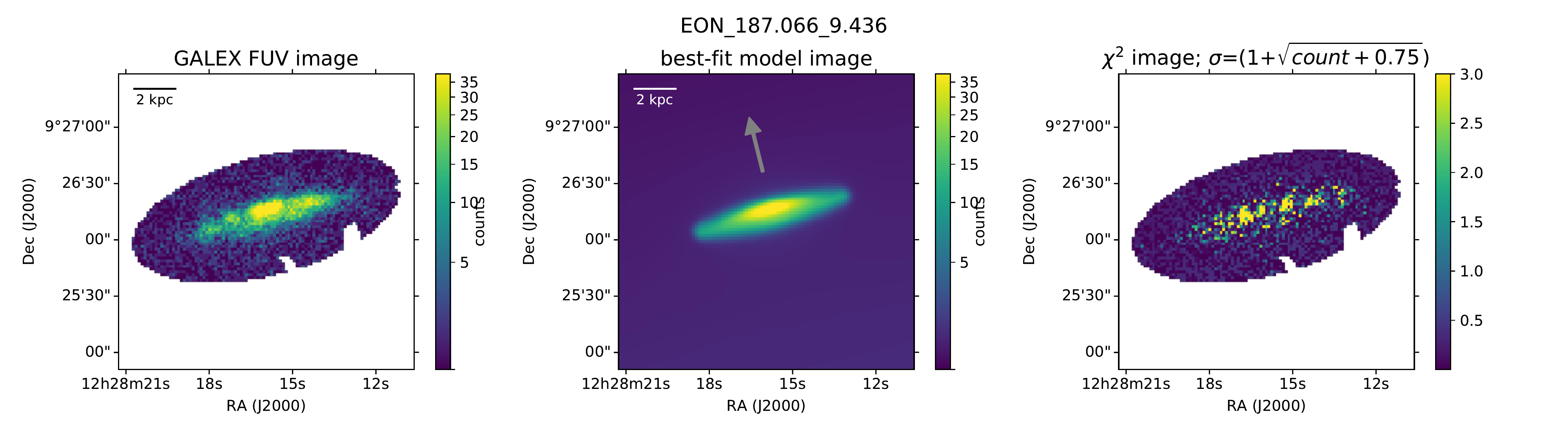}
\includegraphics[scale=0.45]{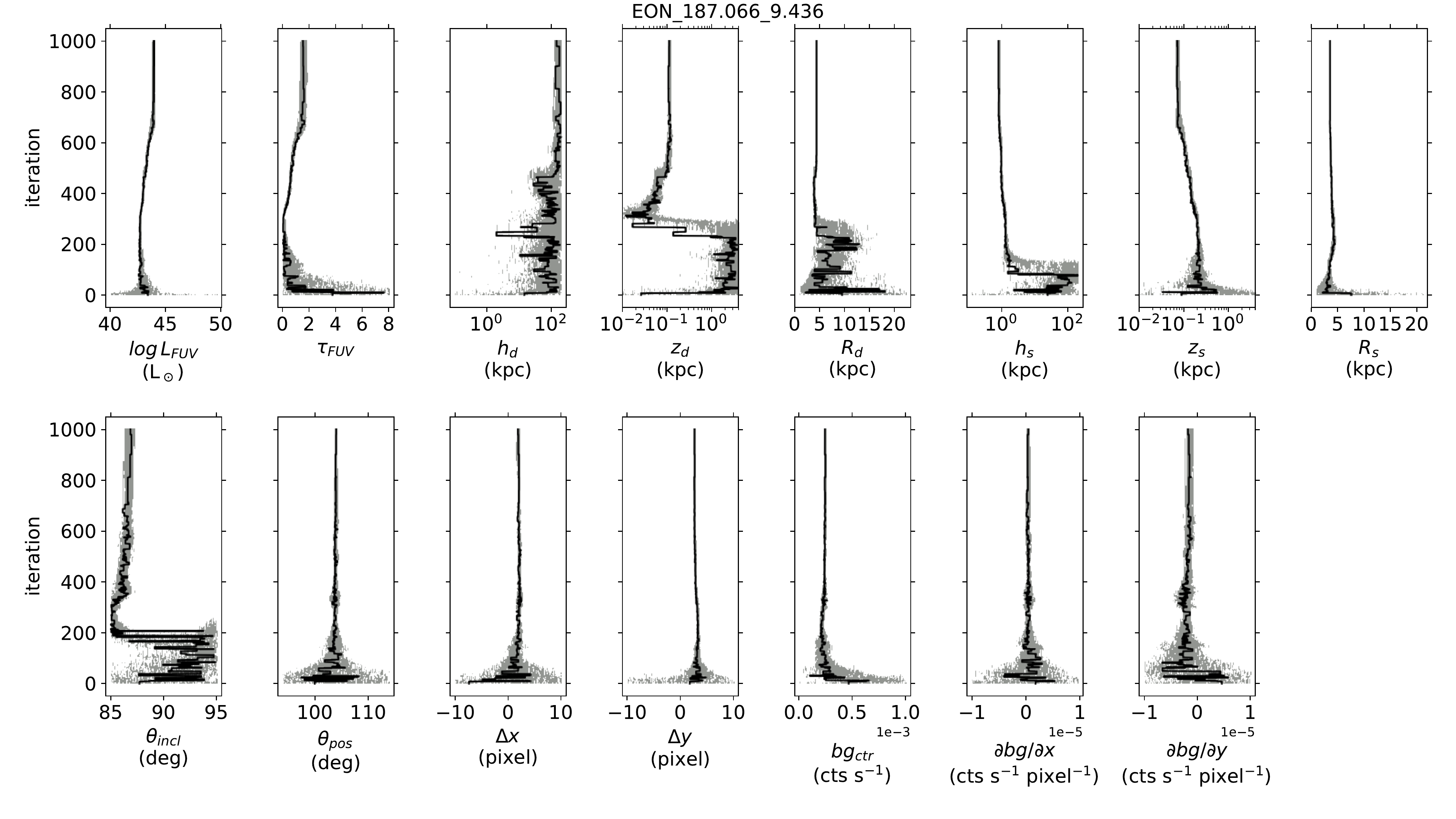}
}
\caption{Fitting results for EON\_187.066\_9.436. \input{caption/fig_fitdata1.tex}} 
\end{figure}

\clearpage
\begin{figure}
\figurenum{18}
\center{
\includegraphics[scale=0.45]{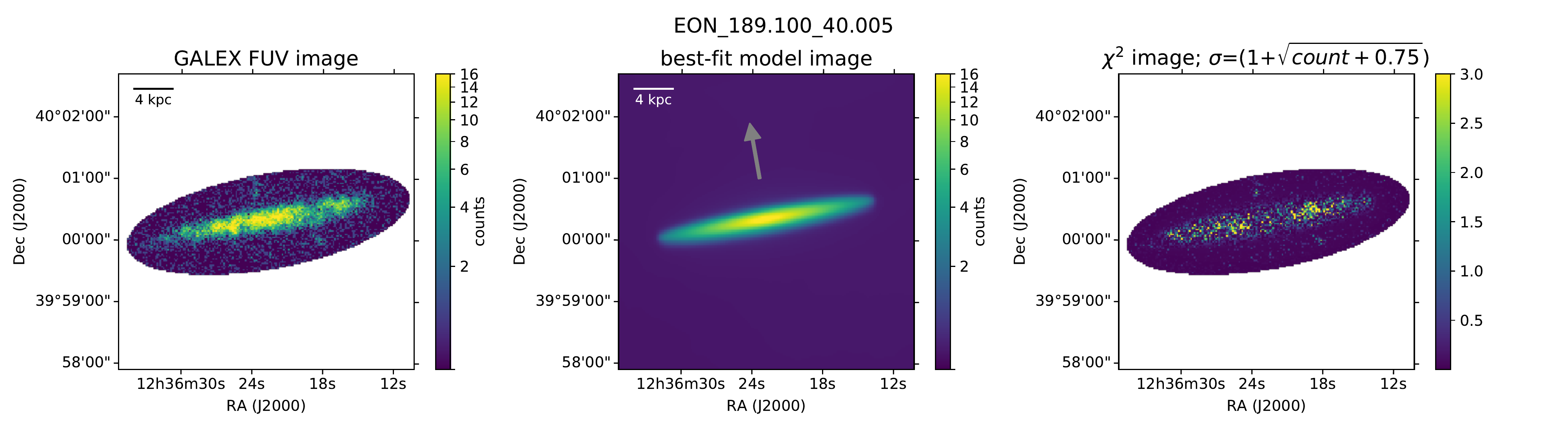}
\includegraphics[scale=0.45]{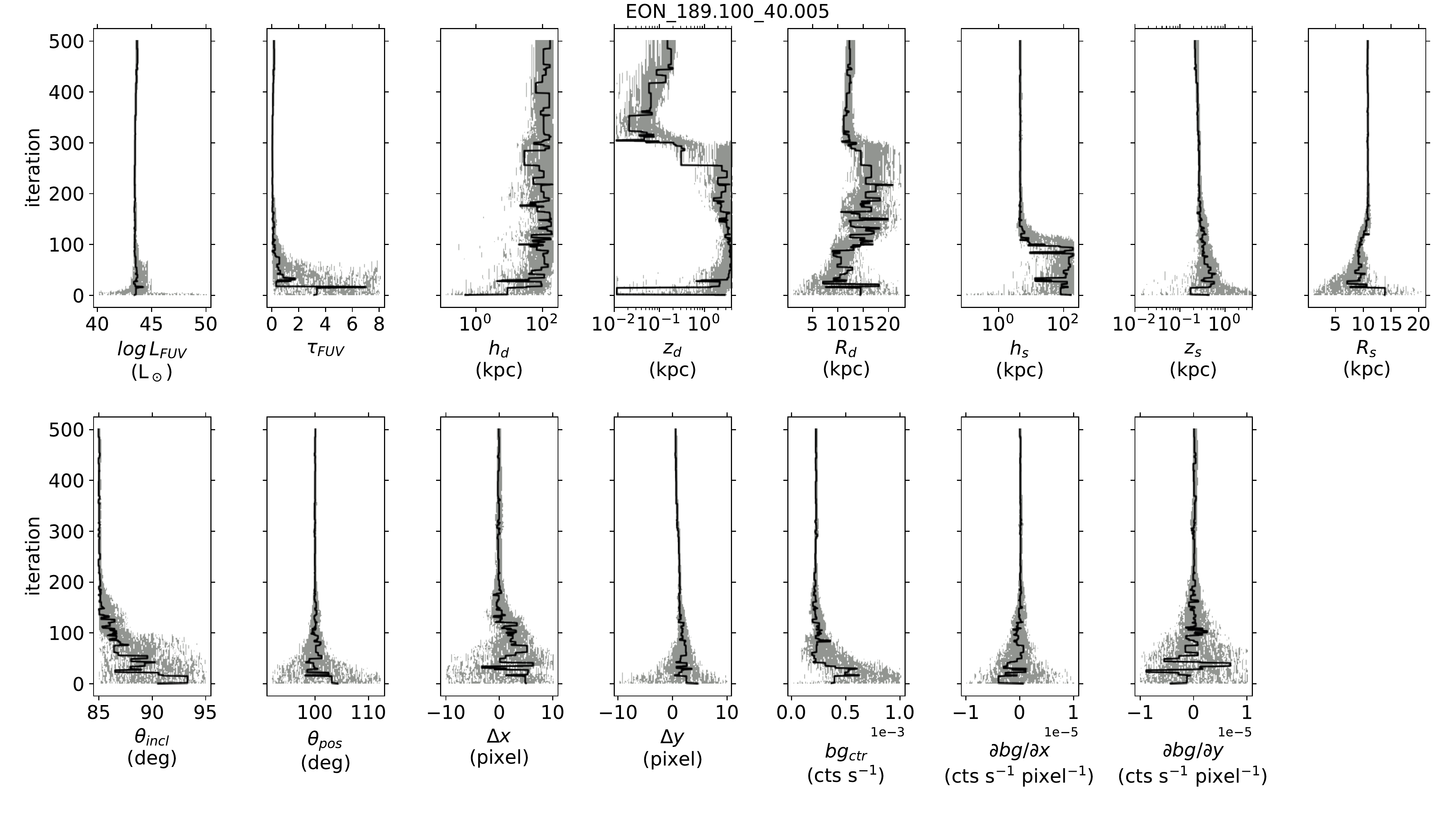}
}
\caption{Fitting results for EON\_189.100\_40.005. \input{caption/fig_fitdata1.tex}} 
\end{figure}

\clearpage
\begin{figure}
\figurenum{19}
\center{
\includegraphics[scale=0.45]{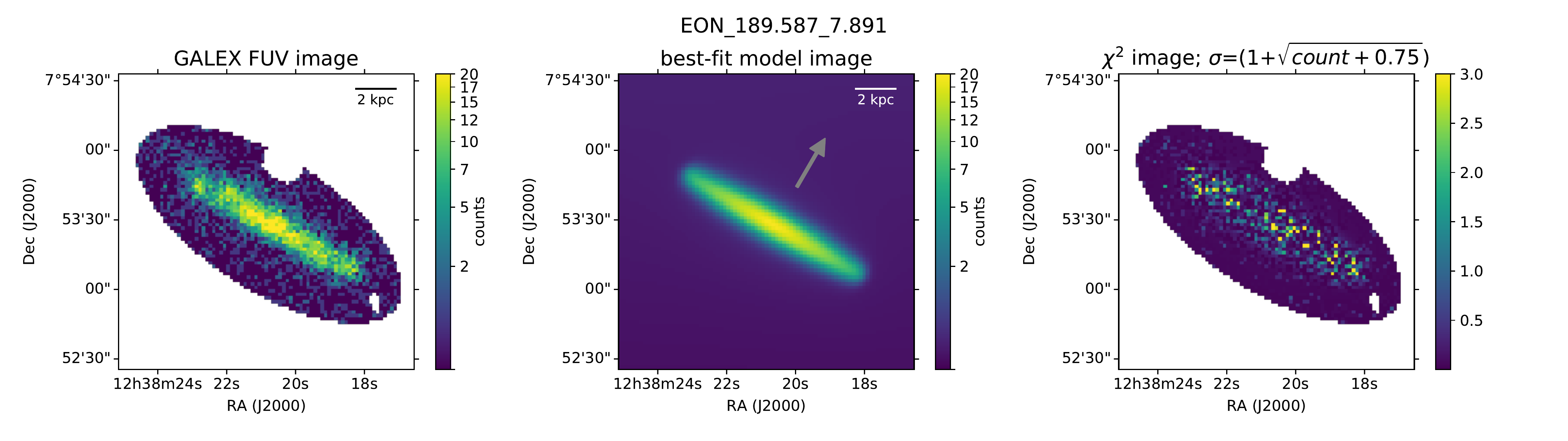}
\includegraphics[scale=0.45]{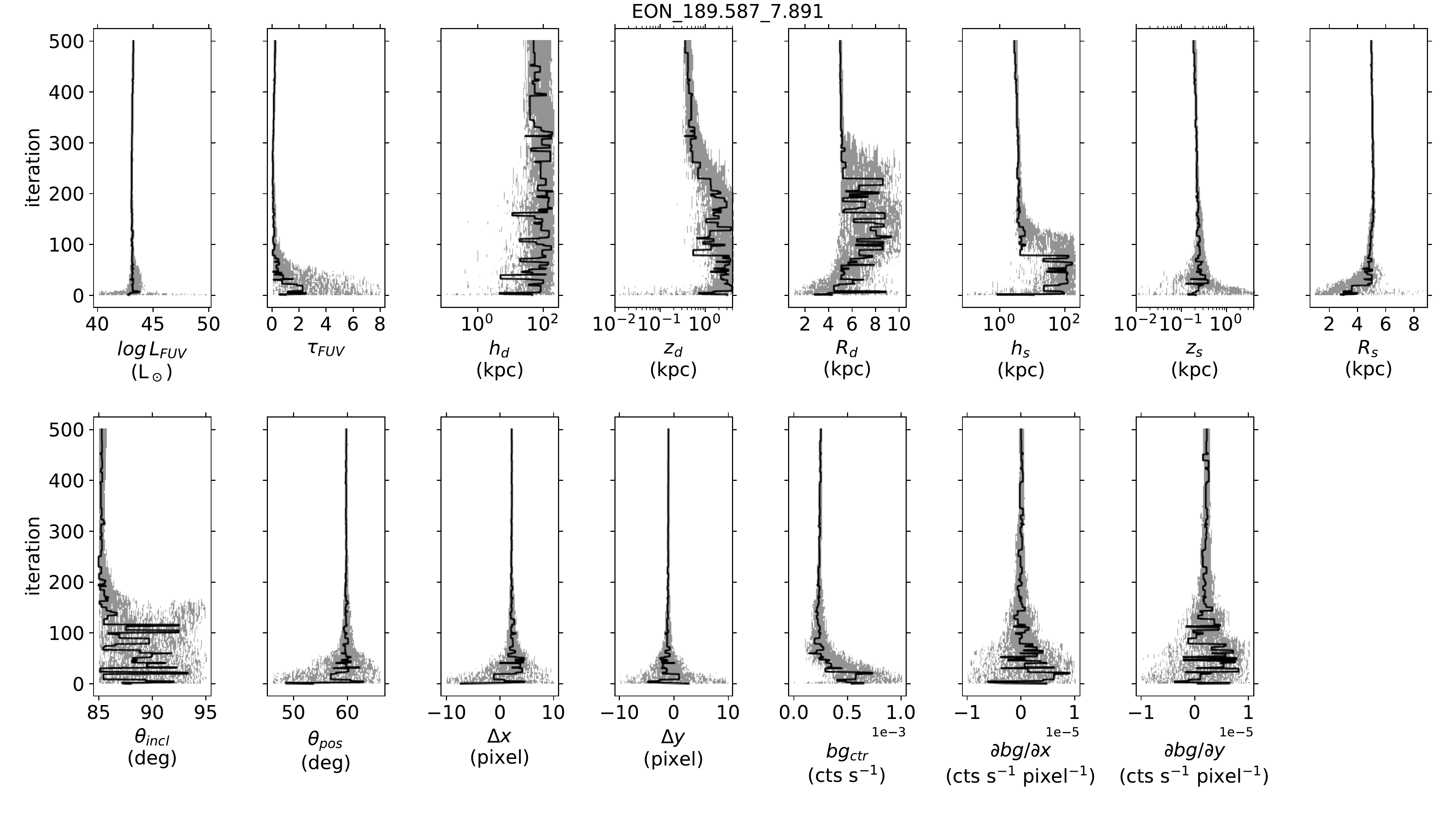}
}
\caption{Fitting results for EON\_189.587\_7.891. \input{caption/fig_fitdata1.tex}} 
\end{figure}

\clearpage
\begin{figure}
\figurenum{20}
\center{
\includegraphics[scale=0.45]{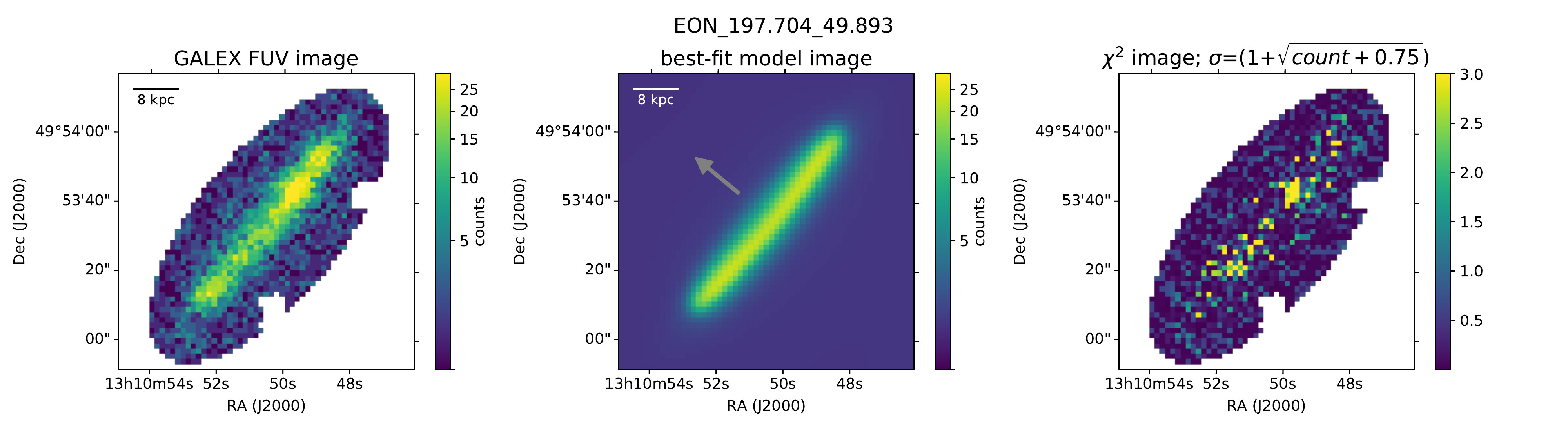}
\includegraphics[scale=0.45]{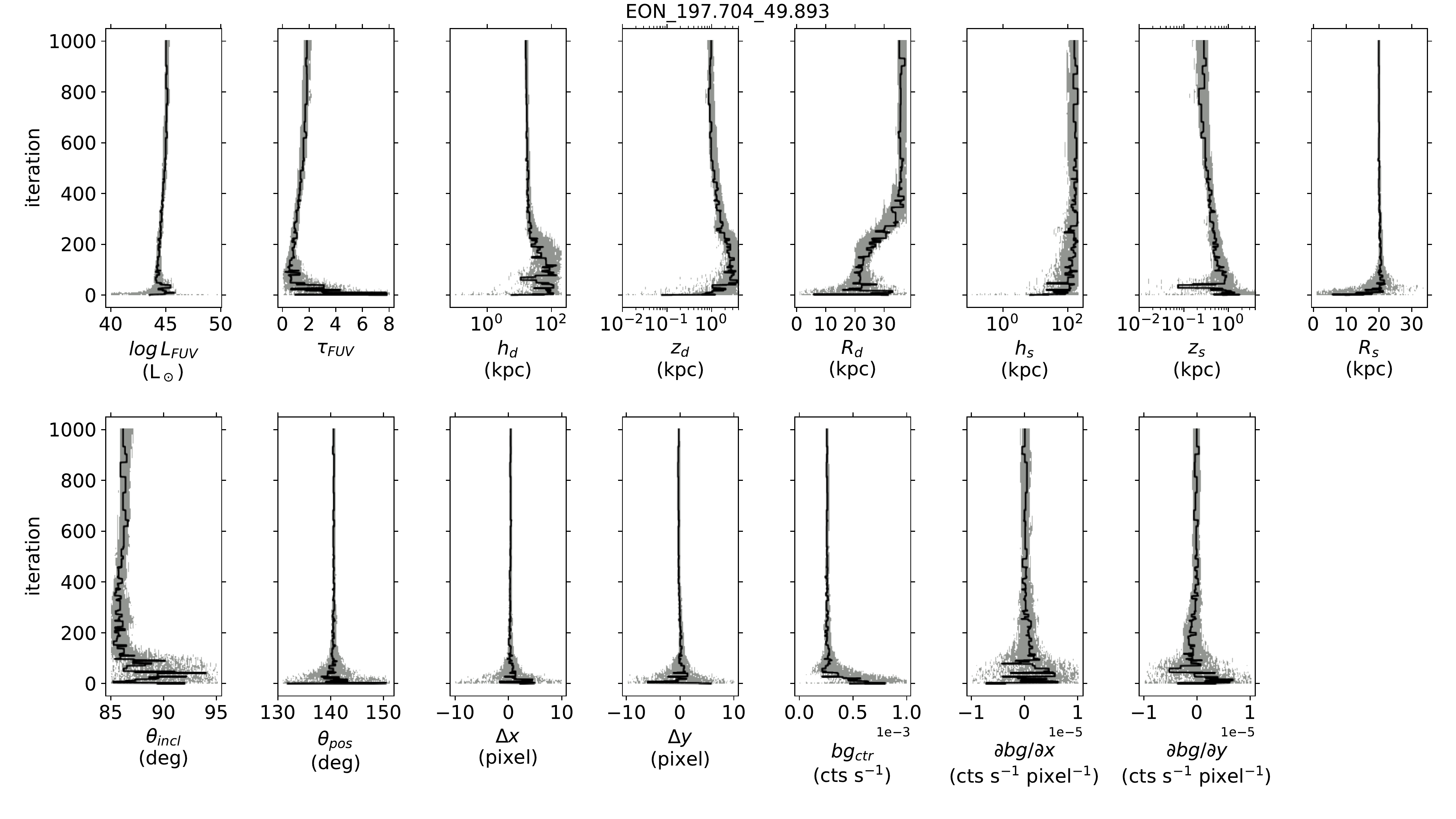}
}
\caption{Fitting results for EON\_197.704\_49.893. \input{caption/fig_fitdata1.tex}} 
\end{figure}

\clearpage
\begin{figure}
\figurenum{21}
\center{
\includegraphics[scale=0.45]{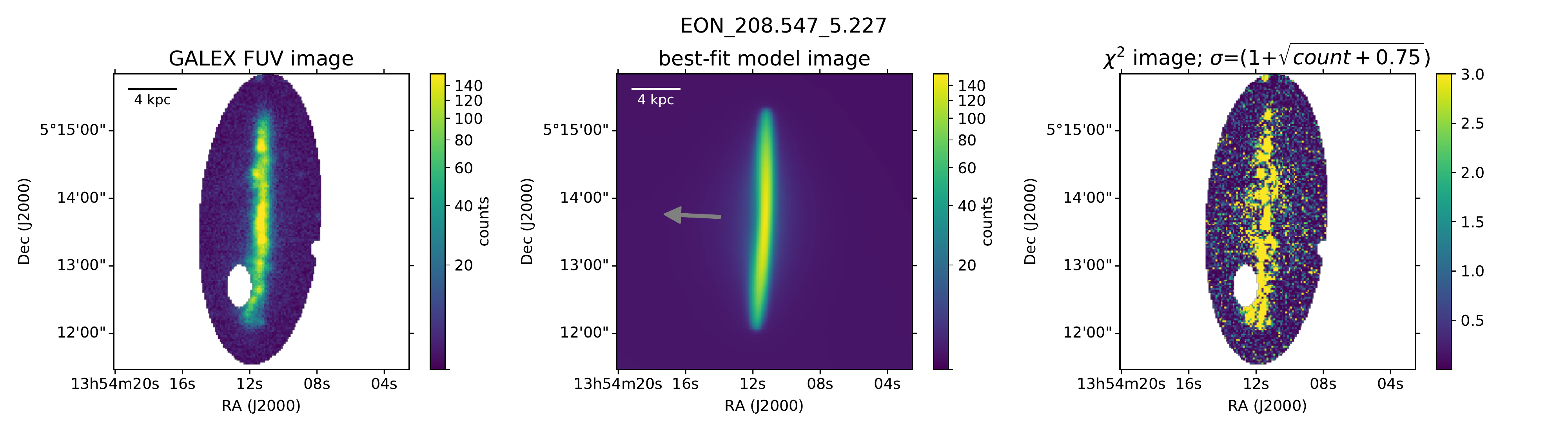}
\includegraphics[scale=0.45]{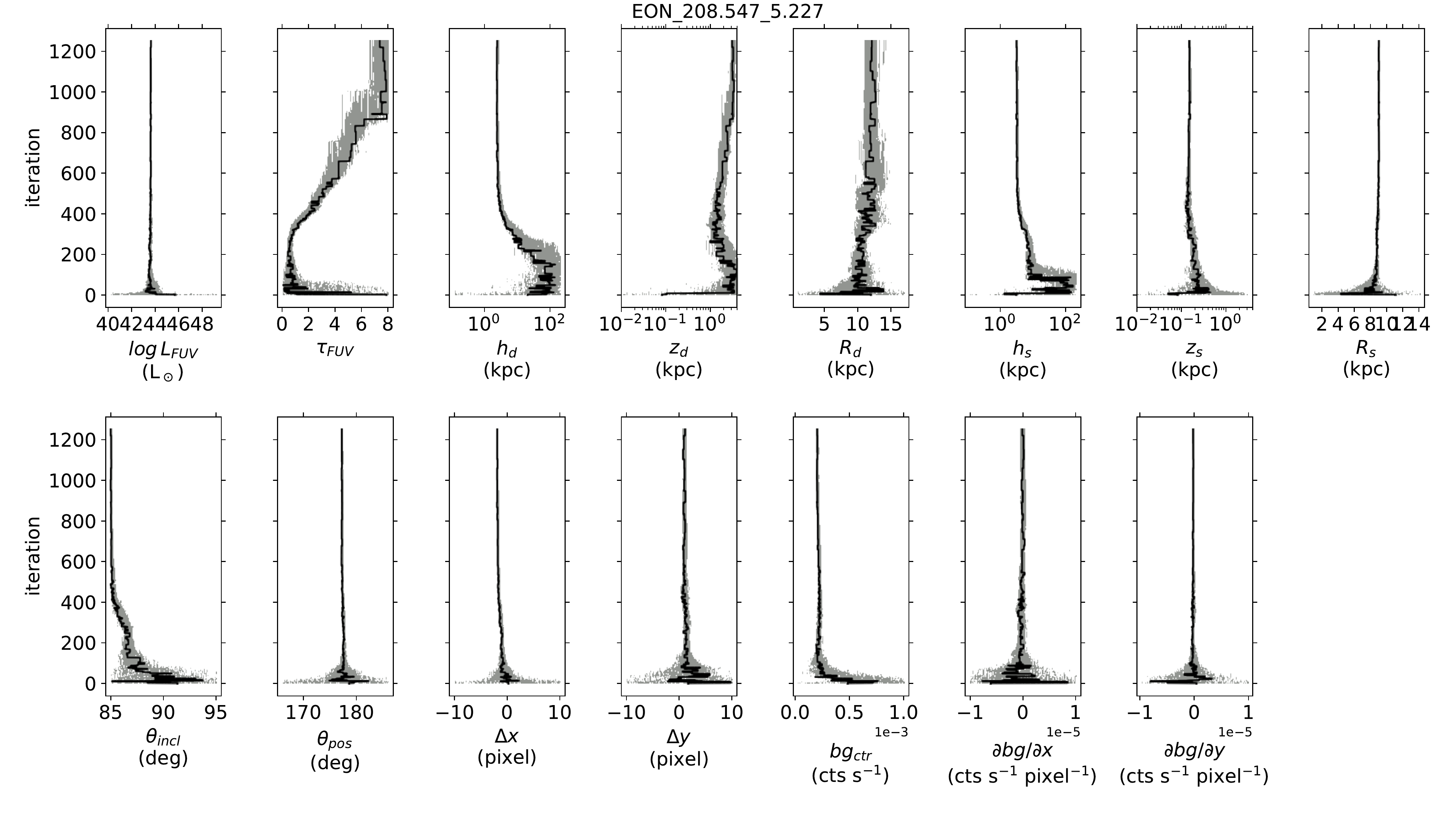}
}
\caption{Fitting results for EON\_208.547\_5.227. \input{caption/fig_fitdata1.tex}} 
\end{figure}

\clearpage
\begin{figure}
\figurenum{22}
\center{
\includegraphics[scale=0.45]{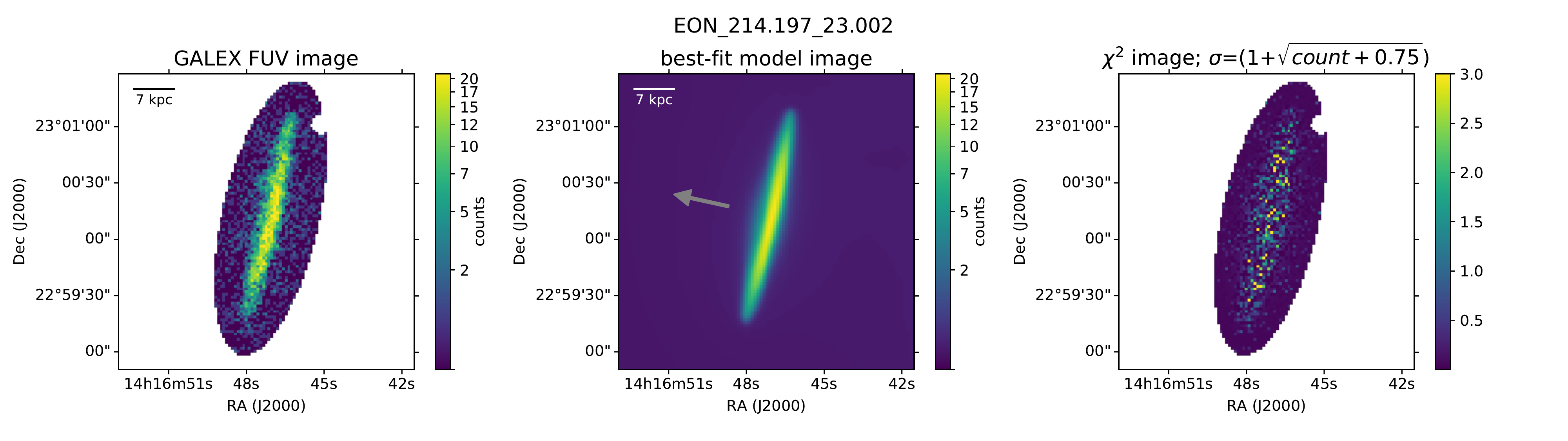}
\includegraphics[scale=0.45]{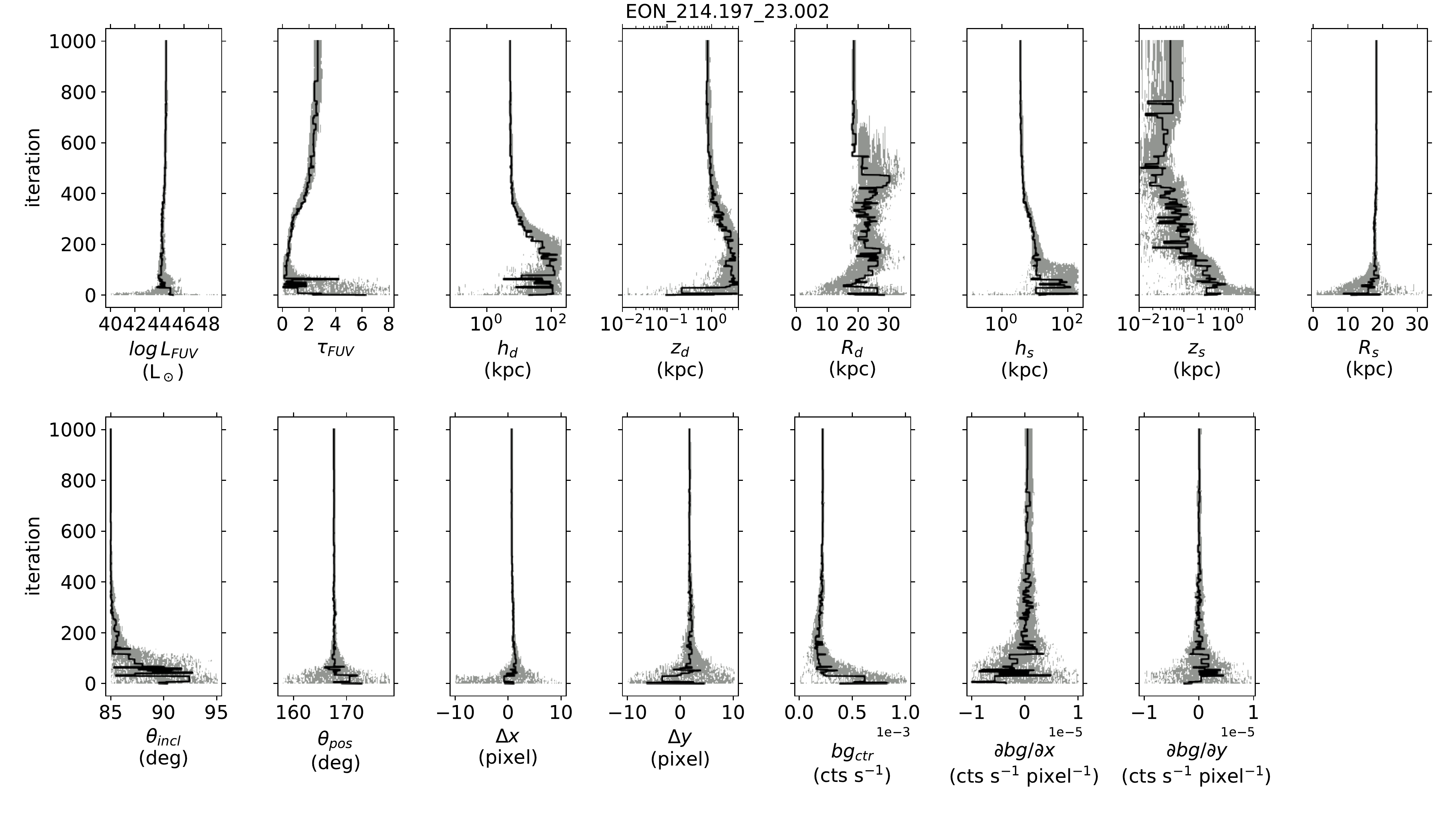}
}
\caption{Fitting results for EON\_214.197\_23.002. \input{caption/fig_fitdata1.tex}} 
\end{figure}

\clearpage
\begin{figure}
\figurenum{23}
\center{
\includegraphics[scale=0.45]{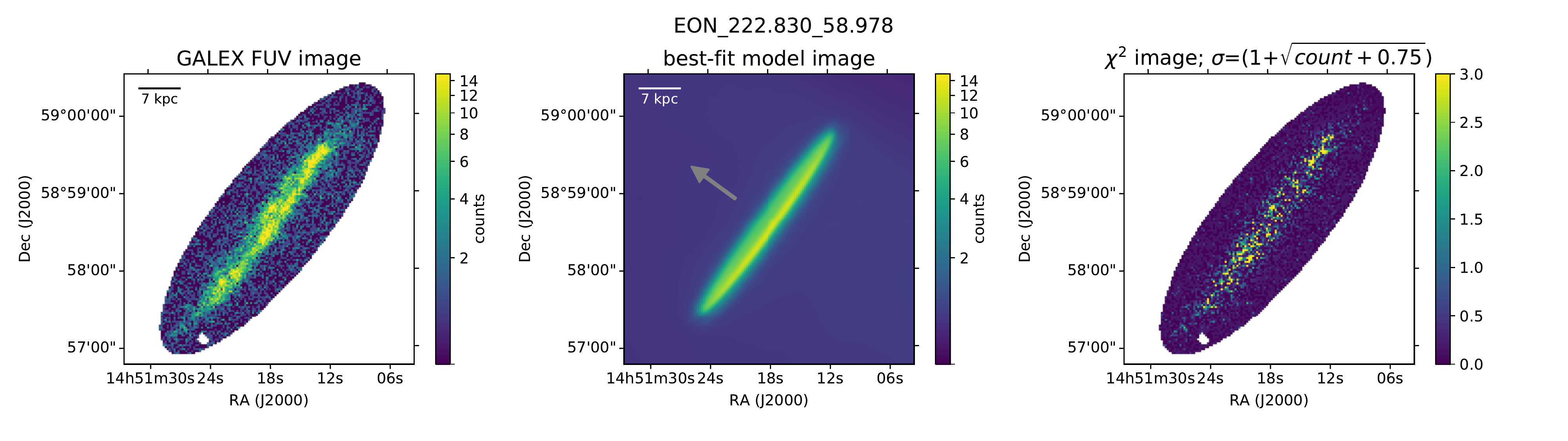}
\includegraphics[scale=0.45]{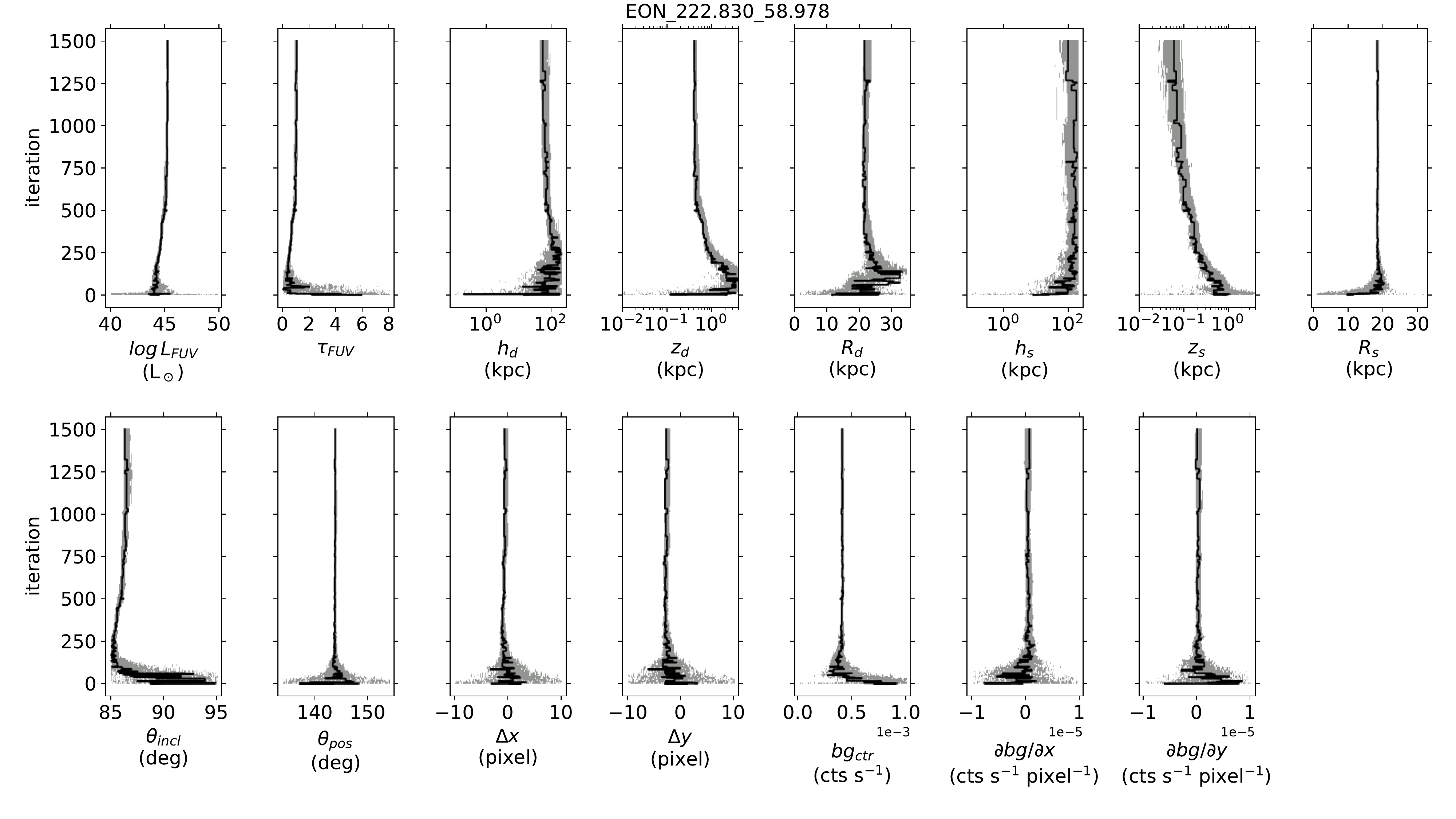}
}
\caption{Fitting results for EON\_222.830\_58.978. \input{caption/fig_fitdata1.tex}} 
\end{figure}

\clearpage
\begin{figure}
\figurenum{24}
\center{
\includegraphics[scale=0.45]{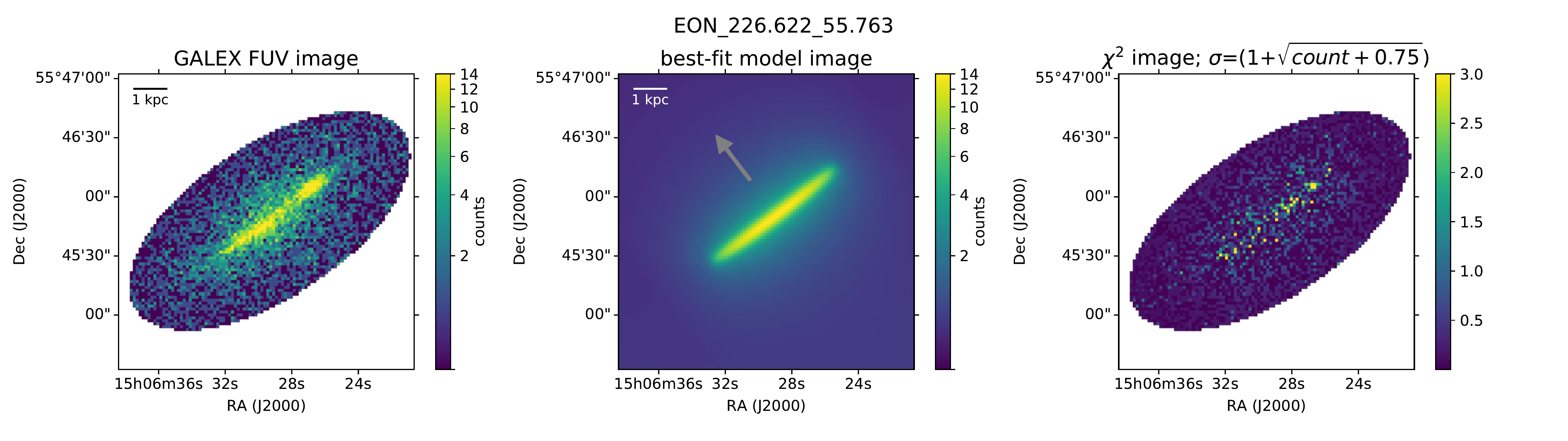}
\includegraphics[scale=0.45]{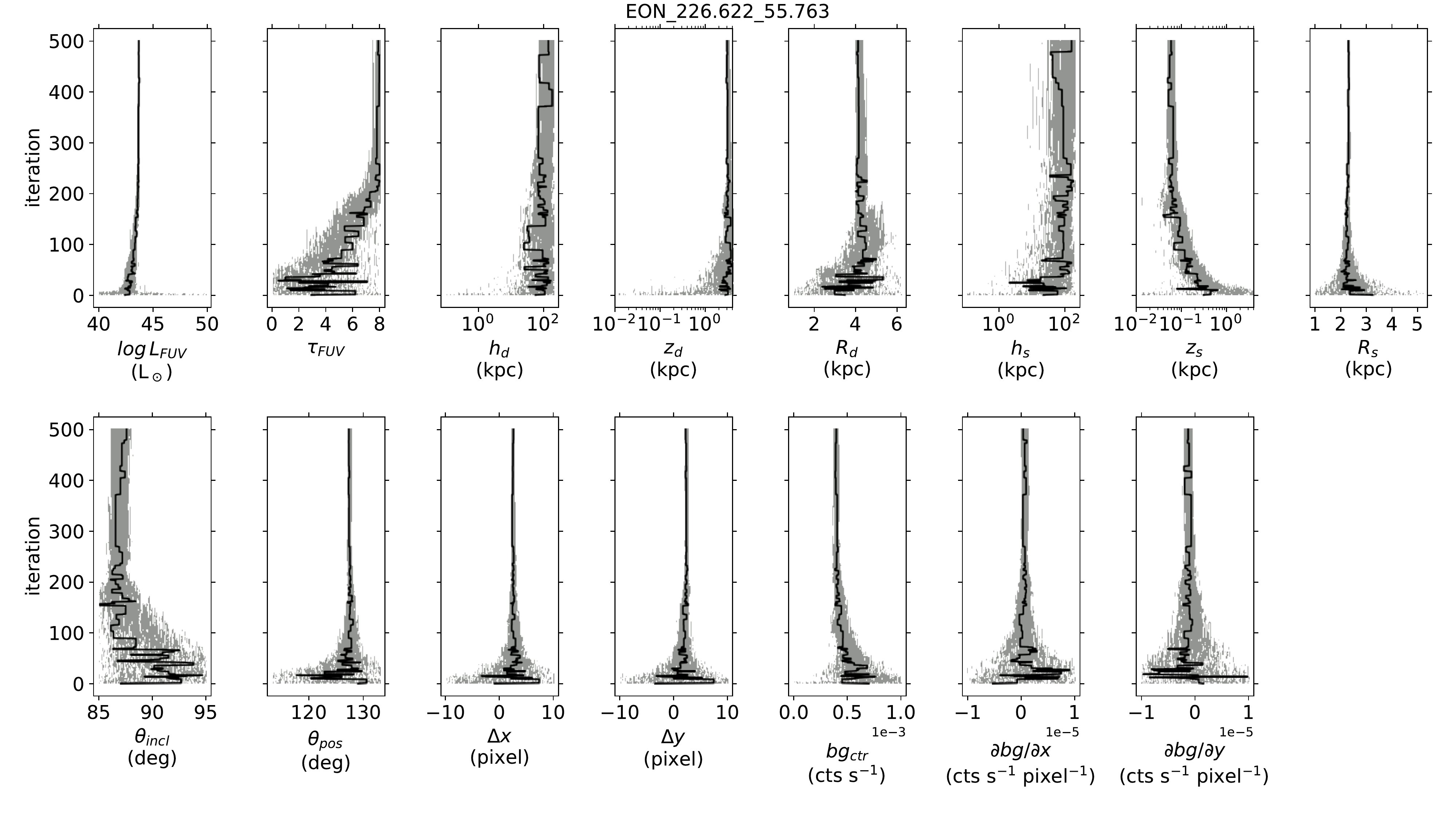}
}
\caption{Fitting results for EON\_226.622\_55.763. \input{caption/fig_fitdata1.tex}} 
\end{figure}

\clearpage
\begin{figure}
\figurenum{25}
\center{
\includegraphics[scale=0.45]{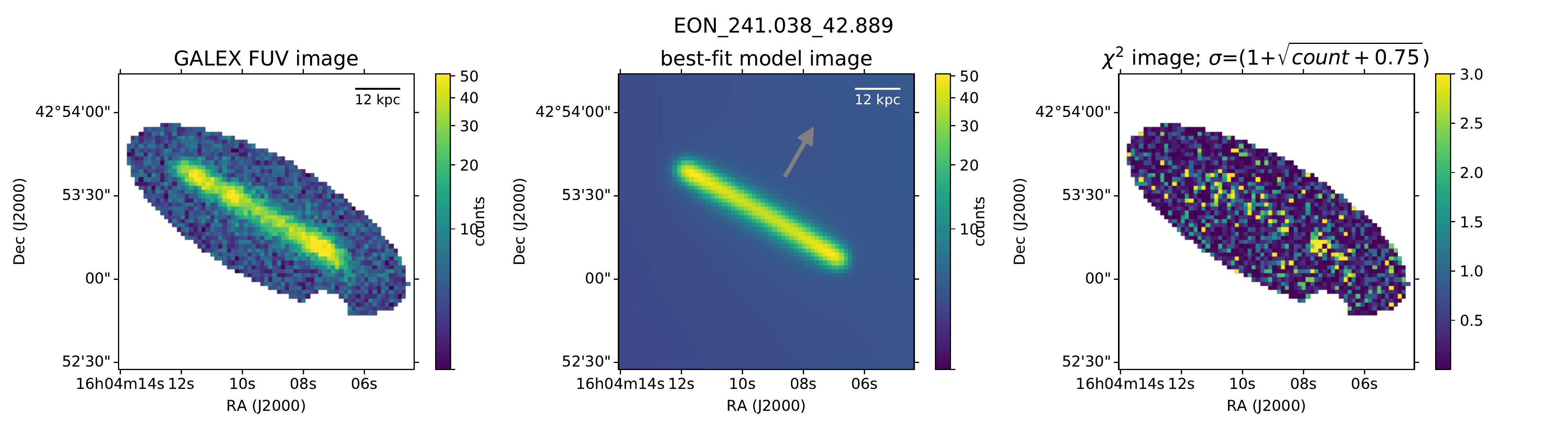}
\includegraphics[scale=0.45]{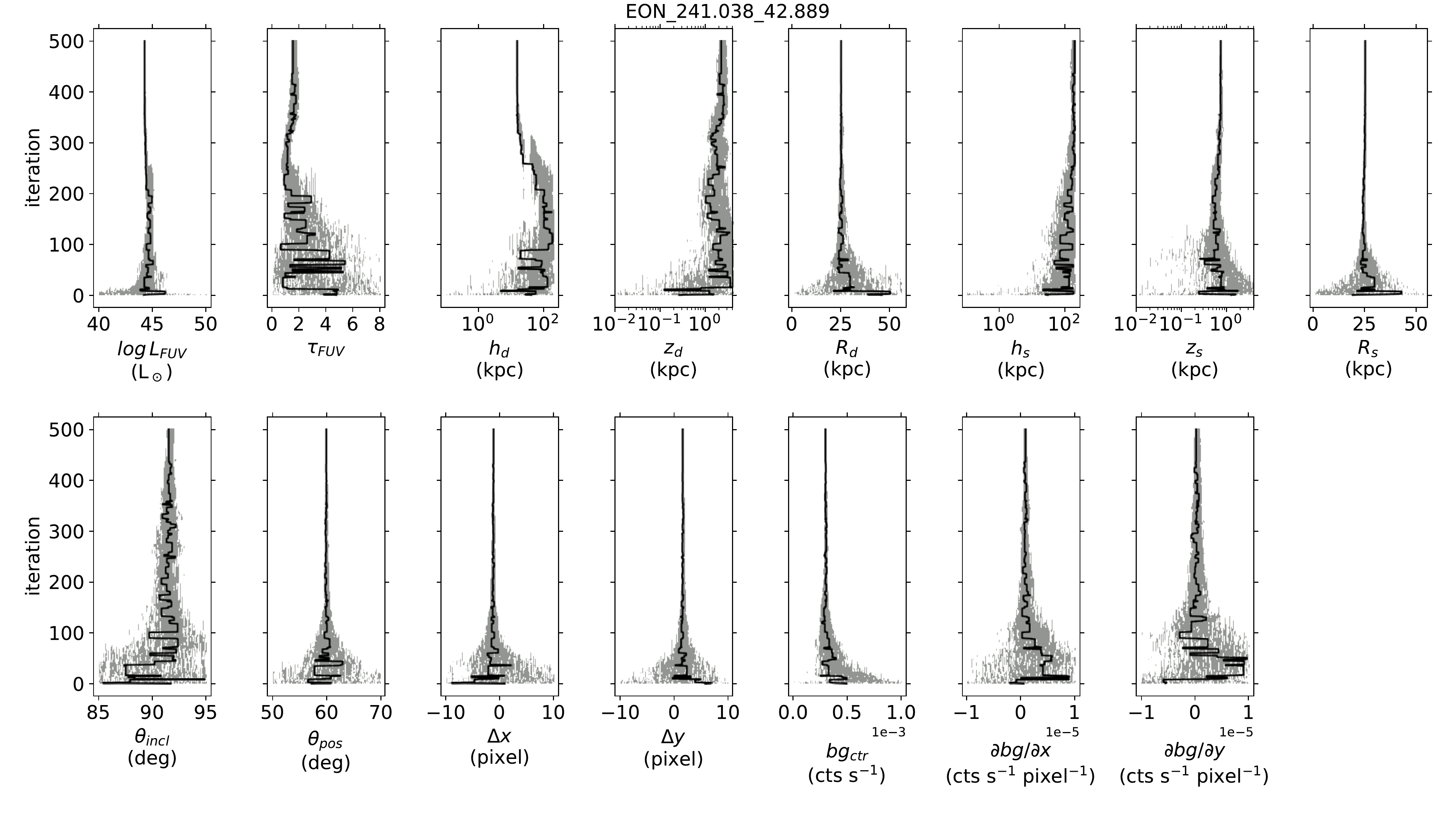}
}
\caption{Fitting results for EON\_241.038\_42.889. \input{caption/fig_fitdata1.tex}} 
\end{figure}

\clearpage
\begin{figure}
\figurenum{26}
\center{
\includegraphics[scale=0.45]{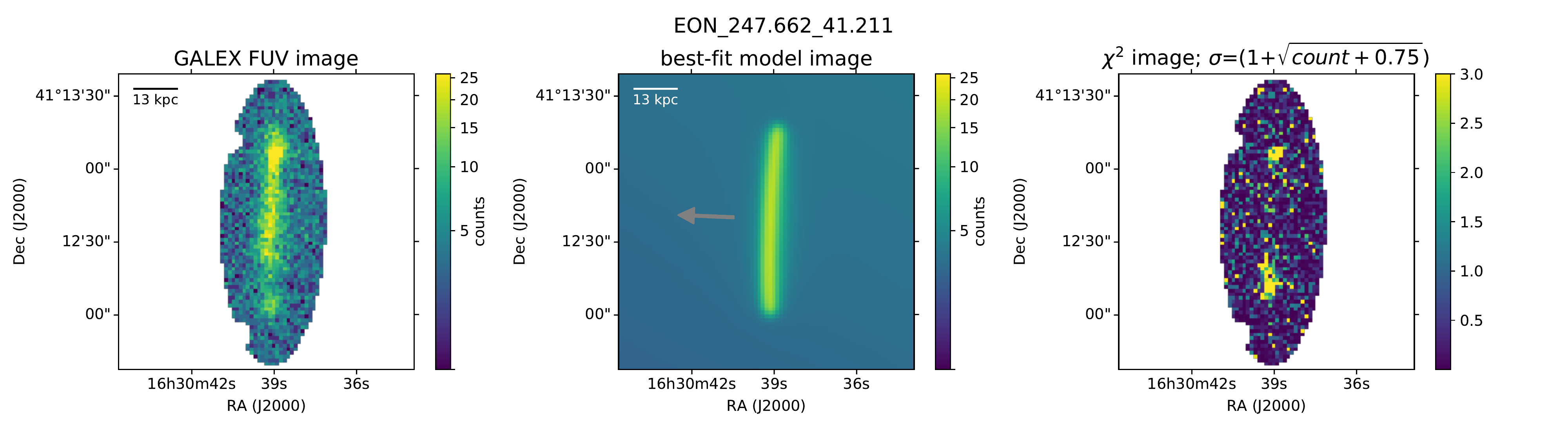}
\includegraphics[scale=0.45]{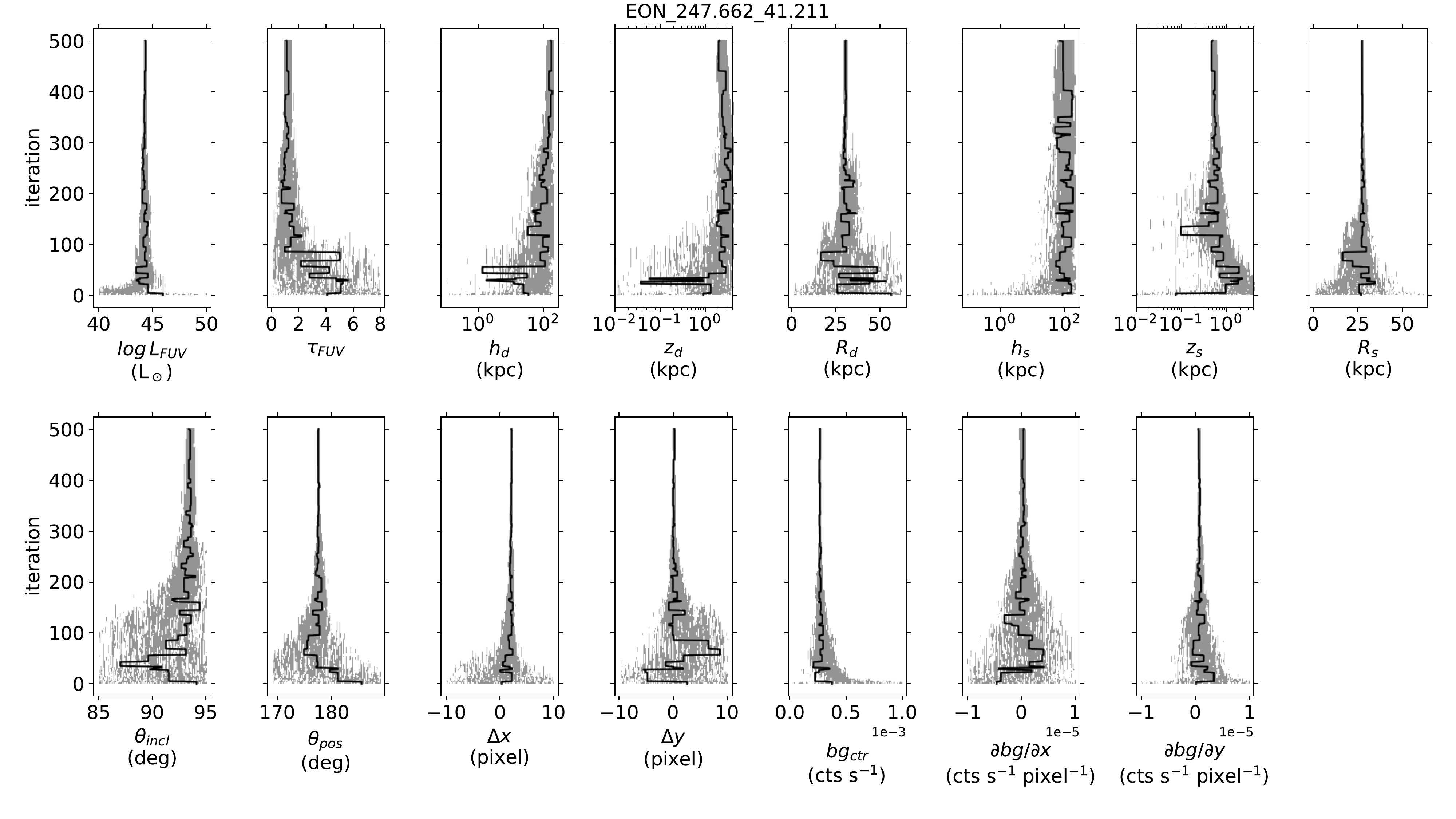}
}
\caption{Fitting results for EON\_247.662\_41.211. \input{caption/fig_fitdata1.tex}} 
\end{figure}

\clearpage
\begin{figure}
\figurenum{27}
\center{
\includegraphics[scale=0.45]{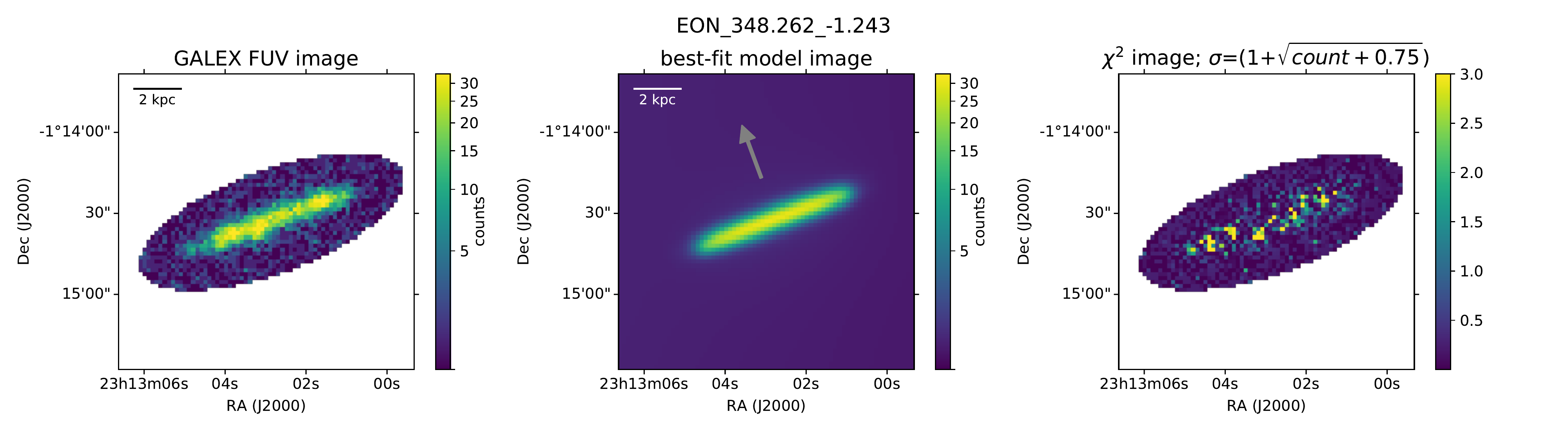}
\includegraphics[scale=0.45]{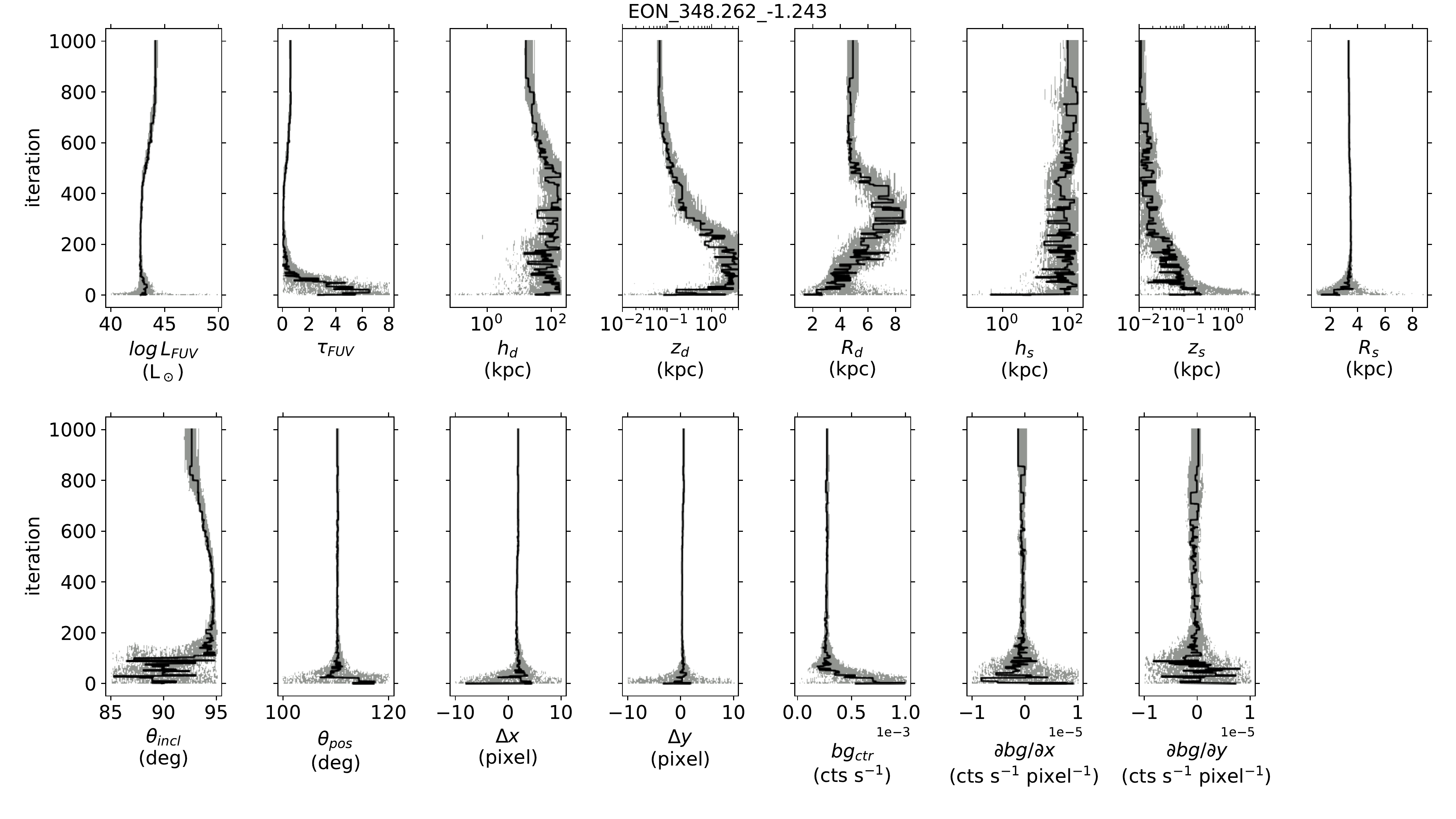}
}
\caption{Fitting results for EON\_348.262\_-1.243. \input{caption/fig_fitdata1.tex}} 
\end{figure}

%% file: caption/fig_fitdata0.tex
The three \emph{top panels} show the \galex{} FUV \emph{count} image, best-fit model image, and the goodness of fit image from left to right, respectively. The colorbar maximum is set as 3 in the goodness of fit image. The white areas are excluded from the model fitting.\added{ The gray arrow on the best-fit model image is to define the inclination angle which increases from the arrow to the direction into the page.} The \emph{middle and bottom panels} show the evolution of fitting parameters over the iteration. The eight \emph{middle panels} are for the galactic structural parameters, while the seven \emph{bottom panels} are for the viewpoints and the background. The \emph{gray points} indicate the parameter values of the population. The \emph{black line} shows the evolution of the best-fit parameter along the iteration. \label{fig-fitdata_st}

%% file: caption/fig_fitdata1.tex
The rest are the same as in Fig.~\ref{fig-fitdata_st} \label{fig-fitdata_en}

%% file: ms_tbl.tex
{\catcode`\&=11
\gdef \Bianchi {\cite{Bianchi_2007_A&A_471_765}}
\gdef \DeGeyter {\cite{DeGeyter_2013_A&A_550_A74}}
\gdef \aXilouris {\cite{Xilouris_1997_A&A_325_135}}
\gdef \bXilouris {\cite{Xilouris_1999_A&A_344_868}}
\gdef \Schechtman {\cite{Schechtman-Rook_2012_ApJ_746_70}}
}

\clearpage
\begin{deluxetable}{llrrrrl}
\tablewidth{0pt}
\tablecaption{Target Edge-on Galaxies with Relevant Information \label{tbl-tg}}
\tablehead{
\colhead{ID} &  \colhead{Other Name} & \colhead{$D_{25}$} & \colhead{$d$} & \colhead{$D_{25,ph}$} & \colhead{Exposure Time} & \colhead{Tile Name} \\
 &  & \colhead{($''$)} & \colhead{(Mpc)} & \colhead{(kpc)} & \colhead{(s)} & \\
\colhead{(1)} &  \colhead{(2)} & \colhead{(3)} & \colhead{(4)} & \colhead{(5)} & \colhead{(6)} & \colhead{(7)}
}
\startdata
\input{tbl/tbl_target.tex}
\enddata
\tablecomments{Columns: (1) ID from the edge-on galaxy catalog of \cite{Bizyaev_2014_ApJ_787_24}, (2) target's other name, (3) major-axis diameter, (4) distance to the target, (5) $D_{25}$ converted to the physical size using the distance $d$, (6) \galex{} FUV exposure time, and (7) \galex{} exposure tile name. Columns (3)-(4) are from the NASA/IPAC Extragalactic Database (\url{https://ned.ipac.caltech.edu/}).}
\end{deluxetable}

\clearpage
\begin{deluxetable}{lrrrrrrrrr}
\tablewidth{0pt}
\rotate
\tablecaption{Best Model Parameters \label{tbl-bestfit}}
\tablehead{
\colhead{ID} &  \colhead{$\log{(L_{FUV}/L_\odot)}$} & \colhead{$\tau_{FUV}$} & \colhead{$h_d$} & \colhead{$z_d$} &  \colhead{$R_d$} & \colhead{$h_s$} & \colhead{$z_s$} & \colhead{$R_s$} &  \colhead{$i$} \\
 & \colhead{} &  & \colhead{(kpc)} & \colhead{(kpc)} & \colhead{(kpc)} & \colhead{(kpc)} & \colhead{(kpc)} & \colhead{(kpc)} & \colhead{($^\circ$)} \\
\colhead{(1)} &  \colhead{(2)} & \colhead{(3)} & \colhead{(4)} &\colhead{(5)} &  \colhead{(6)} & \colhead{(7)} & \colhead{(8)} & \colhead{(9)} &  \colhead{(10)}
}
\startdata
\input{tbl/tbl_bestfit.tex}
\enddata
\tablecomments{Model parameters for the viewpoint and the background, except the inclination, are excluded for simplicity. The errors are set as the standard deviation of the best-fit parameters obtained from three independent fitting runs, in a similar way to \DeGeyter{} and \cite{DeGeyter_2014_MNRAS_441_869}.}
\added{\tablenotetext{a}{The model parameters for these targets should be handled with cautions. The two prominent clumps of emission, seen along the galactic disk at the symmetric positions, might be real emitting clumps rather than the effect of radiative transfer.}}
\end{deluxetable}

\clearpage
\begin{deluxetable}{lrrc}
\tablewidth{0pt}
\tablecaption{Ratio of Scale-height to Galactic Diameter \label{tbl-z_nrm}}
\tablehead{
\colhead{ID} &  \colhead{{\large$\frac{z_{s}}{D_{25,ph}}$}$\times100$} & \colhead{{\large$\frac{z_{d}}{D_{25,ph}}$}$\times100$} & \colhead{symbol\tablenotemark{a}} \\
\colhead{(1)} &  \colhead{(2)} & \colhead{(3)}
}
\startdata
\input{tbl/tbl_zz.tex}
\enddata
\tablenotetext{a}{The corresponding symbol in Fig.~\ref{fig-zz}. $\bullet$ is ``low-group'', while $\circ$ is ``high-group.''}
\added{\tablenotetext{b}{These targets should be handled with cautions since there are some ambiguity in the best-fit model parameters. See Table \ref{tbl-bestfit}.}}
\end{deluxetable}

\clearpage
\begin{deluxetable}{lccccrrrl}
\tablewidth{0pt}
\tablecaption{Ratio of Scale-height to Galactic Diameter from Optical Radiative Transfer Studies \label{tbl-zD_ref}}
\tablehead{ \\
\colhead{Name} & \colhead{{\large$\frac{z_{s}}{D_{25,ph}}$}$\times100$} & \colhead{{\large$\frac{z_{d}}{D_{25,ph}}$}$\times100$} & \colhead{$z_s$} & \colhead{$z_d$} & \colhead{$D_{25,ph}$} & \colhead{$D_{25}$} & \colhead{$d$} & \colhead{Reference} \\
 &  &  & \colhead{(kpc)} & \colhead{(kpc)} & \colhead{(kpc)} & \colhead{($''$)} & \colhead{(Mpc)} \\
\colhead{(1)} &  \colhead{(2)} & \colhead{(3)} & \colhead{(4)} &  \colhead{(5)} & \colhead{(6)} & \colhead{(7)} & \colhead{(8)} & \colhead{(9)}
}
\startdata
\input{tbl/tbl_zD_ref.tex}
\enddata
\tablecomments{Columns: (1) galaxy name, (2) ratio for light source, (3) ratio for dust, (4) light source scale-height, (5) dust scale-height, (6) major-axis diameter in kpc, (7) major-axis diameter in arcsec which is obtained from the NASA/IPAC Extragalactic Database (\url{https://ned.ipac.caltech.edu/}), (8) distance to the galaxy, and (9) the reference which provides $z_s$, $z_d$, and $d$. Columns (2)-(3) are calculated from Columns (4)-(6). For Columns (4)-(5), we quote a weight-averaged mean and its error when multiple estimations are available. Column (6) is calculated from Columns (7)-(8).}
\end{deluxetable}

%% file: tbl/tbl_target.tex
EON\_10.477\_41.954  &FGC 0079             &        60 &       160 &      46.5 &      7418 &           PS\_M31\_MOS07 \\
EON\_12.715\_0.851   &FGC 0096             &        70 &        71 &      24.1 &      3223 &    MISWZS01\_29059\_0266 \\
EON\_17.154\_1.641   &UGC 00711            &       228 &        24 &      26.5 &      1682 &    MISWZS01\_30937\_0266 \\
EON\_20.349\_-1.863  &FGC 0155             &        80 &        41 &      15.9 &      3262 &    MISWZS01\_17362\_0269 \\
EON\_24.788\_-10.504 &MCG -02-05-030       &        80 &        72 &      27.9 &      3268 &   GI1\_013020\_IRASF0136 \\
EON\_123.495\_45.742 &IC 2233              &       281 &        13 &      17.7 &      3207 &             NGA\_NGC2537 \\
EON\_163.623\_17.344 &NGC 3454             &       125 &        27 &      16.4 &      1600 &  GI2\_121006\_LGG225\_PO \\
EON\_168.614\_17.260 &NGC 3592             &       107 &        25 &      13.0 &      3715 &   GI3\_103009\_Abell1204 \\
EON\_183.320\_43.699 &NGC 4183             &       335 &        18 &      29.2 &      1676 &    GI1\_047063\_UGC07271 \\
EON\_184.175\_46.079 &UGC 07301            &       129 &        23 &      14.4 &     13756 &       PS\_NGC4258\_MOS26 \\
EON\_184.392\_22.540 &UGC 07321            &       330 &        18 &      28.8 &      1683 &    GI4\_095024\_UGC07321 \\
EON\_185.427\_14.598 &NGC 4302             &       330 &        21 &      33.6 &     21177 &   GI2\_017001\_J121754p1 \\
EON\_187.066\_9.436  &NGC 4445             &       158 &        18 &      13.8 &      4383 &     GI5\_057003\_NGC4445 \\
EON\_189.100\_40.005 &UGC 07774            &       218 &        21 &      22.2 &      1664 &     GI4\_095042\_UGC7774 \\
EON\_189.587\_7.891  &UGC 07802            &        97 &        23 &      10.8 &      1661 &   GI2\_125005\_AGESstrip \\
EON\_197.704\_49.893 &UGC 08257            &        66 &       126 &      40.3 &      6255 &         WDST\_J1308p4930 \\
EON\_208.547\_5.227  &NGC 5348             &       213 &        19 &      19.6 &     16943 &         PS\_VISTA\_MOS04 \\
EON\_214.197\_23.002 &UGC 09138            &       117 &        65 &      36.9 &      1974 &      GI4\_016007\_DDO187 \\
EON\_222.830\_58.978 &NGC 5777             &       185 &        44 &      39.5 &      2488 &      MISDR1\_10146\_0610 \\
EON\_226.622\_55.763 &NGC 5866             &       281 &        12 &      16.3 &      1526 &             NGA\_NGC5866 \\
EON\_241.038\_42.889 &UGC 10171            &        66 &       152 &      48.6 &     19658 &        HRC\_CLJ1604p4321 \\
EON\_247.662\_41.211 &UGC 10432            &        81 &       147 &      57.7 &     12687 &              ELAISN2\_01 \\
EON\_348.262\_-1.243 &FGC 2468             &        80 &        23 &       8.9 &      3152 &      MISDR1\_29149\_0381 \\

%% file: tbl/tbl_bestfit.tex
EON\_10.477\_41.954  &   44.230$\pm$0.022   &    0.990$\pm$0.049   &  104.903$\pm$18.309  &    2.076$\pm$0.003   &   21.890$\pm$0.170   &   28.347$\pm$1.466   &    0.768$\pm$0.003   &   21.600$\pm$0.036   &   85.907$\pm$0.182   \\
EON\_12.715\_0.851   &   44.730$\pm$0.029   &    0.300$\pm$0.022   &   76.877$\pm$30.903  &    0.294$\pm$0.015   &   14.400$\pm$0.285   &  105.503$\pm$39.169  &    0.022$\pm$0.003   &    9.150$\pm$0.078   &   89.857$\pm$0.119   \\
EON\_17.154\_1.641   &   44.250$\pm$0.156   &    0.217$\pm$0.034   &  183.633$\pm$6.298   &    0.183$\pm$0.014   &   14.610$\pm$0.422   &    4.057$\pm$0.054   &    0.100$\pm$0.012   &   14.283$\pm$0.090   &   88.457$\pm$0.125   \\
EON\_20.349\_-1.863\tablenotemark{a} &   43.073$\pm$0.005   &    5.660$\pm$0.150   &    0.487$\pm$0.005   &    0.164$\pm$0.024   &   13.860$\pm$2.080   &  137.353$\pm$27.286  &    0.197$\pm$0.003   &    6.217$\pm$0.005   &   88.093$\pm$0.141   \\
EON\_24.788\_-10.504\tablenotemark{a} &   43.993$\pm$0.012   &    7.960$\pm$0.016   &    1.420$\pm$0.008   &    0.083$\pm$0.004   &   17.777$\pm$1.392   &    2.940$\pm$0.075   &    0.010$\pm$0.000   &   13.407$\pm$0.012   &   93.583$\pm$0.045   \\
EON\_123.495\_45.742 &   44.580$\pm$0.079   &    0.533$\pm$0.059   &   91.040$\pm$32.152  &    0.205$\pm$0.014   &    9.843$\pm$0.226   &  143.187$\pm$37.819  &    0.091$\pm$0.008   &    8.537$\pm$0.024   &   89.407$\pm$0.090   \\
EON\_163.623\_17.344 &   43.793$\pm$0.029   &    1.660$\pm$0.170   &    4.833$\pm$1.144   &    1.002$\pm$0.067   &    8.280$\pm$0.311   &  147.863$\pm$21.804  &    0.082$\pm$0.006   &    5.613$\pm$0.009   &   86.820$\pm$0.126   \\
EON\_168.614\_17.260 &   43.587$\pm$0.104   &    0.323$\pm$0.033   &  123.300$\pm$10.831  &    0.742$\pm$0.141   &   18.780$\pm$4.370   &    2.127$\pm$0.005   &    0.219$\pm$0.020   &    6.330$\pm$0.024   &   86.153$\pm$0.774   \\
EON\_183.320\_43.699 &   44.480$\pm$0.008   &    0.413$\pm$0.009   &  139.617$\pm$32.116  &    0.504$\pm$0.031   &   22.993$\pm$1.827   &   18.147$\pm$1.519   &    0.173$\pm$0.010   &   11.940$\pm$0.062   &   85.527$\pm$0.021   \\
EON\_184.175\_46.079 &   44.810$\pm$0.091   &    0.143$\pm$0.034   &  154.507$\pm$32.109  &    0.162$\pm$0.007   &   17.983$\pm$3.164   &  132.530$\pm$60.717  &    0.012$\pm$0.001   &    5.317$\pm$0.060   &   89.997$\pm$0.005   \\
EON\_184.392\_22.540 &   45.120$\pm$0.022   &    7.940$\pm$0.037   &  158.907$\pm$21.919  &    1.170$\pm$0.055   &   14.220$\pm$0.157   &  166.520$\pm$11.402  &    0.130$\pm$0.004   &   13.953$\pm$0.117   &   91.060$\pm$0.028   \\
EON\_185.427\_14.598 &   43.400$\pm$0.008   &    7.840$\pm$0.102   &    1.757$\pm$0.059   &    0.652$\pm$0.087   &   16.427$\pm$0.154   &   16.883$\pm$2.552   &    0.465$\pm$0.005   &   13.223$\pm$0.097   &   89.960$\pm$0.043   \\
EON\_187.066\_9.436  &   43.970$\pm$0.051   &    1.603$\pm$0.099   &  160.400$\pm$11.501  &    0.110$\pm$0.002   &    4.427$\pm$0.005   &    0.807$\pm$0.005   &    0.071$\pm$0.003   &    3.597$\pm$0.005   &   87.003$\pm$0.103   \\
EON\_189.100\_40.005 &   43.657$\pm$0.066   &    0.177$\pm$0.037   &  126.160$\pm$33.766  &    0.134$\pm$0.073   &   12.383$\pm$0.026   &    4.663$\pm$0.065   &    0.226$\pm$0.013   &   10.857$\pm$0.026   &   85.027$\pm$0.031   \\
EON\_189.587\_7.891  &   43.230$\pm$0.014   &    0.250$\pm$0.014   &   65.120$\pm$21.432  &    0.393$\pm$0.018   &    5.003$\pm$0.026   &    2.950$\pm$0.094   &    0.190$\pm$0.002   &    4.993$\pm$0.012   &   85.323$\pm$0.012   \\
EON\_197.704\_49.893 &   45.197$\pm$0.131   &    1.837$\pm$0.098   &   16.977$\pm$0.682   &    0.846$\pm$0.118   &   35.133$\pm$0.878   &  146.487$\pm$15.682  &    0.195$\pm$0.073   &   20.043$\pm$0.040   &   86.470$\pm$0.191   \\
EON\_208.547\_5.227  &   43.667$\pm$0.005   &    4.263$\pm$2.221   &    2.970$\pm$0.332   &    2.056$\pm$0.756   &   12.877$\pm$0.524   &    3.403$\pm$0.163   &    0.137$\pm$0.010   &    9.003$\pm$0.034   &   85.070$\pm$0.045   \\
EON\_214.197\_23.002 &   44.547$\pm$0.045   &    2.273$\pm$0.304   &    6.180$\pm$0.587   &    0.826$\pm$0.055   &   17.733$\pm$0.662   &    3.877$\pm$0.256   &    0.040$\pm$0.011   &   17.563$\pm$0.429   &   85.007$\pm$0.005   \\
EON\_222.830\_58.978 &   45.253$\pm$0.019   &    1.113$\pm$0.077   &   44.593$\pm$7.967   &    0.410$\pm$0.018   &   21.847$\pm$0.152   &  121.420$\pm$14.962  &    0.046$\pm$0.017   &   18.507$\pm$0.068   &   86.487$\pm$0.095   \\
EON\_226.622\_55.763 &   43.700$\pm$0.008   &    7.910$\pm$0.022   &  129.710$\pm$30.591  &    3.170$\pm$0.142   &    4.280$\pm$0.167   &  136.867$\pm$28.485  &    0.061$\pm$0.002   &    2.317$\pm$0.005   &   87.280$\pm$0.361   \\
EON\_241.038\_42.889 &   44.327$\pm$0.031   &    1.470$\pm$0.259   &   24.327$\pm$12.541  &    2.235$\pm$0.044   &   25.287$\pm$0.135   &  192.220$\pm$4.596   &    0.721$\pm$0.028   &   25.283$\pm$0.139   &   91.693$\pm$0.120   \\
EON\_247.662\_41.211 &   44.397$\pm$0.042   &    1.140$\pm$0.073   &  125.353$\pm$46.002  &    1.931$\pm$0.086   &   30.410$\pm$0.792   &   89.987$\pm$17.540  &    0.446$\pm$0.037   &   27.917$\pm$0.316   &   93.260$\pm$0.166   \\
EON\_348.262\_-1.243 &   44.097$\pm$0.037   &    0.603$\pm$0.025   &   36.630$\pm$15.969  &    0.068$\pm$0.004   &    4.810$\pm$0.177   &  130.910$\pm$22.734  &    0.013$\pm$0.001   &    3.373$\pm$0.024   &   92.970$\pm$0.227   \\

%% file: tbl/tbl_zz.tex
EON\_10.477\_41.954  &      1.650$\pm$0.006      &      4.461$\pm$0.006      &    $\circ$ \\
EON\_12.715\_0.851   &      0.091$\pm$0.012      &      1.220$\pm$0.062      &  $\bullet$ \\
EON\_17.154\_1.641   &      0.377$\pm$0.045      &      0.690$\pm$0.053      &    $\circ$ \\
EON\_20.349\_-1.863\tablenotemark{b} &      1.239$\pm$0.019      &      1.031$\pm$0.151      &    $\circ$ \\
EON\_24.788\_-10.504\tablenotemark{b} &      0.036$\pm$0.000      &      0.297$\pm$0.014      &  $\bullet$ \\
EON\_123.495\_45.742 &      0.514$\pm$0.045      &      1.158$\pm$0.079      &    $\circ$ \\
EON\_163.623\_17.344 &      0.501$\pm$0.037      &      6.125$\pm$0.410      &    $\circ$ \\
EON\_168.614\_17.260 &      1.689$\pm$0.154      &      5.721$\pm$1.087      &    $\circ$ \\
EON\_183.320\_43.699 &      0.592$\pm$0.034      &      1.724$\pm$0.106      &    $\circ$ \\
EON\_184.175\_46.079 &      0.083$\pm$0.007      &      1.127$\pm$0.049      &  $\bullet$ \\
EON\_184.392\_22.540 &      0.451$\pm$0.014      &      4.062$\pm$0.191      &    $\circ$ \\
EON\_185.427\_14.598 &      1.384$\pm$0.015      &      1.940$\pm$0.259      &    $\circ$ \\
EON\_187.066\_9.436  &      0.515$\pm$0.022      &      0.798$\pm$0.015      &    $\circ$ \\
EON\_189.100\_40.005 &      1.018$\pm$0.059      &      0.604$\pm$0.329      &    $\circ$ \\
EON\_189.587\_7.891  &      1.756$\pm$0.018      &      3.632$\pm$0.166      &    $\circ$ \\
EON\_197.704\_49.893 &      0.484$\pm$0.181      &      2.098$\pm$0.293      &    $\circ$ \\
EON\_208.547\_5.227  &      0.698$\pm$0.051      &     10.479$\pm$3.853      &    $\circ$ \\
EON\_214.197\_23.002 &      0.108$\pm$0.030      &      2.240$\pm$0.149      &  $\bullet$ \\
EON\_222.830\_58.978 &      0.117$\pm$0.043      &      1.039$\pm$0.046      &  $\bullet$ \\
EON\_226.622\_55.763 &      0.373$\pm$0.012      &     19.388$\pm$0.869      &    $\circ$ \\
EON\_241.038\_42.889 &      1.482$\pm$0.058      &      4.595$\pm$0.090      &    $\circ$ \\
EON\_247.662\_41.211 &      0.773$\pm$0.064      &      3.345$\pm$0.149      &    $\circ$ \\
EON\_348.262\_-1.243 &      0.146$\pm$0.011      &      0.762$\pm$0.045      &  $\bullet$ \\

%% file: tbl/tbl_zD_ref.tex
IC 2098    & 1.48$\pm$0.21        & 0.69$\pm$0.07        & 0.430$\pm$0.060      & 0.200$\pm$0.020      & 29.06      & 125.40     & 47.80      & \cite{DeGeyter_2014_MNRAS_441_869} \\
IC 2461    & 0.41$\pm$0.03        & 0.32$\pm$0.08        & 0.150$\pm$0.010      & 0.120$\pm$0.030      & 36.97      & 140.70     & 54.20      & \cite{DeGeyter_2014_MNRAS_441_869} \\
IC 2531    & 0.95$\pm$0.01        & 0.53$\pm$0.01        & 0.422$\pm$0.006      & 0.236$\pm$0.006      & 44.27      & 415.10     & 22.00      & \bXilouris \\
IC 3203    & 1.58$\pm$0.04        & 0.42$\pm$0.04        & 0.830$\pm$0.020      & 0.220$\pm$0.020      & 52.47      & 90.80      & 119.20     & \cite{DeGeyter_2014_MNRAS_441_869} \\
IC 4225    & 3.73$\pm$0.18        & 1.07$\pm$0.09        & 0.840$\pm$0.040      & 0.240$\pm$0.020      & 22.50      & 62.80      & 73.90      & \cite{DeGeyter_2014_MNRAS_441_869} \\
NGC 3650   & 2.10$\pm$0.07        & 0.51$\pm$0.03        & 0.620$\pm$0.020      & 0.150$\pm$0.010      & 29.49      & 101.90     & 59.70      & \cite{DeGeyter_2014_MNRAS_441_869} \\
NGC 3987   & 2.70$\pm$0.13        & 1.00$\pm$0.05        & 1.080$\pm$0.050      & 0.400$\pm$0.020      & 39.98      & 134.30     & 61.40      & \cite{DeGeyter_2014_MNRAS_441_869} \\
NGC 4013   & 1.15$\pm$0.03        & 0.69$\pm$0.03        & 0.203$\pm$0.006      & 0.123$\pm$0.006      & 17.71      & 314.90     & 11.60      & \bXilouris \\
NGC 4013   & 1.01$\pm$0.37        & 0.68$\pm$0.06        & 0.287$\pm$0.104      & 0.192$\pm$0.016      & 28.40      & 314.90     & 18.60      & \DeGeyter  \\
NGC 4013   & 1.70                 & 0.66                 & 0.376                & 0.145                & 22.14      & 314.90     & 14.50      & \Bianchi   \\
NGC 4175   & 1.48$\pm$0.18        & 1.12$\pm$0.31        & 0.330$\pm$0.040      & 0.250$\pm$0.070      & 22.34      & 109.20     & 42.20      & \cite{DeGeyter_2014_MNRAS_441_869} \\
NGC 4217   & 0.77                 & 1.28                 & 0.199                & 0.331                & 25.80      & 314.90     & 16.90      & \Bianchi   \\
NGC 4302   & 3.17                 & 0.56                 & 1.023                & 0.181                & 32.29      & 329.70     & 20.20      & \Bianchi   \\
NGC 4565   & 0.86                 & 0.26                 & 0.672                & 0.205                & 77.91      & 950.90     & 16.90      & \cite{deLooze_2012_MNRAS_427_2797} \\
NGC 5166   & 1.58$\pm$0.10        & 0.81$\pm$0.14        & 0.660$\pm$0.040      & 0.340$\pm$0.060      & 41.73      & 137.50     & 62.60      & \cite{DeGeyter_2014_MNRAS_441_869} \\
NGC 5529   & 0.80$\pm$0.01        & 0.65$\pm$0.01        & 0.425$\pm$0.007      & 0.345$\pm$0.007      & 53.10      & 370.00     & 29.60      & \bXilouris \\
NGC 5529   & 0.77                 & 0.29                 & 0.581                & 0.218                & 75.16      & 370.00     & 41.90      & \Bianchi   \\
NGC 5746   & 1.51                 & 0.88                 & 0.865                & 0.505                & 57.36      & 444.80     & 26.60      & \Bianchi   \\
NGC 5907   & 0.83$\pm$0.01        & 0.28$\pm$0.01        & 0.333$\pm$0.006      & 0.113$\pm$0.006      & 40.29      & 755.40     & 11.00      & \bXilouris \\
NGC 5908   & 0.88$\pm$0.04        & 0.24$\pm$0.08        & 0.440$\pm$0.020      & 0.120$\pm$0.040      & 50.28      & 194.20     & 53.40      & \cite{DeGeyter_2014_MNRAS_441_869} \\
NGC 5965   & 1.34                 & 0.27                 & 0.969                & 0.194                & 72.36      & 314.90     & 47.40      & \Bianchi   \\
NGC 891    & 1.07$\pm$0.01        & 0.72$\pm$0.02        & 0.400$\pm$0.004      & 0.267$\pm$0.006      & 37.28      & 809.40     & 9.50       & \bXilouris \\
NGC 891    & 1.23$\pm$0.35        & 0.64$\pm$0.27        & 0.460$\pm$0.130      & 0.240$\pm$0.100      & 37.28      & 809.40     & 9.50       & \Schechtman \\
UGC 1082   & 1.63$\pm$0.05        & 1.01$\pm$0.05        & 0.452$\pm$0.013      & 0.278$\pm$0.013      & 27.66      & 154.20     & 37.00      & \bXilouris \\
UGC 12518  & 1.50$\pm$0.26        & 0.72$\pm$0.10        & 0.290$\pm$0.050      & 0.140$\pm$0.020      & 19.34      & 77.30      & 51.60      & \cite{DeGeyter_2014_MNRAS_441_869} \\
UGC 2048   & 1.37$\pm$0.02        & 0.85$\pm$0.01        & 0.933$\pm$0.012      & 0.577$\pm$0.006      & 68.08      & 222.90     & 63.00      & \aXilouris \\
UGC 4136   & 1.81$\pm$0.11        & 0.91$\pm$0.11        & 0.660$\pm$0.040      & 0.330$\pm$0.040      & 36.45      & 82.80      & 90.80      & \cite{DeGeyter_2014_MNRAS_441_869} \\
UGC 4277   & 1.28                 & 0.28                 & 1.104                & 0.245                & 86.56      & 233.40     & 76.50      & \Bianchi   \\
UGC 5481   & 1.45$\pm$0.15        & 1.26$\pm$0.22        & 0.590$\pm$0.060      & 0.510$\pm$0.090      & 40.63      & 92.90      & 90.20      & \cite{DeGeyter_2014_MNRAS_441_869} \\

%% file: ms.bbl
\begin{thebibliography}{}
\expandafter\ifx\csname natexlab\endcsname\relax\def\natexlab#1{#1}\fi
\providecommand{\url}[1]{\href{#1}{#1}}
\providecommand{\dodoi}[1]{doi:~\href{http://doi.org/#1}{\nolinkurl{#1}}}
\providecommand{\doeprint}[1]{\href{http://ascl.net/#1}{\nolinkurl{http://ascl.net/#1}}}
\providecommand{\doarXiv}[1]{\href{https://arxiv.org/abs/#1}{\nolinkurl{https://arxiv.org/abs/#1}}}

\bibitem[{Abazajian {et~al.}(2009)Abazajian, Adelman-McCarthy, Ag{\"u}eros,
  Allam, Allende~Prieto, An, Anderson, Anderson, Annis, Bahcall, Bailer-Jones,
  Barentine, Bassett, Becker, Beers, Bell, Belokurov, Berlind, Berman,
  Bernardi, Bickerton, Bizyaev, Blakeslee, Blanton, Bochanski, Boroski,
  Brewington, Brinchmann, Brinkmann, Brunner, Budav{\'a}ri, Carey, Carliles,
  Carr, Castander, Cinabro, Connolly, Csabai, Cunha, Czarapata, Davenport,
  de~Haas, Dilday, Doi, Eisenstein, Evans, Evans, Fan, Friedman, Frieman,
  Fukugita, G{\"a}nsicke, Gates, Gillespie, Gilmore, Gonzalez, Gonzalez,
  Grebel, Gunn, Gy{\"o}ry, Hall, Harding, Harris, Harvanek, Hawley, Hayes,
  Heckman, Hendry, Hennessy, Hindsley, Hoblitt, Hogan, Hogg, Holtzman, Hyde,
  Ichikawa, Ichikawa, Im, Ivezi{\'c}, Jester, Jiang, Johnson, Jorgensen,
  Juri{\'c}, Kent, Kessler, Kleinman, Knapp, Konishi, Kron, Krzesinski,
  Kuropatkin, Lampeitl, Lebedeva, Lee, Lee, French~Leger, L{\'e}pine, Li, Lima,
  Lin, Long, Loomis, Loveday, Lupton, Magnier, Malanushenko, Malanushenko,
  Mandelbaum, Margon, Marriner, Mart{\'{\i}}nez-Delgado, Matsubara, McGehee,
  McKay, Meiksin, Morrison, Mullally, Munn, Murphy, Nash, Nebot, Neilsen,
  Newberg, Newman, Nichol, Nicinski, Nieto-Santisteban, Nitta, Okamura,
  Oravetz, Ostriker, Owen, Padmanabhan, Pan, Park, Pauls, Peoples, Percival,
  Pier, Pope, Pourbaix, Price, Purger, Quinn, Raddick, Re~Fiorentin, Richards,
  Richmond, Riess, Rix, Rockosi, Sako, Schlegel, Schneider, Scholz, Schreiber,
  Schwope, Seljak, Sesar, Sheldon, Shimasaku, Sibley, Simmons, Sivarani,
  Allyn~Smith, Smith, Smol{\v c}i{\'c}, Snedden, Stebbins, Steinmetz,
  Stoughton, Strauss, SubbaRao, Suto, Szalay, Szapudi, Szkody, Tanaka, Tegmark,
  Teodoro, Thakar, Tremonti, Tucker, Uomoto, Vanden~Berk, Vandenberg, Vidrih,
  Vogeley, Voges, Vogt, Wadadekar, Watters, Weinberg, West, White, Wilhite,
  Wonders, Yanny, Yocum, York, Zehavi, Zibetti, \&
  Zucker}]{Abazajian_2009_ApJS_182_543}
Abazajian, K.~N., Adelman-McCarthy, J.~K., Ag{\"u}eros, M.~A., {et~al.} 2009,
  ApJS, 182, 543, \dodoi{10.1088/0067-0049/182/2/543}

\bibitem[{{Baes} \& {Viaene}(2016)}]{Baes_2016_A&A_587_A86}
{Baes}, M., \& {Viaene}, S. 2016, A\&A, 587, A86,
  \dodoi{10.1051/0004-6361/201527812}

\bibitem[{{Bahcall} \& {Soneira}(1980)}]{Bahcall_1980_ApJS_44_73}
{Bahcall}, J.~N., \& {Soneira}, R.~M. 1980, ApJS, 44, 73,
  \dodoi{10.1086/190685}

\bibitem[{Bianchi {et~al.}(2014)Bianchi, Conti, \&
  Shiao}]{Bianchi_2014_AdSpR_53_900}
Bianchi, L., Conti, A., \& Shiao, B. 2014, AdSpR, 53, 900,
  \dodoi{10.1016/j.asr.2013.07.045}

\bibitem[{{Bianchi}(2007)}]{Bianchi_2007_A&A_471_765}
{Bianchi}, S. 2007, A\&A, 471, 765, \dodoi{10.1051/0004-6361:20077649}

\bibitem[{{Bizyaev} {et~al.}(2014){Bizyaev}, {Kautsch}, {Mosenkov},
  {Reshetnikov}, {Sotnikova}, {Yablokova}, \&
  {Hillyer}}]{Bizyaev_2014_ApJ_787_24}
{Bizyaev}, D.~V., {Kautsch}, S.~J., {Mosenkov}, A.~V., {et~al.} 2014, ApJ, 787,
  24, \dodoi{10.1088/0004-637X/787/1/24}

\bibitem[{{Blanton} \& {Moustakas}(2009)}]{Blanton_2009_ARA&A_47_159}
{Blanton}, M.~R., \& {Moustakas}, J. 2009, ARA\&A, 47, 159,
  \dodoi{10.1146/annurev-astro-082708-101734}

\bibitem[{Cash(1979)}]{Cash_1979_ApJ_228_939}
Cash, W. 1979, ApJ, 228, 939, \dodoi{10.1086/156922}

\bibitem[{{Conroy}(2013)}]{Conroy_2013_ARA&A_51_393}
{Conroy}, C. 2013, ARA\&A, 51, 393, \dodoi{10.1146/annurev-astro-082812-141017}

\bibitem[{{De Geyter} {et~al.}(2014){De Geyter}, {Baes}, {Camps}, {Fritz}, {De
  Looze}, {Hughes}, {Viaene}, \& {Gentile}}]{DeGeyter_2014_MNRAS_441_869}
{De Geyter}, G., {Baes}, M., {Camps}, P., {et~al.} 2014, MNRAS, 441, 869,
  \dodoi{10.1093/mnras/stu612}

\bibitem[{{De Geyter} {et~al.}(2013){De Geyter}, {Baes}, {Fritz}, \&
  {Camps}}]{DeGeyter_2013_A&A_550_A74}
{De Geyter}, G., {Baes}, M., {Fritz}, J., \& {Camps}, P. 2013, A\&A, 550, A74,
  \dodoi{10.1051/0004-6361/201220126}

\bibitem[{{de Looze} {et~al.}(2012){de Looze}, {Baes}, {Bendo}, {Ciesla},
  {Cortese}, {de Geyter}, {Groves}, {Boquien}, {Boselli}, {Brondeel}, {Cooray},
  {Eales}, {Fritz}, {Galliano}, {Gentile}, {Gordon}, {Hony}, {Law}, {Madden},
  {Sauvage}, {Smith}, {Spinoglio}, \&
  {Verstappen}}]{deLooze_2012_MNRAS_427_2797}
{de Looze}, I., {Baes}, M., {Bendo}, G.~J., {et~al.} 2012, MNRAS, 427, 2797,
  \dodoi{10.1111/j.1365-2966.2012.22045.x}

\bibitem[{{Freeman} \& {Bland-Hawthorn}(2002)}]{Freeman_2002_ARA&A_40_487}
{Freeman}, K., \& {Bland-Hawthorn}, J. 2002, ARA\&A, 40, 487,
  \dodoi{10.1146/annurev.astro.40.060401.093840}

\bibitem[{{Gehrels}(1986)}]{Gehrels_1986_ApJ_303_336}
{Gehrels}, N. 1986, ApJ, 303, 336, \dodoi{10.1086/164079}

\bibitem[{{Haffner} {et~al.}(2009){Haffner}, {Dettmar}, {Beckman}, {Wood},
  {Slavin}, {Giammanco}, {Madsen}, {Zurita}, \&
  {Reynolds}}]{Haffner_2009_RvMP_81_969}
{Haffner}, L.~M., {Dettmar}, R.-J., {Beckman}, J.~E., {et~al.} 2009, RvMP, 81,
  969, \dodoi{10.1103/RevModPhys.81.969}

\bibitem[{{Hodges-Kluck} \& {Bregman}(2014)}]{Hodges-Kluck_2014_ApJ_789_131}
{Hodges-Kluck}, E., \& {Bregman}, J.~N. 2014, ApJ, 789, 131,
  \dodoi{10.1088/0004-637X/789/2/131}

\bibitem[{Hodges-Kluck {et~al.}(2016)Hodges-Kluck, Cafmeyer, \&
  Bregman}]{Hodges-Kluck_2016_ApJ_833_58}
Hodges-Kluck, E., Cafmeyer, J., \& Bregman, J.~N. 2016, ApJ, 833, 58,
  \dodoi{10.3847/1538-4357/833/1/58}

\bibitem[{{Hollenbach} \& {Tielens}(1999)}]{Hollenbach_1999_RvMP_71_173}
{Hollenbach}, D.~J., \& {Tielens}, A.~G.~G.~M. 1999, RvMP, 71, 173,
  \dodoi{10.1103/RevModPhys.71.173}

\bibitem[{{Howk}(1999)}]{Howk_1999_Ap&SS_269_293}
{Howk}, J.~C. 1999, Ap\&SS, 269, 293, \dodoi{10.1023/A:1017032610206}

\bibitem[{{Howk} \& {Savage}(1997)}]{Howk_1997_AJ_114_2463}
{Howk}, J.~C., \& {Savage}, B.~D. 1997, AJ, 114, 2463, \dodoi{10.1086/118660}

\bibitem[{{Howk} \& {Savage}(1999)}]{Howk_1999_AJ_117_2077}
---. 1999, AJ, 117, 2077, \dodoi{10.1086/300857}

\bibitem[{{Irwin} {et~al.}(2007){Irwin}, {Kennedy}, {Parkin}, \&
  {Madden}}]{Irwin_2007_A&A_474_461}
{Irwin}, J.~A., {Kennedy}, H., {Parkin}, T., \& {Madden}, S. 2007, A\&A, 474,
  461, \dodoi{10.1051/0004-6361:20077729}

\bibitem[{{Irwin} \& {Madden}(2006)}]{Irwin_2006_A&A_445_123}
{Irwin}, J.~A., \& {Madden}, S.~C. 2006, A\&A, 445, 123,
  \dodoi{10.1051/0004-6361:20053233}

\bibitem[{{Kennicutt} \& {Evans}(2012)}]{Kennicutt_2012_ARA&A_50_531}
{Kennicutt}, R.~C., \& {Evans}, N.~J. 2012, ARA\&A, 50, 531,
  \dodoi{10.1146/annurev-astro-081811-125610}

\bibitem[{Martig {et~al.}(2014)Martig, Minchev, \&
  Flynn}]{Martig_2014_MNRAS_442_2474}
Martig, M., Minchev, I., \& Flynn, C. 2014, MNRAS, 442, 2474,
  \dodoi{10.1093/mnras/stu1003}

\bibitem[{{Meiksin}(2009)}]{Meiksin_2009_RvMP_81_1405}
{Meiksin}, A.~A. 2009, RvMP, 81, 1405, \dodoi{10.1103/RevModPhys.81.1405}

\bibitem[{{Morrissey} {et~al.}(2007){Morrissey}, {Conrow}, {Barlow}, {Small},
  {Seibert}, {Wyder}, {Budav{\'a}ri}, {Arnouts}, {Friedman}, {Forster},
  {Martin}, {Neff}, {Schiminovich}, {Bianchi}, {Donas}, {Heckman}, {Lee},
  {Madore}, {Milliard}, {Rich}, {Szalay}, {Welsh}, \&
  {Yi}}]{Morrissey_2007_ApJS_173_682}
{Morrissey}, P., {Conrow}, T., {Barlow}, T.~A., {et~al.} 2007, ApJS, 173, 682,
  \dodoi{10.1086/520512}

\bibitem[{Popescu {et~al.}(2004)Popescu, Tuffs, Kylafis, \&
  Madore}]{Popescu_2004_A&A_414_45}
Popescu, C.~C., Tuffs, R.~J., Kylafis, N.~D., \& Madore, B.~F. 2004, A\&A, 414,
  45, \dodoi{10.1051/0004-6361:20031581}

\bibitem[{{Putman} {et~al.}(2012){Putman}, {Peek}, \&
  {Joung}}]{Putman_2012_ARA&A_50_491}
{Putman}, M.~E., {Peek}, J.~E.~G., \& {Joung}, M.~R. 2012, ARA\&A, 50, 491,
  \dodoi{10.1146/annurev-astro-081811-125612}

\bibitem[{{Sandin}(2014)}]{Sandin_2014_A&A_567_A97}
{Sandin}, C. 2014, A\&A, 567, A97, \dodoi{10.1051/0004-6361/201423429}

\bibitem[{{Sandin}(2015)}]{Sandin_2015_A&A_577_A106}
---. 2015, A\&A, 577, A106, \dodoi{10.1051/0004-6361/201425168}

\bibitem[{Schechtman-Rook {et~al.}(2012)Schechtman-Rook, Bershady, \&
  Wood}]{Schechtman-Rook_2012_ApJ_746_70}
Schechtman-Rook, A., Bershady, M.~A., \& Wood, K. 2012, ApJ, 746, 70,
  \dodoi{10.1088/0004-637X/746/1/70}

\bibitem[{{Seon} \& {Witt}(2012)}]{Seon_2012_ApJ_758_109}
{Seon}, K.-I., \& {Witt}, A.~N. 2012, ApJ, 758, 109,
  \dodoi{10.1088/0004-637X/758/2/109}

\bibitem[{{Seon} {et~al.}(2014){Seon}, {Witt}, {Shinn}, \&
  {Kim}}]{Seon_2014_ApJ_785_L18}
{Seon}, K.-I., {Witt}, A.~N., {Shinn}, J.-H., \& {Kim}, I.-J. 2014, ApJ, 785,
  L18, \dodoi{10.1088/2041-8205/785/1/L18}

\bibitem[{Sharma(2017)}]{Sharma_2017_ARA&A_55_213}
Sharma, S. 2017, ARA\&A, 55, 213, \dodoi{10.1146/annurev-astro-082214-122339}

\bibitem[{{Shinn} \& {Seon}(2015)}]{Shinn_2015_ApJ_815_133}
{Shinn}, J.-H., \& {Seon}, K.-I. 2015, ApJ, 815, 133,
  \dodoi{10.1088/0004-637X/815/2/133}

\bibitem[{{Spitzer}(1942)}]{Spitzer_1942_ApJ_95_329}
{Spitzer}, Jr., L. 1942, ApJ, 95, 329, \dodoi{10.1086/144407}

\bibitem[{{Steinacker} {et~al.}(2013){Steinacker}, {Baes}, \&
  {Gordon}}]{Steinacker_2013_ARA&A_51_63}
{Steinacker}, J., {Baes}, M., \& {Gordon}, K.~D. 2013, ARA\&A, 51, 63,
  \dodoi{10.1146/annurev-astro-082812-141042}

\bibitem[{Storn \& Price(1997)}]{Storn_1997_J.GlobalOptim._11_341}
Storn, R., \& Price, K. 1997, J. Global Optim., 11, 341,
  \dodoi{10.1023/A:1008202821328}

\bibitem[{{Thompson} {et~al.}(2004){Thompson}, {Howk}, \&
  {Savage}}]{Thompson_2004_AJ_128_662}
{Thompson}, T.~W.~J., {Howk}, J.~C., \& {Savage}, B.~D. 2004, AJ, 128, 662,
  \dodoi{10.1086/422485}

\bibitem[{{Veilleux} {et~al.}(2005){Veilleux}, {Cecil}, \&
  {Bland-Hawthorn}}]{Veilleux_2005_ARA&A_43_769}
{Veilleux}, S., {Cecil}, G., \& {Bland-Hawthorn}, J. 2005, ARA\&A, 43, 769,
  \dodoi{10.1146/annurev.astro.43.072103.150610}

\bibitem[{{Wainscoat} {et~al.}(1992){Wainscoat}, {Cohen}, {Volk}, {Walker}, \&
  {Schwartz}}]{Wainscoat_1992_ApJS_83_111}
{Wainscoat}, R.~J., {Cohen}, M., {Volk}, K., {Walker}, H.~J., \& {Schwartz},
  D.~E. 1992, ApJS, 83, 111, \dodoi{10.1086/191733}

\bibitem[{{Xilouris} {et~al.}(1999){Xilouris}, {Byun}, {Kylafis}, {Paleologou},
  \& {Papamastorakis}}]{Xilouris_1999_A&A_344_868}
{Xilouris}, E.~M., {Byun}, Y.~I., {Kylafis}, N.~D., {Paleologou}, E.~V., \&
  {Papamastorakis}, J. 1999, A\&A, 344, 868

\bibitem[{{Xilouris} {et~al.}(1997){Xilouris}, {Kylafis}, {Papamastorakis},
  {Paleologou}, \& {Haerendel}}]{Xilouris_1997_A&A_325_135}
{Xilouris}, E.~M., {Kylafis}, N.~D., {Papamastorakis}, J., {Paleologou}, E.~V.,
  \& {Haerendel}, G. 1997, A\&A, 325, 135

\bibitem[{{York} {et~al.}(2000){York}, {Adelman}, {Anderson}, {Anderson},
  {Annis}, {Bahcall}, {Bakken}, {Barkhouser}, {Bastian}, {Berman}, {Boroski},
  {Bracker}, {Briegel}, {Briggs}, {Brinkmann}, {Brunner}, {Burles}, {Carey},
  {Carr}, {Castander}, {Chen}, {Colestock}, {Connolly}, {Crocker}, {Csabai},
  {Czarapata}, {Davis}, {Doi}, {Dombeck}, {Eisenstein}, {Ellman}, {Elms},
  {Evans}, {Fan}, {Federwitz}, {Fiscelli}, {Friedman}, {Frieman}, {Fukugita},
  {Gillespie}, {Gunn}, {Gurbani}, {de Haas}, {Haldeman}, {Harris}, {Hayes},
  {Heckman}, {Hennessy}, {Hindsley}, {Holm}, {Holmgren}, {Huang}, {Hull},
  {Husby}, {Ichikawa}, {Ichikawa}, {Ivezi{\'c}}, {Kent}, {Kim}, {Kinney},
  {Klaene}, {Kleinman}, {Kleinman}, {Knapp}, {Korienek}, {Kron}, {Kunszt},
  {Lamb}, {Lee}, {Leger}, {Limmongkol}, {Lindenmeyer}, {Long}, {Loomis},
  {Loveday}, {Lucinio}, {Lupton}, {MacKinnon}, {Mannery}, {Mantsch}, {Margon},
  {McGehee}, {McKay}, {Meiksin}, {Merelli}, {Monet}, {Munn}, {Narayanan},
  {Nash}, {Neilsen}, {Neswold}, {Newberg}, {Nichol}, {Nicinski}, {Nonino},
  {Okada}, {Okamura}, {Ostriker}, {Owen}, {Pauls}, {Peoples}, {Peterson},
  {Petravick}, {Pier}, {Pope}, {Pordes}, {Prosapio}, {Rechenmacher}, {Quinn},
  {Richards}, {Richmond}, {Rivetta}, {Rockosi}, {Ruthmansdorfer}, {Sandford},
  {Schlegel}, {Schneider}, {Sekiguchi}, {Sergey}, {Shimasaku}, {Siegmund},
  {Smee}, {Smith}, {Snedden}, {Stone}, {Stoughton}, {Strauss}, {Stubbs},
  {SubbaRao}, {Szalay}, {Szapudi}, {Szokoly}, {Thakar}, {Tremonti}, {Tucker},
  {Uomoto}, {Vanden Berk}, {Vogeley}, {Waddell}, {Wang}, {Watanabe},
  {Weinberg}, {Yanny}, {Yasuda}, \& {SDSS
  Collaboration}}]{York_2000_AJ_120_1579}
{York}, D.~G., {Adelman}, J., {Anderson}, Jr., J.~E., {et~al.} 2000, AJ, 120,
  1579, \dodoi{10.1086/301513}

\end{thebibliography}
